\documentclass[sigplan, 10pt, natbib=false]{acmart}
\usepackage{mypackage}

\begin{document}
%-------------------------------------------------------------------------------

%don't want date printed
\date{}

\title[\sysname{}]{\sysname{}: A Preemptive Buffer Management for On-chip Shared-memory Switches}

\author{
    Danfeng Shan$^1$, Yunguang Li$^1$, Jinchao Ma$^1$, Zhenxing Zhang$^2$, Zeyu Liang$^1$, Xinyu Wen$^1$,\\
    Hao Li$^1$, Wanchun Jiang$^3$, Nan Li$^2$, Fengyuan Ren$^4$
    \\$^1${\it Xi'an Jiaotong University}
    \hspace{0.1in} $^2${\it Huawei}
    \hspace{0.1in} $^3${\it Central South University}
    \hspace{0.1in} $^4${\it Tsinghua University}
}
\renewcommand{\shortauthors}{Danfeng Shan et al.}
\begin{abstract}
    % v1.1
    % Today's high-speed switches employ on-chip packet buffer for low access latency
    % and make it globally shared for higher buffer utilization.
    Today's high-speed switches employ an on-chip shared packet buffer.
    The buffer is becoming increasingly insufficient
    as it cannot scale with the growing switching capacity.
    Nonetheless, the buffer needs to face highly intense bursts
    and meet stringent performance requirements for datacenter applications. 
    % In recent years, the link speed and switching capacity have been dramatically increasing.
    % However, the buffer size cannot scale with the network bandwidth.
    % As a result, the buffer becomes increasingly insufficient.
    % The inadequate buffer capacity, highly intense bursts, stringent performance requirements.
    This imposes rigorous demand on the Buffer Management (BM) scheme,
    which dynamically allocates the buffer across queues.
    However, the de facto BM scheme, designed over two decades ago,
    is ill-suited to meet the requirements of today's network.

    % In this paper, we rethink how to design BM schemes
    % under today's network environments.
    In this paper, we argue that shallow-buffer switches, intense bursts,
    along with dynamic traffic
    call for a \emph{highly agile} BM
    that can quickly adjust the buffer allocation as traffic changes.
    However, the agility of the current BM is fundamentally limited by its \emph{non-preemptive nature}.
    Nonetheless, we find that preemptive BM, considered unrealizable in history,
    is now feasible on modern switch chips.
    We propose \sysname{}\footnotemark, a preemptive BM that can quickly adjust buffer allocation.
    \sysname{} utilizes the redundant memory bandwidth
    to actively reclaim and reallocate the over-allocated buffer.
    % \sysname{} mainly harnesses the existing switch features
    % and does not embody sophisticated mechanisms.
    % We have implemented \sysname{} with P4 and DPDK.
    Testbed experiments and large-scale simulations show that
    \sysname{} can improve the end-to-end performance by up to $\sim$55\%.
    % in terms of burst absorption, performance isolation, and buffer choking mitigation.
    % approaching the performance of the optimal BM.

    \footnotetext{Named after a fantastic beast in the Wizarding World
    that can quickly grow or shrink to fit the available space.}

    % v1.0
    % Today's high-speed switches employ on-chip shared buffers
    % for higher buffer utilization.
    % In recent years, the link speed and switching capacity have been dramatically increasing.
    % However, the buffer size cannot scale with the network bandwidth.
    % As a result, the buffer becomes increasingly insufficient.
    % Buffer Management (BM) scheme, which dynamically allocates the buffer across queues,
    % is the key to the performance of buffer sharing.
    % However, the de facto BM scheme, designed two decades ago,
    % is ill-suited for today's network.

    % In this paper, we rethink how to design BM schemes
    % under the today's network environments.
    % We argue that the extreme shallow-buffer switch and intense bursts
    % call for a highly adaptive BM.
    % However, the adaptivity of the current BM scheme is bounded by the port speed.
    % To break through this limitation,
    % We propose \sysname{},
    % which can achieve highly adaptive buffer sharing by packet vacation.
    % \sysname{} is simple in which
    % it mainly harnesses the existing switch features
    % and does not embody sophisticated mechanisms.
    % We have implemented \sysname{} with P4 and DPDK.
    % Testbed experiments and large-scale simulations show that
    % \sysname{} can absorb 68.2\% more bursts
    % and improve the end-to-end performance by up to 6.4$\times$,
    % approaching the optimal performance.
\end{abstract}

\maketitle

\section{Introduction}\label{intr}
% The buffer is becoming increasingly insufficient today.
To support the ever-growing datacenter traffic,
switching capacity has rapidly increased over the past decade,
roughly doubling every two years~\cite{BroadcomDCSwitchChip21, BroadcomTomahawk5Ship22}.
By 2024, leading switch ASIC vendors
had increased their switching capacity to 51.2Tbps~\cite{BroadcomTomahawk5Ship23, CiscoSiliconOneG202Announce, NVIDIASpectrum4Announce, MarvellTeralynx10Announce}.
To enable fast packet access at such high speeds,
the packet buffers are embedded onto the switch chip~\cite{BroadcomSmartBuffer, CiscoSiliconOneG202, NVIDIASpectrum4Datasheet}.
As a result, the buffer size is limited by the chip area
and does not scale with the increasing switching capacity.
% However, the size of switch buffers does not scale with the increasing switching capacity.
Over the last decade, the buffer size (relative to capacity) has shrunk
by a factor of 4$\times$~\cite{INFOCOM20BCC, ICNP21FlashPass, NSDI22BFC, ICDCS23PFC}.
Consequently, buffers are becoming increasingly insufficient
and this trend inclines to continue.

Despite the deficiency of buffer space,
the performance requirements and traffic characteristics of datacenter networks (DCNs)
place stringent demands on packet buffers.
Modern networked applications require ultra-low latency
(\eg~<10$\mu$s for resource disaggregation~\cite{CACM17Latency, OSDI16DDC})
and high throughput~(\eg~>100Gbps for NLP model training~\cite{ASPLOS23Optimus-CC}) from the network.
Additionally, workflows such as partition-aggregation and MapReduce can
aggregate thousands of flows at a single point~\cite{SIGCOMM20Swift},
resulting in highly intense traffic bursts.
% Novel end-to-end transport protocols~\cite{SIGCOMM19HPCC, SIGCOMM20Swift, NSDI22PowerTCP, NSDI23Poseidon}
% can keep persistent queue length low.
Moreover, most bursts occur at a sub-RTT level~\cite{IMC17Burst, IMC22Burst},
and are beyond the control of end-to-end mechanisms.
To avoid packet loss, enough buffer space is required to absorb these bursts.
Hence, the buffer on the switch is a key factor in performance.
% Recently, this phenomenon has been noticed by many literatures~\cite{SIGCOMM10DCTCP, INFOCOM20BCC}.

% As the buffer size is fundamentally unscalable,
% the only way to performance assurance is
% to improve the buffer architecture and buffer management.
Modern datacenter switches have multiple ports,
and each port contains several queues,
each of which is dedicated to a traffic class~\cite{BCM88800TM, JuniperQFXTM}.
The on-chip buffer is typically shared across all queues
to improve buffer utilization~\cite{BroadcomSmartBuffer, BroadcomTomahawk5, CiscoSiliconOneG202, Arista7050X3, NVIDIASpectrum4}.
The Buffer Management (BM) scheme,
which dynamically allocates the shared buffer to queues,
is expected to be efficient for burst absorption~\cite{IMC22Burst} and high throughput~\cite{INFOCOM20BCC, TON21BCC},
as well as fair for performance isolation~\cite{SIGCOMM22ABM}.
However, the de facto BM (\ie~Dynamic Threshold or DT~\cite{INFOCOM96DT, TON98DT}),
designed over two decades ago, is far from satisfactory
in terms of both burst absorption~\cite{IMC17Burst, INFOCOM15Burst, BS19FAB, IMC22Burst}
and performance isolation~\cite{FB, SIGCOMM22ABM}.

In this paper, we argue that
the underlying cause of these problems is that
the \emph{non-preemptive nature} of the current BM
confines its agility to adjust buffer allocations when facing dynamic traffic.
By non-preemption, we mean that a previously accepted packet
is not allowed to be expelled.
As a result,
the BM can only adjust the buffer allocation by
passively waiting for the over-allocated queue to naturally release the buffer
(\ie~by sending out the packets).
This incurs two issues.

(1)~Inefficient:
Due to the limited agility in adjusting buffer allocations,
non-preemptive BMs have to proactively reserve a portion of free buffer space
to protect newly active queues from starving for buffer.
As a result, non-preemptive BMs fail to efficiently utilize the scarce buffer,
which is undesirable given that datacenter switch buffers are already insufficient
and need to be fully utilized to absorb bursts
when facing bursty short flows~\cite{SIGCOMM15Jupiter, IMC22Burst}
and achieve high throughput
when facing long flows~\cite{INFOCOM20BCC, TON21BCC, ICNP21FlashPass}.

(2)~Unfair:
In DCN, the traffic arrival rate can be much higher than the departure rate
as a result of highly bursty traffic.
In such cases, non-preemptive BMs fail to adjust the buffer allocation in time,
resulting in \emph{anomalous behaviors}
where packets are reluctantly dropped before obtaining the deserved buffer.
In particular,
we find that non-preemptive BM can suffer from the \emph{buffer choking problem} (\textsection\ref{sec:motv:problem}),
in which high-priority (and usually delay-sensitive) traffic thirsts for buffer,
whereas low-priority traffic holds considerable over-allocated buffer
while draining slowly.

If BM could support preemption
(\ie~actively drop packets residing in the buffer),
to some extent, the above issues can be easily addressed.
Preemptive BMs had been studied a lot in history,
and a kind of BM called Pushout
had been proven optimal in terms of throughput and loss probability~\cite{1981Pushout,
TCOM84Pushout, GLOBECOM91Pushout, JSAC91Pushout, GLOBECOM93Pushout, TON94Pushout, JHSN94Pushout, JSAC95Pushout, CC96Pushout}.
Pushout accepts the arriving packets whenever there is free buffer space,
and when the buffer becomes full,
Pushout expels packets from the longest queue
to make room for the incoming packets.
However, implementing Pushout was historically considered challenging
within the traditional architecture of switching fabric~\cite{CC96Pushout, TON98DT}
(details in \textsection\ref{sec:back:bm}).
Nonetheless,
over the past two decades,
the buffer architecture has evolved a lot.
Notably, the packet buffer has been embedded into the switch chip,
significantly increasing the memory bandwidth.
Thus, it is necessary to ask:
\emph{is it now feasible to support preemptive operations
in today's on-chip shared-memory switch?}

In this paper, we answer the above question with \sysname{},
a simple preemptive BM that
can quickly adjust buffer allocation by actively expelling packets for the over-allocated queues
(\textsection\ref{sec:design}).
\sysname{} has two key differences from Pushout, which makes it simple to be implemented.
(1) Unlike Pushout whose enqueue operations may need to wait for the completion of expelling packets,
\sysname{} decouples packet expulsion from packet enqueue,
keeping enqueue operations simple.
(2) Unlike Pushout that drops packets from the longest queue,
\sysname{} expels packets from all over-allocated queues in a round-robin manner,
avoiding the prohibitive costs of tracking the longest queue.
% Furthermore, to avoid buffer starvation while keeping it agile,

Combining the above two ideas,
we design \sysname{} with two components: a proactive component and a reactive component.
The proactive component can avoid buffer starvation while achieving high buffer efficiency.
As the enqueue operations do not wait for packet expulsion,
\sysname{} proactively reserves a \emph{small} fraction of free buffer
to protect newly active queues from starving for buffer.
The reactive component endows \sysname{} with agility.
When traffic changes,
\sysname{} reactively adjusts the buffer allocation
by actively expelling packets for the over-allocated queues.
The proactive component can be realized
by directly utilizing the off-the-shelf BM in commodity switch chips
with adjusted parameters,
while the reactive component can be achieved by several simple circuit components.

We implement the core components of \sysname{} with 286 lines of Verilog code,
and evaluate its hardware cost with both the Vivado Design Suite~\cite{Vivado}
and an open-source 45nm ASIC technology library~\cite{45nmASICLib}.
The results show that \sysname{} requires only $\sim$1300 LUTs, $\sim$50 Flip-Flops,
less than 0.03$mm^2$ ASIC area, and consumes just 1$mW$ power.
Additionally, we implement a proof-of-concept hardware prototype using P4  (\textsection\ref{sec:impl:p4})
and a software prototype based on DPDK (\textsection\ref{sec:impl:dpdk}).
% We show that \sysname{} only introduces at most 2.35\% more hardware resources.
% We evaluate \sysname{} with both testbed experiments (\textsection\ref{sec:eval:p4}, \textsection\ref{sec:eval:dpdk}, and \textsection\ref{sec:eval:params})
% and large-scale simulations (\textsection\ref{sec:eval:sim}).
% we show that \sysname{} can achieve near-optimal performance as Pushout.
Experiments on an Intel Tofino switch demonstrate that
\sysname{} improves burst absorption (\ie~maximum burst size without packet drops) by up to $\sim$57\% (\textsection\ref{sec:eval:p4}).
Experiments using a DPDK-based software switch
show that \sysname{} improves the average Query Completion Time (QCT) by up to $\sim$55\%,
while effectively ensuring performance isolation between different traffic classes
and avoiding buffer choking (\textsection\ref{sec:eval:dpdk}).
Furthermore,
large-scale simulations show that \sysname{} reduces the average QCT
by up to $\sim$44\% with web-search workload
and by up to $\sim$48\% with all-reduce traffic (\textsection\ref{sec:eval:sim}).
% by quickly adjusting buffer allocation.

We believe \sysname{} showcases a step towards preemptive BM on modern switch chips.
Since two decades ago, the BM studies have been concentrating on
innovating non-preemptive BMs.
In contrast, our exploration demonstrates the potential and feasibility of utilizing preemption
to approach the ideal performance.
While actually implementing preemptive BM in high-speed commodity switch ASICs
may still require lots of effort,
we hope our exploration can inspire the research and industry community
to embrace the preemptive approaches.
% This work does not raise any ethical issues.

In summary, our key contributions are:
\begin{itemize}[label=$\blacksquare$, leftmargin=*]
    \item We show that, with insufficient buffer and dynamic traffic,
        today's DCN requires a highly agile BM,
        while the agility of the current BM is
        fundamentally limited by its non-preemptive nature.
    \item We design \sysname{},
        showing that preemptive BMs are feasible to be realized
        with several simple components.
    \item We use extensive experiments to show that \sysname{}
        outperforms non-preemptive BMs in terms of
        burst absorption and flow completion time.
\end{itemize}

The code of \sysname{} is available at \url{https://github.com/ants-xjtu/Occamy}.

\section{Background}\label{sec:back}
In this section, we provide background on buffer architecture
and buffer management schemes in modern switches.

\begin{figure}[!t]
    \centering
    \resizebox{\linewidth}{!}{\begin{tikzpicture}[font=\Large]
    \tikzset{
        every node/.style={
            text=dadiannaotextgray,
        },
        data label/.style={
            text=dadiannaolinegray, font=\Large\it
        },
        control block/.style={
            fill=none, draw=black, rectangle,
            text=black, font=\Large\bf,
        },
        admission block/.style={
            control block,
        },
        vacation block/.style={
            control block, fill=dadiannaored,
        },
        pd/.style={
            draw=darkgray, rectangle, minimum width=3mm, minimum height=5mm,
            font=\normalsize, very thick,
        },
        pd head/.style={
            draw, circle, minimum width=1mm, very thick,
        },
        pd tail/.style={
            draw, circle, minimum width=1mm, very thick,
        },
        packet/.style={
            fill=darkgray, rectangle, minimum width=6mm, minimum height=1cm,
            text=white,
            font=\large,
        },
        arbiter/.style={
             trapezium,
             trapezium angle=135,
             trapezium stretches=true,
        },
        demux/.style={
             trapezium,
             trapezium angle=135,
             trapezium stretches=true,
        },
        control path/.style={
            dashed, line width=2pt, dadiannaolinegray,
        },
        data path/.style={
            line width=3pt, dadiannaolinegray,
        },
        pd queue/.style={
            queue, line width=0pt, minimum width=2cm, minimum height=5mm,
        },
    }

    % pds
    \node[pd queue] at (0, 0) (q0) {};
    \node[below=2mm of q0.center] (q1) {};
    % \node[below=2mm of q0.center] (q1) {\bf\huge $\vdots$};
    \node[below=5mm of q1.center, pd queue] (q2) {};
    \foreach \qid in {0, 2} {
        \foreach \pos in {0.2, 0.4, 0.6} {
            \path[line width=2pt] ($(q\qid.north east)!\pos!(q\qid.north west)$) -- ($(q\qid.south east)!\pos!(q\qid.south west)$);
        }
    }

    % pd module border
    % \coordinate (pd-list-ne) at ([xshift=3mm, yshift=3.9mm] q0.north east);
    % \coordinate (pd-list-sw) at ([xshift=-3mm, yshift=-3.9mm] q2.south west);
    \coordinate (pd-list-ne) at (0, 0);
    \coordinate (pd-list-sw) at (-2cm, -2cm);
    \coordinate (pd-list-nw) at (pd-list-ne-|pd-list-sw);
    \coordinate (pd-list-se) at (pd-list-ne|-pd-list-sw);
    \coordinate (pd-list-e) at ($(pd-list-ne)!0.5!(pd-list-se)$);
    \coordinate (pd-list-w) at ($(pd-list-nw)!0.5!(pd-list-sw)$);
    \coordinate (pd-list-n) at ($(pd-list-nw)!0.5!(pd-list-ne)$);
    % \node[below, dadiannaotextgray] at ($(pd-list-sw)!0.5!(pd-list-se)$) {PD Memory (Queues)};
    \node[dadiannaotextgray] at ($(pd-list-sw)!0.5!(pd-list-ne)$) {\bf Packet\\\bf Buffer};

    \draw[data path, <-, line width=2pt] (pd-list-w) -- +(-8mm, 0)
        % node[pos=0.68, above, data label, darkgreen] {PD}
        node[left, admission block, minimum height=2cm] (admission) {Admission};

    % data path
    \begin{scope}[data path, line width=2pt]
        \node[right=9mm of pd-list-e, control block, minimum height=2cm, minimum width=2cm] (dequeue) {Dequeue};
        \draw[->] (pd-list-e) -- (dequeue) node[above, pos=0.32, data label, darkgreen] {};
    \end{scope}

    % control path
    \begin{scope}[control path]
        \path (admission.north) -- (pd-list-n)
            node[midway, above=1cm, control block, minimum height=1.2cm, minimum width=3cm]
                (qlen) {Statistics \\ \normalsize (\eg~queue length)};
        \coordinate (qlen-read-out) at ($(qlen.south)!0.5!(qlen.south west)$);
        \draw[->] (qlen-read-out) -- (qlen-read-out |- admission.north)
            node[left, pos=0.5, data label] {Read};
        \coordinate (qlen-update-in) at ($(qlen.south)!0.5!(qlen.south east)$);
        \draw[<-] (qlen-update-in) -- (qlen-update-in |- pd-list-n)
            node[right, pos=0.5, data label] {Update};
        \node[control block, minimum height=1cm, minimum height=1.2cm] at (qlen-|dequeue) (scheduler) {Scheduler};
        \draw[->] (scheduler) -- (dequeue);
        % \draw[->] (qlen) |- (selection);
        % \draw[->] (selection.south) -- +(0, -1mm) -| ([xshift=-5mm]arbiter.north);
        % \draw[->] (scheduler.south) -- +(0, -1mm) -| ([xshift=+5mm]arbiter.north);
        % \draw[->] (arbiter) -- (pd-list-n);
    \end{scope}

    % % buffer module
    % \coordinate (buffer-nw) at ([yshift=-1cm] admission.south west);
    % \coordinate (buffer-ne) at (buffer-nw -| dequeue.south east);
    % \coordinate (buffer-se) at ([yshift=-2cm] buffer-ne);
    % \coordinate (buffer-sw) at (buffer-nw|-buffer-se);
    % \node[below right, packet] (pkt-1) at ([xshift=2mm, yshift=-2mm] buffer-nw) {};
    % \foreach[remember=\i as \pi (initially 1)] \i in {2, 3, ..., 9} {
    %     \node[packet, right=2mm of pkt-\pi] (pkt-\i) {};
    % }
    % \node[below] at ($(buffer-se)!0.5!(buffer-sw)$) {Packet Buffer (Packet Data)};
    % \begin{scope}[data path, dadiannaodarkblue]
    %     \draw[->] (admission) -- (admission |- buffer-nw) node[pos=0.39, left, data label, text=dadiannaodarkblue] {Packet Data};
    %     \draw[<-] (dequeue) -- (dequeue |- buffer-nw) node[midway, right, data label, text=dadiannaodarkblue] {Packet Data};
    % \end{scope}
    % % TM module
    \coordinate (tm-nw) at ([shift={(-2mm, 3mm)}] admission.west |- qlen.north west);
    \coordinate (tm-se) at ([shift={(2mm, -8mm)}] dequeue.south east);
    % \coordinate (tm-se) at ([shift={(2mm, -6mm)}] buffer-se);
    \coordinate (tm-sw) at (tm-nw|-tm-se);
    \coordinate (tm-ne) at (tm-nw-|tm-se);
    \coordinate (tm-w) at ($(tm-nw)!0.5!(tm-sw)$);
    \coordinate (tm-e) at ($(tm-ne)!0.5!(tm-se)$);
    \coordinate (tm-s) at ($(tm-se)!0.5!(tm-sw)$);
    \coordinate (tm-n) at ($(tm-ne)!0.5!(tm-nw)$);
    \node[above] at (tm-s) {\bf\Large Traffic Manager};
    \begin{scope}[on background layer]
        \fill[dadiannaoyellow, line width=2pt] (tm-nw) rectangle (tm-se);
        \draw[draw=white, fill=gray, fill opacity=0.3, line width=2pt] (pd-list-nw) rectangle (pd-list-se);
        % \draw[draw=white, fill=dadiannaodarkblue, fill opacity=0.3, line width=2pt] (buffer-nw) rectangle (buffer-se);
    \end{scope}
    % ipp
    \coordinate (ipp-ne) at ([xshift=-0.5cm] tm-nw);
    \coordinate (ipp-se) at ([xshift=-0.5cm] tm-sw);
    \coordinate (ipp-nw) at ([xshift=-0.8cm] ipp-ne);
    \coordinate (ipp-sw) at ([xshift=-0.8cm] ipp-se);
    \coordinate (ipp-e) at ($(ipp-se)!0.5!(ipp-ne)$);
    \draw[dadiannaolinegray, line width=1pt] (ipp-ne) rectangle (ipp-sw);
    \node[rotate=-90] at ($(ipp-nw)!0.5!(ipp-se)$) (ipp) {Ingress Packet Processing};
    % epp
    \coordinate (epp-nw) at ([xshift=0.8cm] tm-ne);
    \coordinate (epp-sw) at ([xshift=0.8cm] tm-se);
    \coordinate (epp-ne) at ([xshift=0.8cm] epp-nw);
    \coordinate (epp-se) at ([xshift=0.8cm] epp-sw);
    \coordinate (epp-w) at ($(epp-sw)!0.5!(epp-nw)$);
    \draw[dadiannaolinegray, line width=1pt] (epp-ne) rectangle (epp-sw);
    \node[rotate=-90] at ($(epp-nw)!0.5!(epp-se)$) (epp) {Egress Packet Processing};

    \draw[data path, ->, line width=3.5pt] (dequeue.east) --   ([xshift=-1mm] epp.south|- dequeue.east);
    \draw[data path, <-, line width=3.5pt] (admission.west) -- ([xshift=+1mm] ipp.north |- admission.west);

    \draw[control path, line width=1pt, <-] ([xshift=-1mm, yshift=3mm] tm-ne) -- +(-5mm, 0) node[left] (control path label) {control path};
    \draw[data path,    line width=1pt, <-] ([xshift=-5mm] control path label.west) -- +(-5mm, 0) node[left] {data path};
\end{tikzpicture}}
    \caption{Structure of on-chip shared-memory switch}\label{fig:tm-structure}
\end{figure}
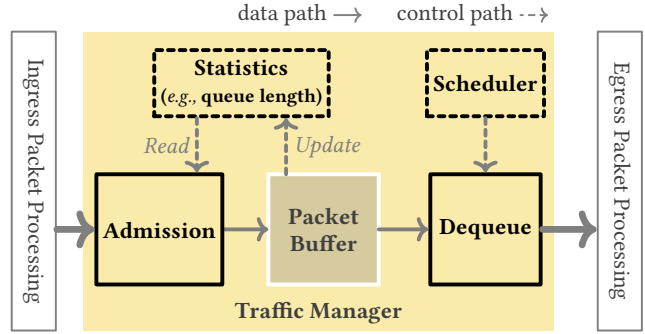
\subsection{Architecture of Shared-memory Switches}\label{sec:back:buffer}
% \reffig{fig:switch-packet-flow} shows a typical packet flow in the commodity switch~\cite{CiscoNexus3100}.
% Packet buffer stays at the heart of a switch
% (as shown in \reffig{fig:switch-packet-flow})
% and is essential to accommodate short-term congestion and maintain high throughput.
% Historically, packet buffer is placed outside the switch chip,
% providing sufficient memory with external DRAM chips.

\mypara{Switch structure} \figurename~\ref{fig:tm-structure}
depicts the high-level structure of modern shared-memory switch chips~\cite{
    BroadcomTomahawk4, CiscoNexus3100, NVIDIASpectrum4, HotChips20Tofino2, Arista7050X3,
}.
A typical switch chip mainly consists of three parts:
Ingress Packet Processing, Traffic Manager (TM),
and Egress Packet Processing.
% In today's high-speed switch, the packet buffer usually resides on the switch chip
% to achieve fast packet readings and writings~\cite{BroadcomSmartBuffer, CiscoSiliconOneG202, NVIDIASpectrum4Datasheet}.
In this paper, we mainly focus on TM,
which accommodates the packets and dynamically allocates the buffer across queues.
% The switch chip utilizes a Memory Management Unit (MMU) or Traffic Manager (TM) to allocate buffer for each arriving packet.
% \footnotetext{MMU is known as Traffic Manager (TM)~\cite{TofinoNativeArch, BCM88800TM}.}
% In the commodity switch chip,
% MMU usually divides the buffer into two pools:
% private pool and shared pool.
% The buffer in the private pool is dedicated to egress queues,
% while the buffer in the shared pool is shared by egress queues.
% Most of the buffer
% MMU puts a packet into the private pool first
% and tries to put into the shared pool if the private buffer of the egress queue has been used up.
% Most buffer is in the shared pool~\cite{Arista7050X3}
% to improve the buffer efficiency by statistical multiplexing.

For each arrival packet, TM first performs admission control (\eg~BM and active queue management)
to decide whether the packet can be accepted into the buffer.
Once admitted, the packet is written into a packet buffer,
which is a centralized and globally shared on-chip SRAM.
% Meanwhile, a Packet Descriptor (PD), containing the memory address of the packet data and other metadata (\eg, packet length),
% is put into the corresponding queue.
% Each queue is maintained through a linked list of PDs,
% which lies in another piece of on-chip SRAM
% (hereinafter referred to as \emph{PD memory}).
At the egress side,
a scheduler selects a queue to fetch packets
using a scheduling algorithm (\eg~deficit round robin).
% To dequeue a packet, TM first fetches a PD from the PD memory
% and then reads the actual packet data from the packet buffer.

\begin{figure}[!t]
    \centering
    \includegraphics[width=\linewidth]{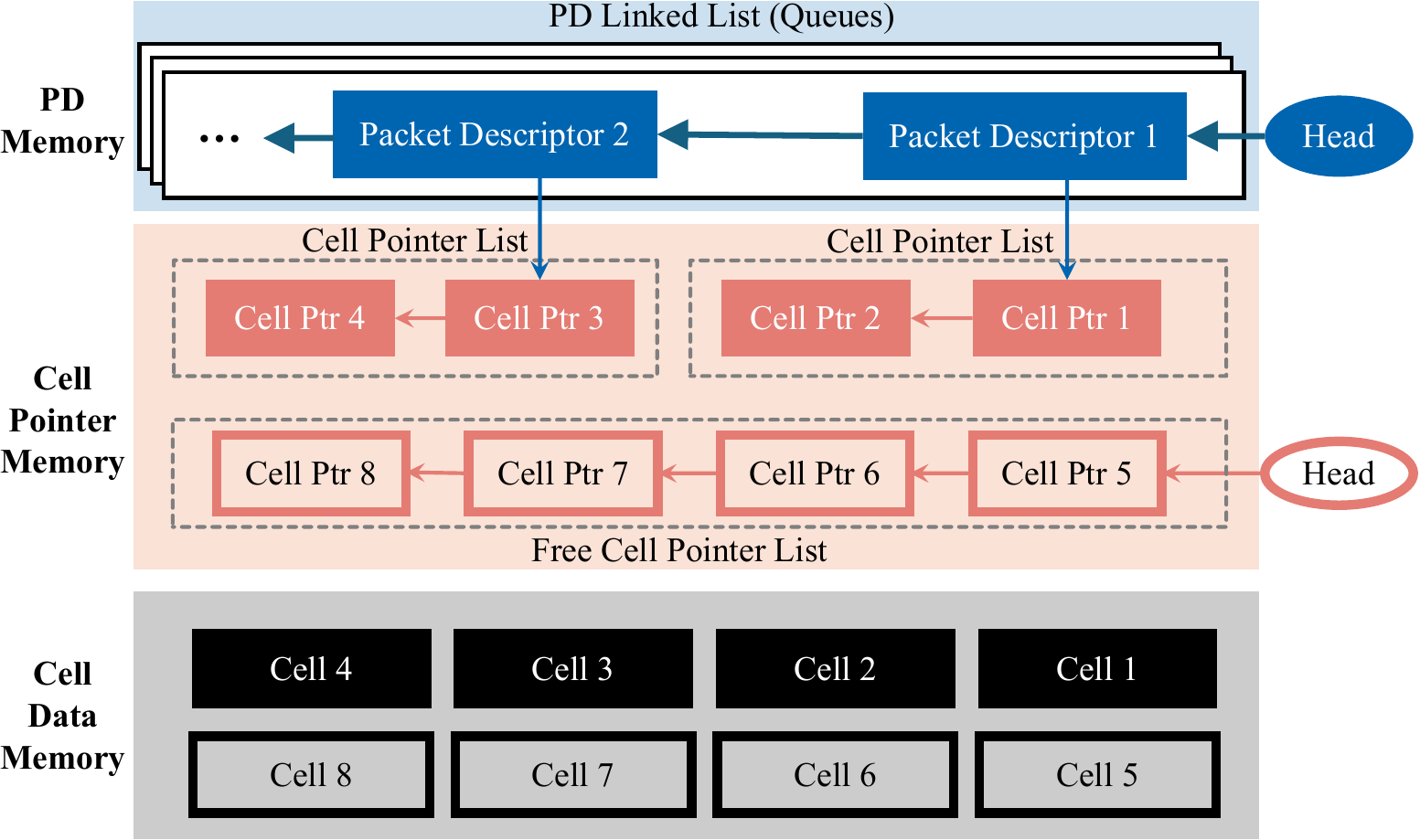}
    \caption{Structure of packet buffer}\label{fig:buffer-structure}
\end{figure}

\mypara{Queue/buffer structure} \figurename~\ref{fig:buffer-structure} dives into the details of
how packets are stored and queues are organized in the packet buffer.
% At the highest-level, each queue is maintained as a PD linked list.
Typically, there are three pieces of memory:
% Packet Descriptor (PD) memory, cell pointer memory, and cell data memory.
cell data memory, cell pointer memory, and Packet Descriptor (PD) memory.
% Packets are usually partitioned into several equal-sized cells.
% The actual cell data is stored in the cell data memory,
% whereas the packet metadata and addresses to the cells
% lie in the PD memory and cell pointer memory, respectively.
% Concretely, at the top level,
% the PD memory contains packet descriptors (PDs).
% each packet has a PD, which contains the packet metadata (\eg, packet length).
% A queue is organized as a linked list of PDs.
% At the middle level,
% each 
At the bottom,
the cell data memory accommodates the actual packet data,
which is divided into equal-sized cells.
In the middle, the cell pointer memory holds the cell pointers to these cells.
For a packet partitioned into multiple cells,
its cell pointers are linked together through a \emph{cell pointer list}.
Moreover,
the cell pointer memory also manages the free cells
through a linked list, referred to as \emph{free cell pointer list}.
The free cell pointer list contains the addresses of all free cells.
Upon the arrival of a packet,
TM allocates memory for the packet by removing several cell pointers from the free cell pointer list.
Upon the departure of a packet,
TM frees the occupied memory by returning all cell pointers of the packet to the free cell pointer list.
Finally, at the top, the PD memory contains packet descriptors (PDs).
Each packet is associated with a PD,
containing the packet metadata (\eg~packet length)
and the head(s) of cell pointer list(s).
A queue is organized as a linked list of PDs.

Modern high-speed switches employ several techniques to speed up packet readings and writings.
For example, the three types of memory are usually in different physical SRAMs,
allowing parallel accesses to PDs, cell pointers, and cells.
% In this way, PDs, cell pointers, and cell data can be accessed in parallel.
Furthermore,
to speed up packet readings,
high-speed switches may divide the cell pointer list into multiple sub-lists
(\eg~dividing Cell Ptr 1 $\to$ Cell Ptr 2 into Cell Ptr 1 and Cell Ptr 2).
% (\eg~dividing $\fillcircle{1} \to \fillcircle{2} \to \fillcircle{3} \to \fillcircle{4}$
% into $\fillcircle{1} \to \fillcircle{3}$ and $\fillcircle{2} \to \fillcircle{4}$),
In this way, multiple cell pointers can be read simultaneously
at the cost of maintaining multiple link headers
(\eg~Cell Ptr 1 and Cell Ptr 2).
% To speed up reading a packet,
% high-speed switches may divide the cell pointer list into multiple parallel linked lists
% (\eg, dividing $\fillcircle{1} \to \fillcircle{2} \to \fillcircle{3} \to \fillcircle{4}$
% into $\fillcircle{1} \to \fillcircle{3}$ and $\fillcircle{2} \to \fillcircle{4}$)
% at the cost of maintaining multiple link headers in the PD (\eg, \fillcircle{1} and \fillcircle{2}).

\subsection{Buffer Management Schemes}\label{sec:back:bm}
TM uses a Buffer Management (BM) scheme to dynamically allocate the shared-buffer across queues.
Depending on whether an accepted packet can be dropped,
BM schemes can be divided into two categories: non-preemptive BMs and preemptive BMs.
% In this part, the overall mechanisms are briefly discussed
% and leave the details of each specific BM scheme to related work (\textsection\ref{sec:rlwk}).
% In this part, we briefly introduce two representative BM schemes.
% and leave other BMs as related work (\textsection\ref{sec:rlwk}).
% BM has three goals:
% \begin{itemize}
%     \setlength\itemsep{0pt}
%     \item[\numcircled{1}] BM should \emph{efficiently} utilize the scarce buffer for high burst absorption.
%     \item[\numcircled{2}] BM should \emph{fairly} allocate buffer across queues for performance isolation.
%     \item[\numcircled{3}] BM should be \emph{simple} enough to facilitate its implementation in switch chip.
% \end{itemize}

% Unfortunately, the above three goals are somewhat conflicted,
% and no existing BM can achieve them all.
% % \inlinetodo{A table listing all BM schemes?}

% \mypara{De facto: Dynamic Threshold (DT)}
\mypara{Non-preemptive BMs}
Non-preemptive BMs are dominant in today's switch chip
due to their simplicity of implementation.
With non-preemptive BMs, a packet won't be expelled or overwritten once accepted into the buffer.
As a result, they have to \emph{proactively} reserve a fraction of free buffer
to accommodate transient bursts arriving at newly active queues.
Furthermore, when the buffer allocation (\ie~the amount of buffer allocated to each queue)
needs to be adjusted,
non-preemptive BMs can only \emph{passively} wait for the over-allocated queue
to naturally release the buffer through packet transmissions.
However, the queue drain rate is limited by the port speed
and can be even lower when bandwidth is shared with other queues.
% Non-preemptive BMs can work \emph{healthily} when the traffic arrival rate is not too high or the packet departure rate is not too low.
Hence, non-preemptive BMs are not agile enough to adjust the amount of buffer allocated to each queue
under the condition of high traffic arrival rate or low departure rate,
leading to anomalous behaviors.
Nevertheless, non-preemptive BMs are straightforward to implement
by limiting the queue length with a threshold
on the admission module.

% deciding whether to accept a packet by comparing the queue length and a threshold.

Next, Dynamic Threshold (DT) is used to exemplify the healthy and anomalous
behaviors of non-preemptive BM schemes.
DT is very prevalent in commodity switch chips~\cite{
    BroadcomSmartBuffer, ExtremeBuffer, BCM88800TM, BroadcomTrident3, MellanoxDT, Arista7050X3, CiscoNexus9000ConfigGuide, SIGCOMM16RDMA,
    SIGCOMM19HPCC, BS19Yahoo, NSDI23RDMA,
}.
It limits queue lengths with a threshold (denoted by $T(t)$),
which is dynamically adjusted based on the buffer occupancy.
Specifically, the threshold is proportional to the current free buffer size, namely,
\begin{equation}
    T(t) = \alpha \left(B - \sum_{i=1}^N q_i(t)\right)
    \label{eq:dt}
\end{equation}
where
$\alpha$ is a control parameter,
$B$ is the shared buffer size,
and $q_i(t)$ is the length of queue $i$ at time $t$.
In practice, $\alpha$ is usually a power of two,
so that the threshold can be simply calculated by shifting the free buffer size.
The intuition of~\refeq{eq:dt} is that,
when the switch is less congested,
DT allocates more buffer to each queue for high efficiency;
when the switch is more congested,
DT limits queue lengths more strictly for fairness.

\begin{figure}[!t]
    \centering
    \begin{subfigure}{.46\linewidth}
        \resizebox{\linewidth}{!}{\begin{tikzpicture}[font=\LARGE]
    \begin{scope}[
        every path/.style={line width=2, draw=gray, ->}
        ]
        \draw (0, 0) -- (0, 5.5) node[midway, sloped, above] {Queue Length};
        \draw (0, 0) -- (6.5, 0) node[midway, below] {Time};
    \end{scope}
    \begin{scope}[
        every path/.style={line width=3}
    ]
        \draw[darkblue]  (0, 5) -- ++(1, 0) -- ++(4, -2) -- ++(1, 0);
        \draw[darkgreen] (0, 0) -- ++(1, 0) -- ++(4, 3)  -- ++(1, 0);
        \draw[darkred, line width=5, dash pattern=on 1pt off 10pt, opacity=0.6] (0, 5) -- ++(1, 0) -- ++(4, -2) -- ++(1, 0);
    \end{scope}
    \begin{scope}[
        every path/.style={line width=1.5, draw=gray}
        ]
        % \draw[<-] ([yshift=1pt] 2, 5) -- +(0.5, 1) node[right] {\textcolor{darkblue}{$q_1(t)$} \texttt{\&\&} \textcolor{darkred}{$T(t)$}};
        \node[above right] at (3, 3.9) {\textcolor{darkblue}{$q_1(t)$} \& \textcolor{darkred}{$T(t)$}};
        % \node[above right] at (5, 3.9) {\textcolor{darkblue}{$q_1(t)$} \texttt{\&\&} \textcolor{darkred}{$T(t)$}};
        % \draw[<-] (5, 1.5) -- +(1, -0.5) node[right] {\textcolor{darkgreen}{$q_2(t)$}};
        \node[below right] at (3, 1.5) {\textcolor{darkgreen}{$q_2(t)$}};
    \end{scope}
\end{tikzpicture}}
        \caption{Healthy behavior}\label{fig:dt-example1}
    \end{subfigure}
    \hfil
    \begin{subfigure}{.46\linewidth}
        \resizebox{\linewidth}{!}{\begin{tikzpicture}[font=\LARGE]
    \begin{scope}[
        every path/.style={line width=2, draw=gray, ->}
        ]
        \draw (0, 0) -- (0, 5.5) node[midway, sloped, above] {Queue Length};
        \draw (0, 0) -- (6.5, 0) node[midway, below] {Time};
    \end{scope}
    \begin{scope}[
        every path/.style={line width=3}
    ]
        % (0, 5), (1, 5), (5, 3), (11, 3)
        \draw[darkblue]  (0, 5) -- ++(1, 0) -- ++(4, -2) -- ++(1, 0);
        \draw[darkgreen] (0, 0) -- ++(1, 0) -- ++(1, 1.5) -- ++(3, 1.5)  -- ++(1, 0);
        \draw[darkred, line width=5, dash pattern=on 1pt off 10pt]
            (0, 5) -- ++(1, 0) -- ++(1, -3.5) -- ++(3, 1.5) -- ++(1, 0);
    \end{scope}
    \begin{scope}[
        every path/.style={line width=1.5, draw=gray}
        ]
        % \draw[<-] (5, 4) -- +(1, 1) node[right] {\textcolor{darkblue}{$q_1(t)$}};
        \node[above right] at (3, 4) {\textcolor{darkblue}{$q_1(t)$}};
        % \draw[<-] (2.5, 1) -- +(1, -0.5) node[right] {\textcolor{darkgreen}{$q_2(t)$}};
        \node[right] at (1.5, 0.5) {\textcolor{darkgreen}{$q_2(t)$}};
        % \draw[<-] (2.5, 3.5) -- +(-0.5, -0.5) node[left] {\textcolor{darkred}{$T(t)$}};
        \node[below left] at (2.9, 3.5) {\textcolor{darkred}{$T(t)$}};
    \end{scope}
    \begin{scope}[on background layer]
        \fill[gray, opacity=0.3] (3, 4) ellipse [x radius=2cm, y radius=7mm, rotate=-25];
        \node[circle, fill=gray, opacity=0.35, inner sep=3mm] (q2-drop-point) at (2, 1.5) {};
    \end{scope}
    \node[gray, right] at (2, 5.2) {$q_1$: over-allocated};
    \node[below right, gray] at ([xshift=3mm, yshift=1mm] q2-drop-point) {$q_2$: drop before fair};
\end{tikzpicture}}
        \caption{Anomalous behavior}\label{fig:dt-example2}
    \end{subfigure}
    \caption{Dynamic behavior of DT}\label{fig:dt-example}
\end{figure}
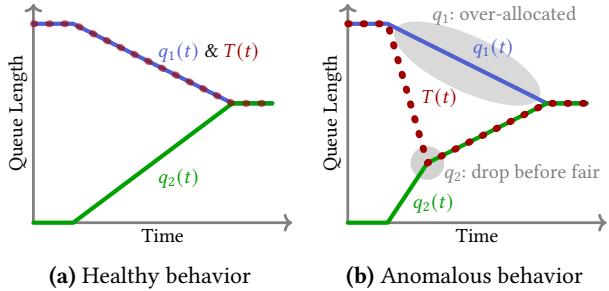

% We use an example to illustrate DT's workings.
To illustrate the behavior of DT,
consider a scenario with two queues.
Only queue 1 is congested and has reached the steady state at the beginning
(\ie~its queue length has reached $T(t)$).
Then some traffic bursts arrive at queue 2, needing to occupy some buffer.
\reffig{fig:dt-example1} shows the evolutions of queue length and threshold
in the healthy case.
% only queue 1 is active at the beginning and the threshold is high,
% allocating more buffer to queue 1 for efficiency.
As the length of queue 2 (\ie~$q_2(t)$) grows,
DT gradually reduces the threshold.
This, in turn, causes the length of queue 1 (\ie~$q_1(t)$) to decrease
by draining the excess buffer occupancy.
Finally, both queue 1 and queue 2 will occupy the same amount of buffer,
achieving fair buffer sharing.

\reffig{fig:dt-example2} illustrates the anomalous behavior,
where queue~1 cannot decrease its length as fast as $T(t)$.
This case occurs either when $q_2(t)$ increases too quickly
due to high traffic arrival rate,
or when $q_1(t)$ reduces too slowly
due to the shared capacity of the egress port with other queues.
Either way, DT is unable to adjust the buffer allocation in time.
As a result,
queue~1 occupies excess buffer,
forcing queue 2 to reluctantly drop packets
before receiving its deserved buffer space.

% Apparently, we would expect DT to work in the healthy case.
% However, there are several reasons that DT can work in the anomalous case in datacenters.
% First, the traffic arrival rate can be very high.
% In datacenters, the traffic patterns
% such as incast~\cite{CoNEXT21Floodgate, SIGCOMM20Swift}
% can create highly bursty traffic.
% As a result, DT is not agile enough to vacate the allocated buffer for queue 1.
% Second, the packet departure rate can be very low.
% Datacenter networks can leverage different queues to
% isolate the performance of different services.
% Latency-sensitive services are usually put into the higher-priority queue
% with strict priority queueing
% or high-weight queue with weighted fair queueing.
% In the above examples, when Queue 1 is 
% Historically, DT is likely to work in the healthy mode
% for two reasons.
% One is that the buffer size is adequate, allowing DT to reserve enough free buffer size
% without sacrificing the throughput and burst absorption.
% The other is that the traffic is not quite bursty as the datacenters
% where thousands of flows aggregate.
% However, in datacenter networks,
% this situation cannot hold as the traffic rate can be very high
% and the buffer is quite insufficient
% (we will show in \textsection\ref{sec:motv}).

% However, DT is not efficient enough for DCN~\cite{INFOCOM15Burst, BS19FAB},
% as it can waste a large fraction of buffer with few active queues.
% For example, with $\alpha=1$, 50\% buffer is always left free with one active queue.

\mypara{Preemptive BMs}
Preemptive BMs can expel or overwrite the accepted packets.
Thus, preemptive BMs usually do not need to proactively reserve buffer for inactive queues
as they can quickly make room by packet expulsion.
Instead, they \emph{reactively} adjust the buffer allocation when facing dynamic traffic.
% Consequently, preemptive BMs have both better ability of buffer adjustment and higher buffer efficiency.
Consequently, preemptive BMs can achieve higher buffer efficiency.

In particular, a kind of preemptive BM, known as Pushout,
had been regarded as optimal~\cite{
    1981Pushout, TCOM84Pushout, GLOBECOM91Pushout,
    JSAC91Pushout, GLOBECOM93Pushout,
    TON94Pushout, JHSN94Pushout,
    JSAC95Pushout, CC96Pushout,
}.
% Different from DT, Pushout is preemptive in that it allows incoming packets
% to overwrite or evict the existing packets in the buffer.
% Pushout does not needs to proactively reserve free buffer for inactive queues.
% Instead, it can \emph{reactively} vacate the buffer for newly active queues.
% Pushout is quite efficient.
% Specifically,
Pushout allows a packet to enter into the buffer whenever there is some free space.
% Pushout is also fair.
When the buffer is out of space,
Pushout expels a packet from the longest queue to make room for the new arrival packet.
% Pushout has been proven to be optimal from the perspective of throughput~\cite{
%     TCOM84Pushout, GLOBECOM91Pushout, TON94Pushout, JSAC95Pushout,
% }.
% It is efficient in that all buffer can be utilized when needed.
% It is also fair in that shorter queues can always obtain buffer.
However, implementing Pushout in switch chips was historically considered difficult
due to the following three reasons~\cite{TON98DT, CC96Pushout, COMST00BM}:

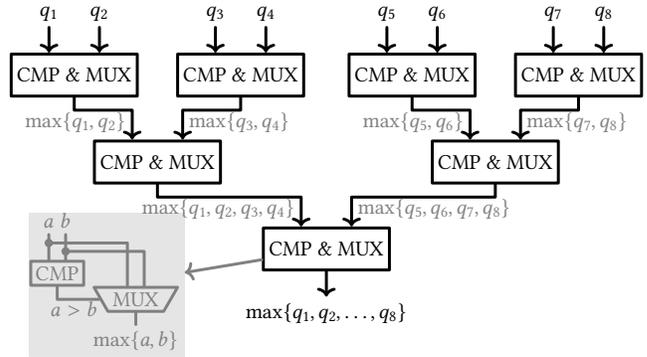
\begin{figure}[!t]
    \centering
    \resizebox{\linewidth}{!}{\begin{tikzpicture}[font=\LARGE]
    \tikzset{
        cmpmux/.style={
            draw, minimum height=1cm, minimum width=3cm,
            label=center:CMP \& MUX,
            font=\LARGE\bf,
            line width=2pt,
        },
        mux/.style={
            draw,
            shape=trapezium,
            trapezium angle=120,
            line width=2pt,
        },
    },
    \path (0, 0) -- (12cm, 0)
        node[cmpmux, pos=0.00] (cmp-1-1) {}
        node[cmpmux, pos=0.33] (cmp-1-2) {}
        node[cmpmux, pos=0.67] (cmp-1-3) {}
        node[cmpmux, pos=1.00] (cmp-1-4) {}
        ;
    \path (cmp-1-1.south) -- (cmp-1-2.south)
        node[cmpmux, midway, below=1cm] (cmp-2-1) {};
    \path (cmp-1-3.south) -- (cmp-1-4.south)
        node[cmpmux, midway, below=1cm] (cmp-2-2) {};
    \path (cmp-2-1.south) -- (cmp-2-2.south)
        node[cmpmux, midway, below=1cm] (cmp-3) {};
    \begin{scope}[line width=2pt, ->]
        \draw[<-] ([xshift=-0.6cm] cmp-1-1.north) -- +(0, 6mm) node[above] {$q_1$};
        \draw[<-] ([xshift=+0.6cm] cmp-1-1.north) -- +(0, 6mm) node[above] {$q_2$};
        \draw[<-] ([xshift=-0.6cm] cmp-1-2.north) -- +(0, 6mm) node[above] {$q_3$};
        \draw[<-] ([xshift=+0.6cm] cmp-1-2.north) -- +(0, 6mm) node[above] {$q_4$};
        \draw[<-] ([xshift=-0.6cm] cmp-1-3.north) -- +(0, 6mm) node[above] {$q_5$};
        \draw[<-] ([xshift=+0.6cm] cmp-1-3.north) -- +(0, 6mm) node[above] {$q_6$};
        \draw[<-] ([xshift=-0.6cm] cmp-1-4.north) -- +(0, 6mm) node[above] {$q_7$};
        \draw[<-] ([xshift=+0.6cm] cmp-1-4.north) -- +(0, 6mm) node[above] {$q_8$};
        \draw (cmp-1-1.south) -- +(0, -3mm) -| ([xshift=-0.6cm] cmp-2-1.north)
            node[pos=0.7, left, gray] {$\max\{q_1, q_2\}$};
        \draw (cmp-1-2.south) -- +(0, -3mm) -| ([xshift=+0.6cm] cmp-2-1.north)
            node[pos=0.7, right, gray] {$\max\{q_3, q_4\}$};
        \draw (cmp-1-3.south) -- +(0, -3mm) -| ([xshift=-0.6cm] cmp-2-2.north)
            node[pos=0.7, left, gray] {$\max\{q_5, q_6\}$};
        \draw (cmp-1-4.south) -- +(0, -3mm) -| ([xshift=+0.6cm] cmp-2-2.north)
            node[pos=0.7, right, gray] {$\max\{q_7, q_8\}$};
        \draw (cmp-2-1.south) -- +(0, -3mm) -| ([xshift=-0.6cm] cmp-3.north)
            node[pos=0.7, left, gray] {$\max\{q_1, q_2, q_3, q_4\}$};
        \draw (cmp-2-2.south) -- +(0, -3mm) -| ([xshift=+0.6cm] cmp-3.north)
            node[pos=0.7, right, gray] {$\max\{q_5, q_6, q_7, q_8\}$};
        \draw (cmp-3.south) -- +(0, -6mm) node[below] {$\max\{q_1, q_2, \dots, q_8\}$};
    \end{scope}
    \begin{scope}[line width=2pt, gray]
        \node[mux] (mux) at ([xshift=-3cm, yshift=-1.2cm] cmp-3.west) {MUX};
        \draw (mux.left side) -| +(-1cm, 3mm)
            node[pos=0.3, below=-1mm, gray] {$a>b$}
            node[above, draw, line width=2pt] (cmp) {CMP};
        \draw ([xshift=-2mm] cmp.north) -- +(0, 6mm) node[above] (cmp-in-1) {$a$};
        \draw ([xshift=+2mm] cmp.north) -- +(0, 6mm) node[above] (cmp-in-2) {$b$};
        \draw ([yshift=-2mm] cmp-in-1.south) -| ([xshift=-2mm] mux.north)
            node[pos=0, fill, circle, inner sep=2pt] {};
        \draw ([yshift=-4mm] cmp-in-2.south) -| ([xshift=+2mm] mux.north)
            node[pos=0, fill, circle, inner sep=2pt] {};
        \draw (mux) -- +(0, -6mm) node[below] (mux-out) {$\max\{a, b\}$};
        % \path (cmp-in-1) -- (mux-out)
        %     node[pos=0.6, draw, minimum width=5cm, minimum height=3cm] {};
        \fill[gray, opacity=0.2] (cmp-in-1.north -| cmp.west) rectangle (mux-out.south east);
        \draw[->] (cmp-3) -- ([yshift=2cm] mux-out.south east);
    \end{scope}
\end{tikzpicture}}
    \caption{An 8-input Maximum Finder based on binary comparator tree}
    \label{fig:maximum-finder}
\end{figure}

\newcounter{diff}
\refstepcounter{diff}
\textbf{Difficulty \thediff\label{diff:bandwidth}:}
When a packet arrives and the buffer is full,
Pushout needs to evacuate another packet to make room for the new packet.
Thus, accepting a packet may require an extra read operation\footnotemark.
This was unacceptable at that time because memory bandwidth
was the key limiting factor of the forwarding speed for a shared-memory switch~\cite{
    INFOCOM98PingPong, Stanford01SharedMemory, HPSR01SharedMemory,
}.
\footnotetext{
    Note that the packets cannot be dropped without accessing the memory.
    With on-chip buffer, the switch chip has to at least acquire the packet descriptor,
    which is stored in the PD memory.
    With off-chip buffer,
    the switch chip only maintains the head and tail of each queue~\cite{HPSR01SharedMemory},
    and thus needs to access the off-chip buffer to drop packets.
}

Specifically,
The packet buffer was made from off-chip DRAM
due to the requirement of large buffer.
For example, as a rule-of-thumb, TCP requires $C\times RTT$
(where $C$ is the bottleneck capacity and $RTT$ is the round-trip time)
to achieve high throughput.
The Internet $RTT$ was in the order of 200 milliseconds
and $C$ was in the order of 10Gbps~\cite{HPSR01SharedMemory}.
As a result, a 2Gbit packet buffer is required,
while SRAM was too expensive to provide such a memory size.

Worse yet, in the history, off-chip DRAM was slow.
In 1998, the memory access time of the fastest DRAM was $\sim$50ns~\cite{INFOCOM98PingPong}.
If packets are partitioned into 200B cells,
then the memory bandwidth was only 32Gbps,
which is too slow to build a 10Gbps switch~\cite{HPSR01SharedMemory}.
Consequently, the memory bandwidth was a bottleneck
and memory access was at the critical path
in a high-speed shared-memory switch.
There is little space for buffer management to consume memory bandwidth.

\refstepcounter{diff}
% \item[\numcircled{2}]
\textbf{Difficulty \thediff\label{diff:enqueue}:}
\emph{Complex enqueue operations.}
Pushout couples the enqueue operations and dequeue operations
in that enqueuing a packet may require to dequeue and drop a packet from another queue.
To achieve this,
Pushout requires extra buffer for each queue to accommodate the waiting packet.
Moreover, all enqueue operations should be paused.
Otherwise, the vacated buffer room may be occupied by another queue.
Finally, the coordination between the ingress side and egress side is necessary:
the ingress side needs to notify the egress side of the expulsion operation
and the egress side should also notify the ingress side of the completion of the expulsion operation.

\refstepcounter{diff}
% \item[\numcircled{3}]
\textbf{Difficulty \thediff\label{diff:longest}:}
\emph{Monitoring the longest queue in real time.}
To monitor the longest queue,
Pushout requires a circuit called Maximum Finder (MF)~\cite{TC14MaximumFinder}.
\figurename~\ref{fig:maximum-finder} depicts a common MF based on binary comparator tree.
To find the maximum value among $N$ variables,
MF requires $\log_{2}N$ levels of nodes.
Each node finds the maximum value of two numbers,
which is implemented by a multiplexer and a comparator.
The area complexity of such an MF is $O(kN)$ gates,
where $k$ is the bit-width of each compared number.
While the area cost is acceptable,
the time complexity --- $O(\log_2 k \times \log_2 N)$ ---
cannot satisfy the requirement of a high-speed switch chip.
Specifically, several cells can be enqueued or dequeued every clock cycle,
resulting in changes of queue length per cycle,
which is too fast for MF to give the result.
% Our conversations with an industry collaborator
% of commercial high-speed switch chip.
% \begin{itemize}
% \setlength\itemsep{0pt}
% \item[\numcircled{1}]

One may wonder whether it is enough to record the maximum queue id with a register
and update the register whenever the queue length changes.
However, there is an issue for this strawman approach.
For example,
consider a scenario with two queues (\ie~$q_1$ and $q_2$), which are at the same egress port.
They are scheduled by the strict priority algorithm, and $q_1$ has a higher priority.
Initially, $q_1$ has a length of 80KB,
whereas $q_2$ has a length of 60KB.
Thus, the longest queue is $q_1$.
As being drained, the length of $q_1$ falls to 50KB,
while the length of $q_2$ keeps unchanged.
At this time, the longest queue has switched to $q_2$,
while this strawman approach fails to discover it.

% can consume lots of extra hardware resources and raise power consumption
% (we analyze the overhead in details in Appendix~\ref{sec:po-difficult}).

% \end{itemize}

%:
\section{Motivation}\label{sec:motv}
% While working fine for decades,
% DT's performance becomes increasingly unsatisfactory with the recent trends.
% On the other hands, with the advances on switch chips,
% the issues that impede the realization of Pushout become less insurmountable.
% Due to the nature of non-preemption,
% DT can result in several problems in datacenter networks.
% Nevertheless, the advances on switch chip have opened up new opportunities.
In this section, we demonstrate the issues of non-preemptive BMs and analyze the underlying causes.
% We show that DT's problems are intrinsic in its non-preemption.
Then, we discuss the opportunities opened up by modern switch chips,
which motivate us to innovate preemptive BMs.

\begin{figure}[!t]
    \centering
    \resizebox{.9\linewidth}{!}{\begin{tikzpicture}[font=\Huge]
    \colorlet{hp}{dadiannaored}
    \colorlet{lp}{dadiannaodarkblue}
    \tikzset{
        scheduler/.style = {
            fill=dadiannaolinegray, circle, line width=1.5pt,
            minimum width=10mm,
        },
        pkt/.style = {
            minimum width=6mm, minimum height=6mm, inner sep=0,
        },
        hp pkt/.style = {
            pkt, fill=hp,
        },
        lp pkt/.style = {
            pkt, fill=lp,
        },
        port/.style = {
            draw=dadiannaolinegray, line width=5pt,
            cylinder, aspect=2,
            minimum height=2cm,
            minimum width =2cm,
        },
    }
    % buffer
    \fill[dadiannaolinegray] (-1.1, -0.1) rectangle (10.1, 6.1);
    \coordinate (buffer-east)  at (10.1, 3);
    \coordinate (buffer-north) at (5, 6.1);
    % hp/lp buffer occupancy
    \begin{scope}[line width=1.5pt, font=\Huge\bf, text=white]
        \fill[lp, draw=white] (-0.5, 0) rectangle (10, 4.5);
        \fill[hp, draw=white] (2, 4.5) rectangle (10, 6);
        \node at (6, 5.25) (hp-center) {High-priority Traffic};
        \node at (5.25, 2.25) (lp-center) {Low-priority Traffic};
        \coordinate (hp-east) at (10, 5.25);
        \coordinate (lp-east) at (10, 2.25);
    \end{scope}
    % \node[scheduler, right=1cm of buffer-east] (sch) {};
    % output port
    \node[port, right=1.5cm of $(hp-east)!0.5!(lp-east)$] (port) {};
    % bandwidth occupancy in output port
    \begin{scope}[on background layer, line width=1pt]
        \fill[hp, draw=white]
            ([xshift=1mm] $(port.north west)!0.75!(port.south west)$)
            rectangle ([yshift=-1mm] port.north east);
        \fill[lp, draw=white]
            ([xshift=1mm] $(port.north west)!0.75!(port.south west)$)
            rectangle ([yshift=1mm] port.south east);
    \end{scope}
    % departured packets
    \path ([xshift=1cm] port.east) -- +(5, 0)
        foreach \pos in {0.1, 0.3,..., 0.9} {
            node[pos=\pos, hp pkt] {}
        };
    \path ([xshift=1cm, yshift=1cm] port.east) -- +(5, 0)
        foreach \pos in {0.1, 0.3666,..., 0.9} {
            node[pos=\pos, hp pkt] {}
        }
        coordinate[pos=0.5, above=1] (pkt-top)
        ;
    \path ([xshift=1cm, yshift=-1cm] port.east) -- +(5, 0)
        foreach \pos in {0.33, 0.66} {
            node[pos=\pos, lp pkt] {}
        };
    % lines and labels
    \draw[line width=9pt, hp] (hp-center-|buffer-east) to[out=0, in=+180] ([yshift=+3mm] port.west);
    \draw[line width=3pt, lp] (lp-center-|buffer-east) to[out=0, in=-180] ([yshift=-3mm] port.west);
    \path (hp-center-|buffer-east) -- ([yshift=+3mm] port.west)
        coordinate[pos=0.5] (hp-deqline-center);
    \path (lp-center-|buffer-east) -- ([yshift=-3mm] port.west)
        coordinate[pos=0.5] (lp-deqline-center);
    \draw[->, line width=5pt] (port.east) -- +(1, 0);
    \begin{scope}[font=\Huge\bf, text=dadiannaotextgray]
        \node[above=1mm] at (buffer-north) {Packet Buffer};
        \node[above=1mm] at (pkt-top |- buffer-north) {Sent Packets};
        \node[above=-1mm] at (port |- buffer-north) {Egress Port};
        \draw[<-, dadiannaolinegray, line width=2pt]
            ([xshift=1mm, yshift=-5mm] hp-deqline-center) to[out=-120, in=120] +(5mm, -3cm)
            node[right, font=\Huge] {High-priority traffic is scheduled first};
    \end{scope}
    % \draw[->, dadiannaolinegray, line width=2pt]
    %     ([yshift=5mm] hp-deqline-center)
    %     arc[start angle=-90, end angle=90] -- ([yshift=-5mm] lp-deqline-center);
\end{tikzpicture}}
    \caption{The buffer choking problem}\label{fig:buffer-choking}
\end{figure}
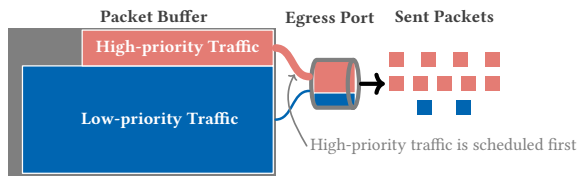

\subsection{Issues of Non-preemptive BMs}\label{sec:motv:problem}
% Under the status quo, DT falls short in meeting the requirements.
% Recall that DT can work in two cases: healthy case and anomalous case.
% Next, we show that DT works either in the healthy case at the expense of low buffer efficiency
% or in the anomalous case at the expense of poor performance isolation.
Despite their dominance in commodity switch chips,
non-preemptive BMs have some intrinsic issues.

\begin{figure*}
\begin{minipage}[b]{.49\linewidth}
    \begin{subfigure}[]{.47\linewidth}
        \includegraphics[width=\linewidth]{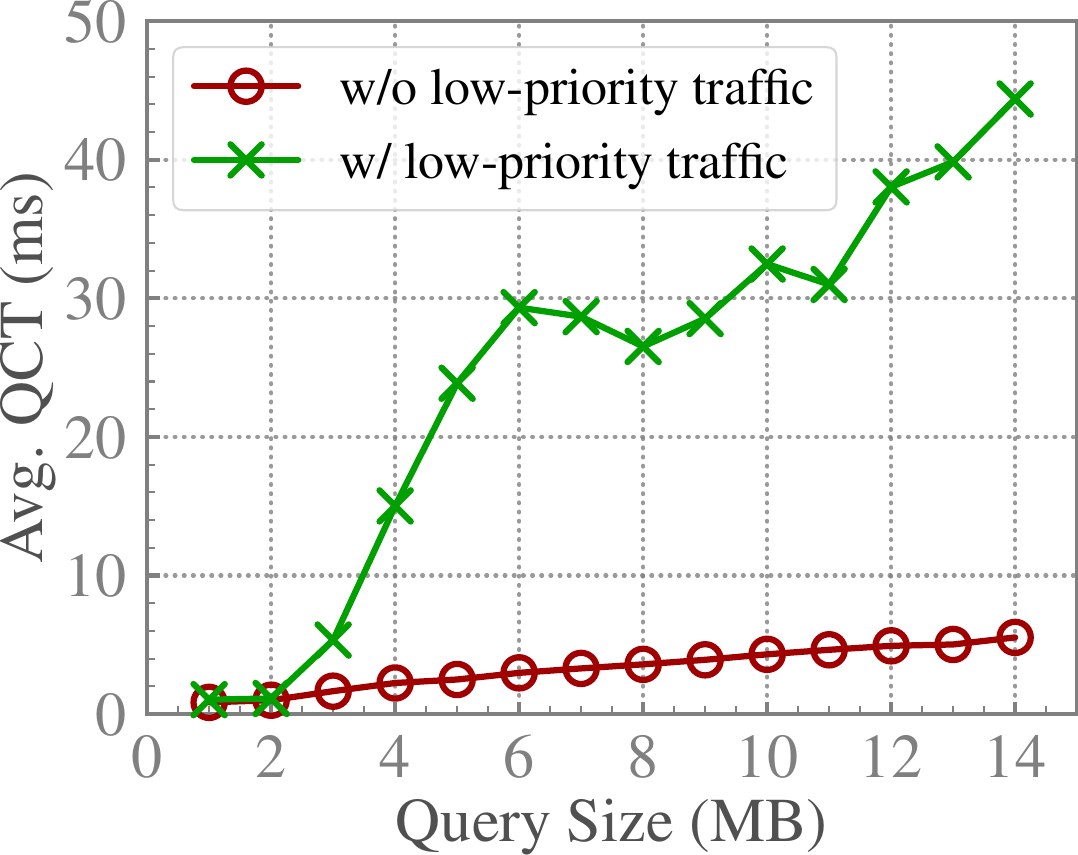}
        \caption{Low-priority influence}
        \label{fig:motv:buffer-choke}
    \end{subfigure}
    \hfil
    \begin{subfigure}[]{.47\linewidth}
        \includegraphics[width=\linewidth]{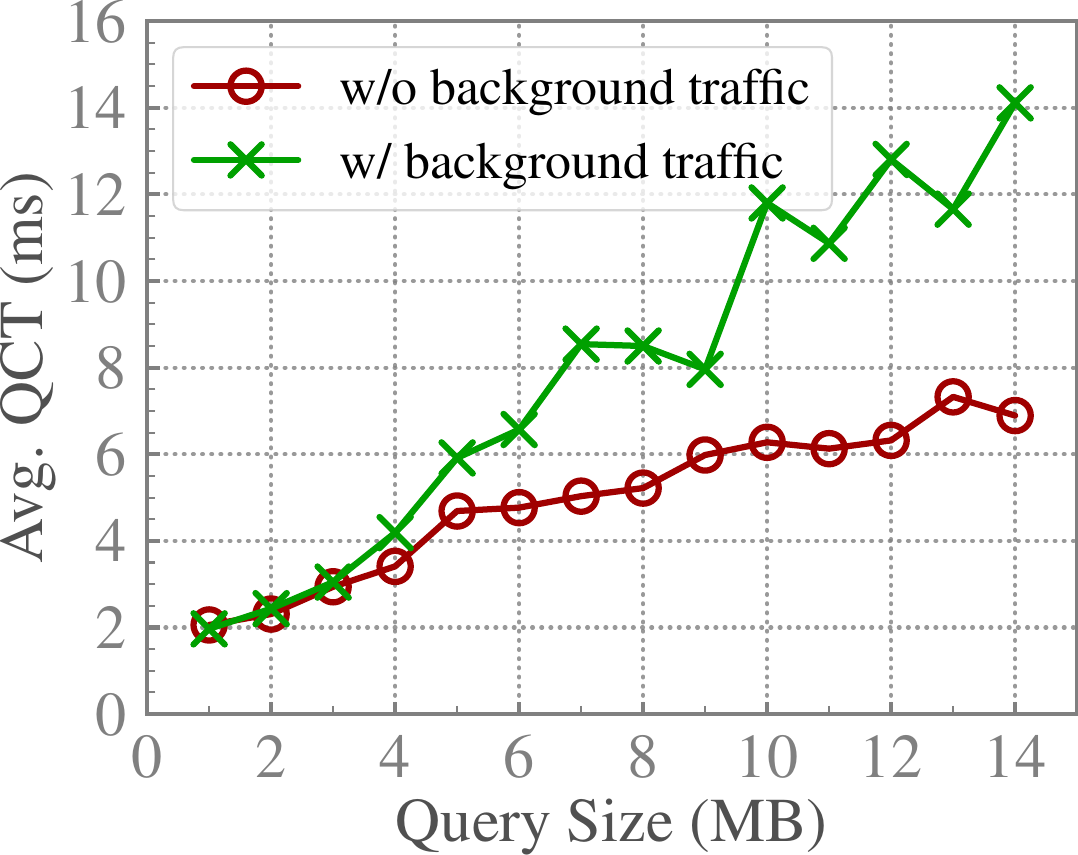}
        \caption{Inter-port influence}
        \label{fig:motv:isolation}
    \end{subfigure}
    \caption{Performance degradation due to anomalous behavior}
    % \caption{$\frac{\textrm{Buffer Occupancy of Background Traffic}}{\textrm{Buffer Occupancy of Incast Traffic}}$}
    \label{fig:motv:pathological}
\end{minipage}
\hfil
\begin{minipage}[b]{.49\linewidth}
    \begin{subfigure}[]{.49\linewidth}
        \includegraphics[width=\linewidth]{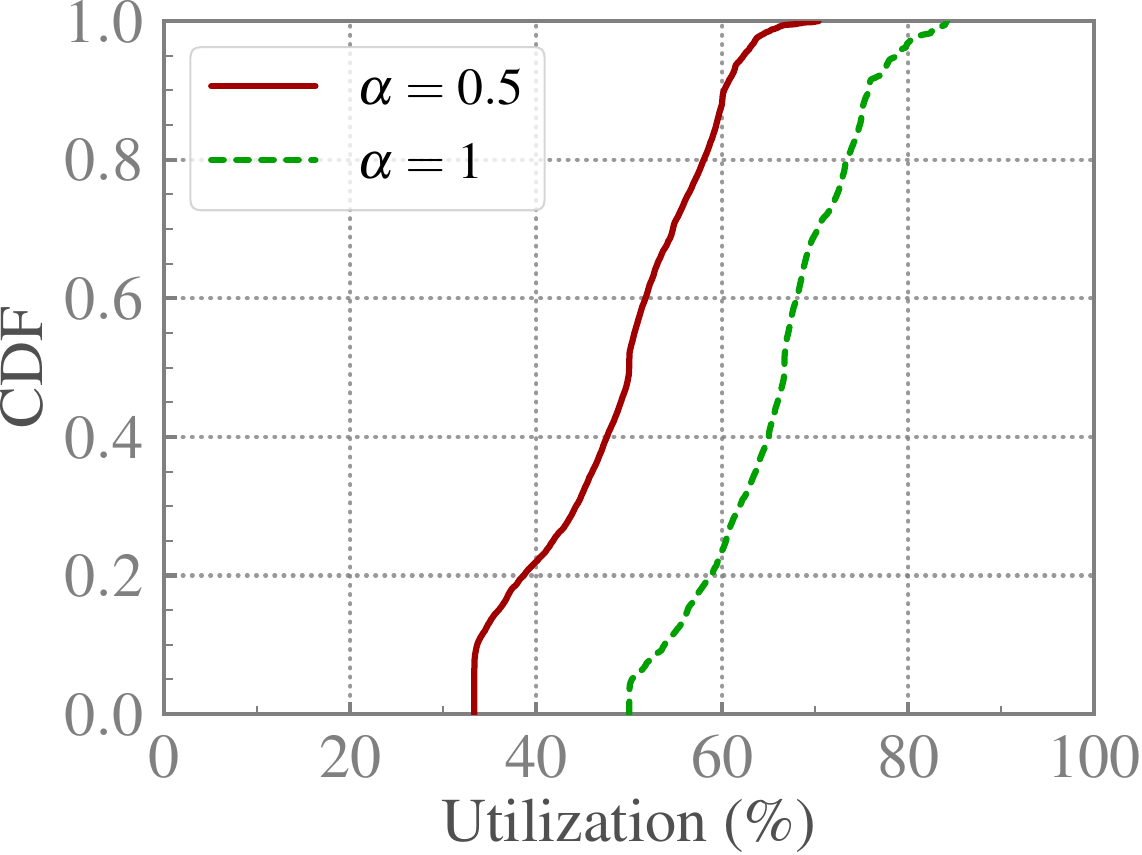}
        \caption{Buffer}
        \label{fig:uti-on-drop:buffer}
    \end{subfigure}
    \hfil
    \begin{subfigure}[]{.49\linewidth}
        \includegraphics[width=\linewidth]{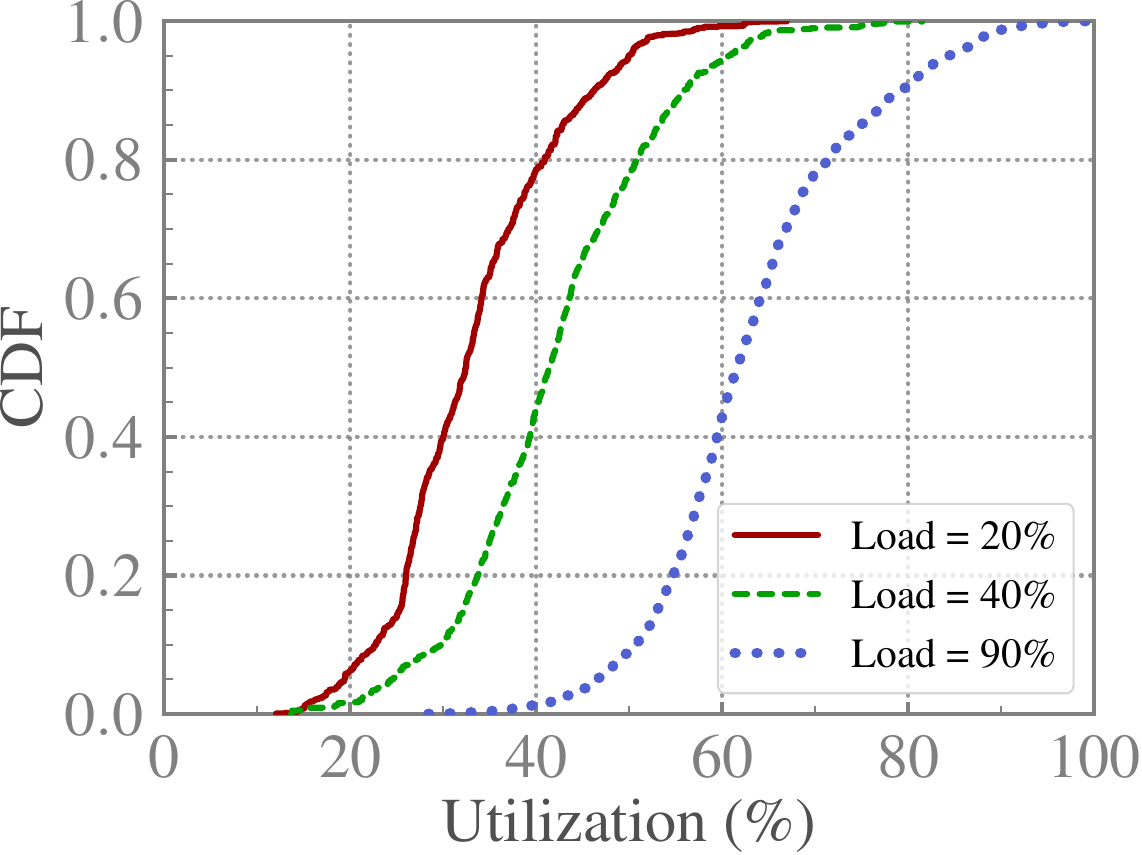}
        \caption{Memory bandwidth}
        \label{fig:uti-on-drop:memory bandwidth}
    \end{subfigure}
    \caption{CDF of buffer/memory bandwidth utilization}
    \label{fig:uti-on-drop}
\end{minipage}
\end{figure*}

\mypara{Non-preemptive BMs can work \emph{anomalously}}
This is due to two reasons.
(1) The queue drain rate can be very low,
causing non-preemptive BMs to suffer from the \emph{buffer choking problem},
as shown in \figurename~\ref{fig:buffer-choking}.
Specifically,
DCN operators can leverage different queues in a port to
isolate the performance of different services~\cite{NSDI16MQECN}.
Latency-sensitive traffic is usually placed in
the high-priority queue with strict priority queueing,
or assigned with higher weight with weighted fair queueing.
Consider a case when the low-priority queues are congested earlier,
and later some bursty traffic arrives at the high-priority queue.
% When different queues in the same port are congested simultaneously,
In this case, the low-priority queues hold considerable buffer.
However, since most of the bandwidth is allocated to the high-priority queue,
the low-priority queues drain slowly.
As a result, the high-priority queue runs out of buffer space soon,
while buffer management (BM) fails to free up enough space for it.
Consequently, high-priority packets are unexpectedly dropped due to insufficient buffer allocation.

To quantitatively demonstrate this issue,
we build a testbed comprising 4 hosts connected to a Huawei CE6865 switch.
Each host is equipped with an Intel XL710 Dual Port 40GbE NIC.
Using network namespaces,
we isolate two ports on each NIC to emulate two separate NICs and end nodes per host
(8 nodes in total).
The switch is configured to dynamically allocate 2MB buffer across eight 40Gbps ports via DT.
% The switch has 2MB buffer shared by eight 40Gbps ports with DT.
% dynamically allocated to eight 40Gbps ports via DT.
The switch supports 8 class-of-service queues,
with one designated as high priority,
and the rest as low priority.
% One is a high-priority queue, whereas the others are low-priority queues.
We employ DCTCP as the congestion control algorithm,
with the ECN threshold set to 300KB as recommended in~\cite{TON21BCC}.
We generate two types of traffic,
which are from different senders to the same receiver:
\numcircled{1} High-priority incast traffic:
The receiver simultaneously sends queries to 5 senders,
each responding with data upon receiving the query.
The incast degree is 40 (\ie~8 flows per sender).
\numcircled{2} Low-priority background traffic:
We generate 14 long-lived flows from 2 other senders to the same receiver,
each of which is classified into one of 7 low-priority queues.
To examine the performance of incast traffic,
we measure the query completion time (QCT),
which is the completion time of all incast flows.
We compare QCT between two scenarios:
one with low-priority traffic and the other without.
For a fair comparison, we configure DT such that
the incast traffic should be allocated
the same amount of buffer in both scenarios.
Specifically, we set $\alpha=8$ for the high-priority queue with low-priority traffic
and $\alpha=1$ without low-priority traffic.
For low-priority queues, we set $\alpha=1$.
In this way, the incast traffic deserves 1MB buffer
either with or without low-priority traffic~\cite{TON98DT, SIGCOMM22ABM},
and ideally, the QCT performance should be unaffected by low-priority traffic.

\figurename~\ref{fig:motv:buffer-choke} shows
the average QCT with different query sizes
(\ie~the total volume of incast traffic),
with and without low-priority traffic.
Although the QCT is expected to remain unaffected,
the presence of low-priority traffic significantly degrades the average QCT
by up to $\sim$8$\times$ compared to the scenario without low-priority traffic.
% the average QCT with low-priority traffic is significantly degraded by up to $\sim$8$\times$
% compared to that without low-priority traffic.
This is because
DT cannot quickly drain the low-priority traffic from the buffer due to buffer choking.
As a result, the high-priority incast traffic receives only a small portion of the buffer
and experiences packet drops before getting its deserved buffer.

(2) The traffic arrival rate can be very high.
The traffic patterns such as incast are very prevalent in datacenters~\cite{CoNEXT21Floodgate}.
As a result, thousands of flows can aggregate at the switch~\cite{SIGCOMM20Swift},
introducing intense traffic bursts.
Without the ability to quickly make room for the highly bursty traffic,
packets can be dropped before enough buffer is allocated to them.

To quantitatively illustrate this issue,
we use the same experimental settings as before,
except that two types of traffic are congested at different ports,
thereby eliminating the impact of buffer choking.
\figurename~\ref{fig:motv:isolation} shows that
the average QCT with background traffic still degrades by up to $\sim$2$\times$
compared to the scenario without background traffic,
despite both receiving the same buffer allocation in theory.
This is because DT is not quick enough to adjust the buffer occupancy
when facing fast incast traffic,
leading to packet drops before the buffer is properly reallocated.

\mypara{Non-preemptive BMs are inefficient}
As non-preemptive BMs cannot immediately make room for newly active queues,
they have to proactively reserve some buffer,
which is essential to absorb transient bursts.
As a result, they cannot fully utilize the buffer.
To show this issue,
we conduct a simulation with a leaf-spine topology
of 8 spine switches, 8 leaf switches, and 128 servers
connected via 100Gbps links.
Flows are generated according to the web-search workload~\cite{SIGCOMM10DCTCP}.
The network load is 40\% (detailed settings in \textsection\ref{sec:eval:sim}).
\figurename~\ref{fig:uti-on-drop:buffer}
shows the buffer utilization on packet drop with DT.
With typical settings (\ie~$\alpha=0.5$) in production datacenters~\cite{NSDI23RDMA},
the 99th percentile buffer utilization is only $\sim$66\%.
% For example, with DT, a packet can be dropped while 66.7\% buffer is free
% with $\alpha=1/2$,
% which is a typical setting in production datacenters~\cite{NSDI23RDMA}\footnotemark.
% This is unacceptable in DCN
Given the situation that on-chip buffer is very scarce
and adequate buffer occupancy is needed
for both low latency~\cite{IMC17Burst, IMC22Burst, NSDI23Burst}
and high throughput~\cite{INFOCOM20BCC, TON21BCC, ICNP21FlashPass},
it is better not to waste the scarce buffer.

\mypara{Analysis of root causes}\label{sec:motv:cause}
% Based on the above analysis, we conclude that there are two reasons for the above problems:
The radical reason for the above issues is that
non-preemptive BMs are \emph{not agile enough} to adjust buffer allocations.
The underlying factor that limits their agility is the \emph{non-preemptive nature}.
Specifically,
without the ability to actively expel packets,
non-preemptive BMs have to passively wait for the over-allocated queues to
naturally release buffer through normal packet transmissions.
Moreover,
due to the limited speed of buffer adjustment,
non-preemptive BMs have to proactively reserve some free buffer
to prevent newly active queues from buffer starvation,
resulting in buffer inefficiency.
Thus, an effective approach to address the above issues
is preemptive BM that can actively expel packets for the over-allocated queues.

\subsection{Opportunities}\label{sec:motv:opportunities}
As shown in \textsection\ref{sec:back:bm},
preemptive BM schemes were historically considered hard to implement
due to three difficulties.
In this part, we argue that
the limitation of bandwidth (\ie~Difficulty~\ref{diff:bandwidth}) is less critical
with the current buffer architecture,
which opens up opportunities to innovate preemptive BMs
by overcoming Difficulty~\ref{diff:enqueue} and Difficulty~\ref{diff:longest}.

% However, the switch chip has evolved a lot since then.
% In this part, we analyze the opportunities opened up by the modern switch chip.
% Based on the above reasons,
% we show that to achieve high adaptivity,
% a BM should quickly vacate the over-allocated buffer,
% % which is in accordance with core idea of Pushout.
% Although some mechanisms of Pushout is still not easy to be implemented
% (\eg~maintaining the longest queue),
% the advances of switch chip has also opened up some opportunities
% that allowing us to innovate BM.

(1) \emph{It is easier to extend the memory bandwidth in modern switches.}
Different from the traditional switches using off-chip buffer,
modern switches have embedded the packet buffer in the switch chip.
Different off-chip buffer,
on-chip buffer has a significantly wider memory data path,
since no external pins (\ie~pins connected to external memory)
are required~\cite{BroadcomSmartBuffer, SwitchBook, CiscoSiliconOneHBM},
thereby reducing the requirements of IC packaging (\ie~without the need for inter-chip connection).
As a result, the memory bandwidth can be significantly increased.
% Second, embedded memory can be clocked much faster
% due to lower lead inductance and reduced drive currents.
% For example, Intel Tofino 2 has a switching capacity of 12.8Tbps,
% and its packet buffer can achieve 25Tbps of write and read throughput~\cite{HotChips20Tofino2}.
% As a result, the packet buffer is not in the critical path any longer.

(2) \emph{Preemptive BM does not consume the bandwidth of cell data memory.}
As explained in \textsection\ref{sec:back:buffer},
with the current buffer structure,
dropping a packet only needs to dequeue the PD
and move the packet's cell pointers to the free cell pointer list.
These operations can be finished within PD memory and cell pointer memory.
There is no need to access the cell data memory,
thereby reducing the memory bandwidth requirement.

(3) \emph{The throughput of reading cell pointers can be improved by batching.}
A potential bottleneck of dequeuing a packet lies in the cell pointer memory.
This is because
cell pointers are very small,
and the throughput of reading cell pointers is limited by the clock rate (rather than the memory bandwidth),
which is unfortunately hard to be increased due to the limits of thermal dissipation.
% Furthermore, reading a packet may require fetching multiple cell pointers.
% Furthermore, as packets are usually partitioned into fixed-size cells,
% reading a packet requires fetching multiple PDs.
% For example, consider a switch chip with a clock rate of 1.2GHz.
% Assume that it can perform 4 reads per clock cycle and each read can fetch 2 cells.
% If the cell size is 200B, then 1600B data can be fetched per clock cycle,
% resulting in a maximum read throughput of $\sim$15.3Tbps.
% % Note that a cell may not be filled with packet data.
% For a 12.8Tbps switch chip, this leaves little excess bandwidth.
Nevertheless,
as described in \textsection\ref{sec:back:buffer},
the read throughput can be increased by reading multiple cell pointers at a time.
For example, a PD can maintain 4 parallel cell pointer lists,
allowing 4 cell pointers to be read simultaneously.

(4) \emph{In practice, the switch does not always use up all memory bandwidth.}
For a commodity switch chip,
the TM is designed to be able to achieve full bisection bandwidth.
However, in practice,
it is unlikely that all ports are sending and receiving packets at line rate,
as the traffic load rarely reaches 100\% for all ports.
% and the traffic from different ingress ports may head for the same egress port.
% Besides, some ports may be left unused for future expansion.
To concretely show this,
we conduct the same large-scale simulation as in \textsection\ref{sec:motv:cause}.
\figurename~\ref{fig:uti-on-drop:memory bandwidth}
shows the CDF of memory bandwidth utilization on packet drop with different network loads,
where utilization = $\frac{\textrm{consumed memory bandwidth}}{\textrm{overall memory bandwidth}}$.
We can see that,
even under 90\% network load,
the median free memory bandwidth is $\sim$38\%.

\section{\sysname{} Design}\label{sec:design}
% In this section, we first introduce the design goals,
% followed by \sysname{}'s high-level ideas and overall structure.
% Finally, we present the design details.
In this section, we first introduce \sysname{}'s high-level ideas and overall structure.
Then, we present the design details.

\subsection{Overview}\label{sec:design:overview}
To maintain simplicity while leveraging redundant memory bandwidth for packet expulsion,
\sysname{} embodies three main ideas:

(1) \emph{\sysname{} decouples admission and expulsion.}
To overcome Difficulty~\ref{diff:enqueue},
% With Pushout, packet admission module
% can trigger the packet vacation operation,
% and packet admission may wait for the completion of vacating buffer.
% Different from Pushout,
\sysname{} makes the admission and expulsion mutually independent.
In this way, the admission module
can enqueue the admitted packets immediately,
keeping the enqueue operations simple.
However, this can result in buffer starvation.
As the enqueue operations do not wait,
any arriving packet is dropped once the buffer is full.
Thus, we use the following idea to avoid this issue.
% The vacation is.
% At the input side, a packet is enqueued immediately once passing the admission control.
% In this way, \sysname{} keeps the simplicity of packet enqueue operations.

(2) \emph{\sysname{} proactively reserves a small fraction of free buffer.}
Since the enqueue operations do not wait,
\sysname{} proactively reserves a portion of free buffer
to protect newly active flows from buffer starvation.
Note that only a small fraction is needed
because \sysname{} can agilely vacate buffer for newly active flows.

(3) \emph{\sysname{} reactively expels packets for all over-allocated queues in a round-robin manner.}
To overcome Difficulty~\ref{diff:longest},
\sysname{} monitors all over-allocated queues,
which is simple as it can be realized by comparing the queue lengths to a threshold.
Besides, rather than only pushing out the longest queue,
\sysname{} expels packets for all over-allocated queues in a round-robin manner.

% (3) \emph{
% \sysname{} reactively adjusts buffer allocation with head drop.}
% When the buffer allocation requires adjustment,
% \sysname{} utilizes the redundant bandwidth
% to actively vacate the over-allocated buffer.
% The packet vacation is achieved by dropping packets at the head of a queue.
% Head drop is no more complex than a dequeue operation,
% making the packet vacations simple to be implemented.

\begin{figure}[!t]
    \centering
    \resizebox{\linewidth}{!}{\begin{tikzpicture}[font=\Large]
    \tikzset{
        every node/.style={
            text=dadiannaotextgray,
        },
        data label/.style={
            text=dadiannaolinegray, font=\Large\it
        },
        control block/.style={
            fill=gray, rectangle,
            text=white, font=\Large\bf,
        },
        admission block/.style={
            control block, fill=dadiannaodarkblue,
        },
        vacation block/.style={
            control block, fill=dadiannaored,
        },
        pd/.style={
            draw=darkgray, rectangle, minimum width=3mm, minimum height=5mm,
            font=\normalsize, very thick,
        },
        pd head/.style={
            draw, circle, minimum width=1mm, very thick,
        },
        pd tail/.style={
            draw, circle, minimum width=1mm, very thick,
        },
        packet/.style={
            fill=darkgray, rectangle, minimum width=6mm, minimum height=1cm,
            text=white,
            font=\large,
        },
        arbiter/.style={
             trapezium,
             trapezium angle=135,
             trapezium stretches=true,
        },
        demux/.style={
             trapezium,
             trapezium angle=135,
             trapezium stretches=true,
        },
        control path/.style={
            dashed, line width=2pt, dadiannaolinegray,
        },
        data path/.style={
            line width=3pt, dadiannaolinegray,
        },
        pd queue/.style={
            queue, line width=2pt, draw=dadiannaolinegray, minimum width=2cm, minimum height=5mm,
        },
    }

    % PD memory
    % pds
    % \node[pd queue, draw=none] at (0, 0) (q0) {};
    % \node[below=2mm of q0.center] (q1) {\bf\huge $\vdots$};
    % \node[below=5mm of q1.center, pd queue] (q2) {};
    % \foreach \qid in {0} {
    %     \foreach \pos in {0.2, 0.4, 0.6} {
    %         \draw[line width=2pt, dadiannaolinegray] ($(q\qid.north east)!\pos!(q\qid.north west)$) -- ($(q\qid.south east)!\pos!(q\qid.south west)$);
    %     }
    % }

    % pd module border
    % \coordinate (pd-list-ne) at ([xshift=3mm, yshift=3.9mm] q0.north east);
    % \coordinate (pd-list-sw) at ([xshift=-3mm, yshift=-3.9mm] q0.south west);
    \coordinate (pd-list-ne) at (2.5cm, 1.2cm);
    \coordinate (pd-list-sw) at (0, 0);
    \coordinate (pd-list-nw) at (pd-list-ne-|pd-list-sw);
    \coordinate (pd-list-se) at (pd-list-ne|-pd-list-sw);
    \coordinate (pd-list-e) at ($(pd-list-ne)!0.5!(pd-list-se)$);
    \coordinate (pd-list-w) at ($(pd-list-nw)!0.5!(pd-list-sw)$);
    \coordinate (pd-list-n) at ($(pd-list-nw)!0.5!(pd-list-ne)$);
    % \node[below, dadiannaotextgray] at ($(pd-list-sw)!0.5!(pd-list-se)$) {PD Memory (Queues)};
    \node[dadiannaotextgray] at ($(pd-list-nw)!0.5!(pd-list-se)$) {PD\\Memory};

    % cell ptr memory border
    \coordinate (cell-ptr-ne) at ([yshift=-2mm] pd-list-se);
    \coordinate (cell-ptr-sw) at ([xshift=-2.5cm, yshift=-1.2cm] cell-ptr-ne);
    \coordinate (cell-ptr-se) at (cell-ptr-ne |- cell-ptr-sw);
    \coordinate (cell-ptr-nw) at (cell-ptr-ne -| cell-ptr-sw);
    \coordinate (cell-ptr-w) at ($(cell-ptr-nw)!0.5!(cell-ptr-sw)$);
    \coordinate (cell-ptr-e) at ($(cell-ptr-ne)!0.5!(cell-ptr-se)$);
    \coordinate (cell-ptr-s) at ($(cell-ptr-se)!0.5!(cell-ptr-sw)$);
    \coordinate (cell-ptr-n) at ($(cell-ptr-ne)!0.5!(cell-ptr-nw)$);
    \node[dadiannaotextgray] at ($(cell-ptr-nw)!0.5!(cell-ptr-se)$) {Cell Ptr\\Memory};

    % \draw[data path, <-, line width=2pt] ($(pd-list-w)!0.5!(cell-ptr-w)$) -- +(-8mm, 0)
    %     node[pos=0.68, above, data label] {PD}
    %     node[left, admission block, minimum height=2.5cm] (admission) {Admission};
    \path ($(pd-list-w)!0.5!(cell-ptr-w)$) -- +(-1.2cm, 0)
        node[left, admission block, minimum width=1cm, minimum height=2.8cm] (admission) {};
    \node[rotate=-90, admission block] at (admission) {Admission};
    \draw[data path, <-, line width=2pt]
        (pd-list-w) -- (admission.east |- pd-list-w)
        node[pos=0.6, above, data label] {PD};
    \draw[data path, <-, line width=2pt]
        (cell-ptr-w) -- (admission.east |- cell-ptr-w)
        node[pos=0.6, above, data label] {Cell\\Ptr};
    % control path
    \begin{scope}[control path]
        \path (admission.north) -- +(0, 8mm)
            node[above, control block, minimum height=1.2cm] (qlen) {Statistics \\ \normalsize (\eg, qlen)};
        \draw[->] (qlen) -- (admission);
        \draw[->] ($(pd-list-n)!0.5!(pd-list-nw)$) -- +(0, 2mm) -| ($(qlen.south)!0.5!(qlen.south east)$)
            node[above, pos=0.25, data label] {Update};
        \node[vacation block, arbiter, minimum height=1cm] at (qlen-|pd-list-n) (arbiter) {Arbiter};
        \node[vacation block, above left=5mm and -3mm of arbiter, minimum height=1.2cm] (selection) {Head-drop\\Selector};
        \node[control block, above right=5mm and -3mm of arbiter, minimum height=1.2cm] (scheduler) {Output\\Scheduler};
        \draw[->] (qlen) |- (selection);
        \draw[->] (selection.south) -- +(0, -1mm) -| ([xshift=-5mm]arbiter.north);
        \draw[->] (scheduler.south) -- +(0, -1mm) -| ([xshift=+5mm]arbiter.north);
        \draw[->] (arbiter) -- (pd-list-n);
    \end{scope}

    % data path
    \begin{scope}[data path, line width=2pt]
        \node[right=1.5cm of $(pd-list-e)!0.5!(cell-ptr-e)$, control block, minimum width=1cm, minimum height=2.8cm, anchor=center] (dequeue) {};
        \node[control block, rotate=-90] at (dequeue) {Dequeue};
        \node[right=8mm of dequeue.east, vacation block, demux, minimum height=8mm, rotate=-90, anchor=center] (demux) {Demux};
        \draw[->] (dequeue) -- (demux);
        \draw[->] (demux.north west) -- +(5mm, 0)
            node[right, vacation block, minimum height=6mm, minimum width=2.4cm] (headdrop) {Head Drop};
        \draw[->] (demux.north east) -- +(5mm, 0)
            node[right, control block, minimum height=6mm, minimum width=2.4cm] (cell-read) {Cell Read};
        \draw[->] (pd-list-e) -- (dequeue.west|-pd-list-e) node[above, pos=0.45, data label] {PD};
        \draw[->] (cell-ptr-e) -- (dequeue.west|-cell-ptr-e) node[above, pos=0.45, data label] {Cell\\Ptr};
    \end{scope}
    \draw[control path, ->] (arbiter.south) -- +(0, -1mm) -| (demux);
    % buffer module
    \coordinate (buffer-nw) at ([yshift=-8mm] admission.south west);
    \coordinate (buffer-ne) at (buffer-nw -| cell-read.south east);
    \coordinate (buffer-se) at ([yshift=-2cm] buffer-ne);
    \coordinate (buffer-sw) at (buffer-nw|-buffer-se);
    \node[below right, packet] (pkt-1) at ([xshift=2mm, yshift=-2mm] buffer-nw) {};
    \foreach[remember=\i as \pi (initially 1)] \i in {2, 3, ..., 9} {
        \node[packet, right=2mm of pkt-\pi] (pkt-\i) {};
    }
    % \node[below] at ($(buffer-se)!0.5!(buffer-sw)$) {Packet Buffer (Packet Data)};
    \node[above] at ($(buffer-se)!0.5!(buffer-sw)$) {Cell Data Memory};
    \begin{scope}[data path]
        \draw[->] (admission) -- (admission |- buffer-nw) node[pos=0.39, right, data label] {Cell Data};
        \draw[<-] (cell-read) -- (cell-read |- buffer-nw) node[midway, left, data label] {Cell Data};
    \end{scope}
    % TM module
    \coordinate (tm-nw) at ([shift={(-2mm, 3mm)}] qlen.west |- selection.north);
    \coordinate (tm-se) at ([shift={(2mm, -8mm)}] buffer-se);
    \coordinate (tm-sw) at (tm-nw|-tm-se);
    \coordinate (tm-ne) at (tm-nw-|tm-se);
    \coordinate (tm-w) at ($(tm-nw)!0.5!(tm-sw)$);
    \coordinate (tm-e) at ($(tm-ne)!0.5!(tm-se)$);
    \coordinate (tm-s) at ($(tm-se)!0.5!(tm-sw)$);
    \coordinate (tm-n) at ($(tm-ne)!0.5!(tm-nw)$);
    \node[above] at (tm-s) {\bf\Large Traffic Manager};
    \begin{scope}[on background layer]
        \fill[dadiannaoyellow, line width=2pt] (tm-nw) rectangle (tm-se);
        \draw[draw=white, fill=gray, fill opacity=0.3, line width=2pt] (pd-list-nw) rectangle (pd-list-se);
        \draw[draw=white, fill=gray, fill opacity=0.3, line width=2pt] (buffer-nw) rectangle (buffer-se);
        \draw[draw=white, fill=gray, fill opacity=0.3, line width=2pt] (cell-ptr-nw) rectangle (cell-ptr-se);
    \end{scope}

    \draw[data path, ->, line width=3.5pt] (cell-read.east)   -- +(+6mm, 0)
        node[right=3mm, anchor=center, rotate=-90, text=gray] {To Egress};
    \draw[data path, <-, line width=3.5pt] (admission.west) -- +(-6mm, 0)
        node[left =3mm, anchor=center, rotate=-90, text=gray] {From Ingress};

    \draw[control path, data label, line width=1pt, <-] ([xshift=-1mm, yshift=-3mm] tm-ne) -- +(-5mm, 0) node[left] {control path};
    \draw[data path,    data label, line width=1pt, <-] ([xshift=-1mm, yshift=-9mm] tm-ne) -- +(-5mm, 0) node[left] {data path};
\end{tikzpicture}}
    \caption{\sysname{} overview}\label{fig:structure}
\end{figure}

Overall, \figurename~\ref{fig:structure}
depicts the high-level structure of \sysname{}.
Different from a non-preemptive BM,
% (\reffig{fig:switch-packet-flow})
\sysname{} adjusts the buffer allocation
with both \textcolor{dadiannaodarkblue}{\bf packet admission}
and \textcolor{dadiannaored}{\bf packet expulsion}.

\mypara{\textcolor{dadiannaodarkblue}{Packet admission (\textsection\ref{sec:design:admission})}}
Packet admission lies at the ingress side of TM.
For each arriving packet,
\sysname{} determines
whether to admit it into the packet buffer
based on queue length statistics.
% After passing the admission control,
% the packet data is written into the packet buffer.
% Meanwhile, a packet descriptor (PD),
% containing the address of the packet's memory location,
% is pushed to the corresponding queue, which lies in a separate PD memory.
% Each queue is implemented as a linked list of PDs.
% The PD linked lists are typically stored in a separate SRAM.

\mypara{\textcolor{dadiannaored}{Packet expulsion (\textsection\ref{sec:design:vacation})}}
Packet expulsion lies at the egress side of TM.
It contains four modules:
head-drop selector, arbiter, demultiplexer, and head-drop executor.
First, a head-drop selector picks an over-allocated queue for expulsion
based on queue length statistics.
Next, an arbiter resolves the read conflicts
between the head-drop selector and the output scheduler.
For each fetched PD,
\sysname{} determines whether to drop or dequeue the packet for transmission
through a demultiplexer,
with the head-drop executor handling the packet drops.
% These two modules work independently.
% The scheduler module is the same as that of the existing switch chip,
% which selects a queue to pop a PD,
% reads the corresponding packet from the packet buffer,
% and sends the packet to egress packet processing (EPP).
% The vacation module iteratively checks whether a queue is over-allocated with buffer.
% If so, the vacation module will remove the packet at the head of the queue.

\subsection{Packet Admission}\label{sec:design:admission}
Similar to non-preemptive BM schemes,
\sysname{} proactively reserves a fraction of free buffer
to avoid buffer starvation for newly active queues.
Unlike non-preemptive BM schemes,
\sysname{} only needs to reserve a \emph{small} fraction of free buffer,
as it can utilize head drop to quickly expel the over-allocated buffer.

To realize this,
\sysname{} leverages DT with adjusted parameter
for both high efficiency and ease of implementation.
As detailed in \textsection\ref{sec:back:bm},
DT is very simple and has already been widely deployed in the commodity switch chip.
Utilizing the existing mechanisms can facilitate the implementation of \sysname{}.
Moreover, the efficiency of DT can be tuned
through its parameter, $\alpha$~\cite{TON98DT}.
Concretely, a larger $\alpha$ enables DT to
make more efficient use of scarce memory resources.
For example, with $\alpha = 8$, DT allows a queue to occupy 88.9\% of the buffer.
Nevertheless, $\alpha$ should not be arbitrarily large,
as \sysname{} requires some buffer reservation to
accommodate the newly arriving bursts
(more details in~\textsection\ref{sec:params}).
% We analyze the settings of $\alpha$ in detail in Appendix~\ref{sec:params}.

Combining the above two ideas,
\sysname{} avoids introducing new mechanisms for the packet admission module.
Instead, we only need to adjust the parameter of DT in the existing switch chip.

% An alternative design is to admit packets whenever there is free buffer size
% and vacate packets from a queue when the buffer becomes full (\ie~like Pushout).
% In this way, we can achieve the optimal buffer efficiency.
% However, we argue that it is not worthwhile to realize this approach for two reasons.
%
% (1) It couples the enqueue operation and dequeue operation.
% An enqueue operation can uncertainly wait for the execution of the dequeue operation.
% This can complicate the hardware design.
%
% (2) It needs to select a queue for packet vacation when the buffer is full.
% Optimally, the longest queue should be selected.
% However, maintaining the longest queue in real time needs some extra logic,
% and thus can make hardware design intricate.
% % And arbitrarily selecting a queue can result in unfairness issues.
% One may argue that we can leverage the recent advances on
% finding the heavy hitters~\cite{NSDI16FlowRadar, SIGCOMM16UnivMon, SOSR17HashPipe, TON20PRECISION}.
% However, these techniques are not free and require some memory resource,
% which is already scarce, to store the hash table.
% Due to the above reasons, we argue that this approach contradicts
% our goal to make \sysname{} as simple as possible.

\subsection{Packet Expulsion}\label{sec:design:vacation}
The core idea of this module is to leverage redundant memory bandwidth
to actively expel the over-allocated buffer.
Next, we will present its design details.

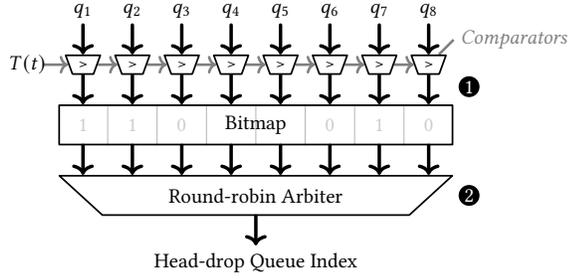
\begin{figure}[!t]
    \centering
    \resizebox{0.9\linewidth}{!}{\begin{tikzpicture}[
    comparator/.style={
        draw,
        shape=trapezium,
        trapezium angle=110,
        line width=1pt,
    },
    font=\Large,
    bitmap/.style = {
        draw,
        minimum height=8mm,
        minimum width=8cm,
        line width=1pt,
    },
    arbiter/.style={
        draw,
        minimum width=8cm,
        minimum height=8mm,
        line width=1pt,
        trapezium,
        trapezium angle=135,
        trapezium stretches=true,
    },
    numcircle/.style={
        fill, text=white, circle, inner sep=1pt, font=\Large,
    },
]
    % bitmap
    \node[bitmap] (bitmap) {Bitmap};
    % comparators
    \foreach \pos/\i in {0.0625/1, 0.1875/2, 0.3125/3, 0.4375/4, 0.5625/5, 0.6875/6, 0.8125/7, 0.9375/8} {
        \draw[<-, line width=2pt]
            ($(bitmap.north west)!\pos!(bitmap.north east)$) -- ++(0, 6mm)
            node[above, comparator, fill=white] (comp-\i) {\normalsize >};
        \draw[<-, line width=2pt]
            (comp-\i.north) -- ++(0, 6mm) node[above] {$q_\i$};
        \draw[->, line width=2pt] ($(bitmap.south west)!\pos!(bitmap.south east)$) -- ++(0, -6mm);
    }
    \begin{scope}[
        on background layer,
        every path/.style={
            gray, line width=1.5,
        }
    ]
            \foreach[remember=\i as \pi (initially 1)] \i in {2, 3,..., 8} {
                \draw[->] (comp-\pi) -- (comp-\i);
            }
            \draw[<-] (comp-1) -- ++(-8mm, 0) node[left=-2mm, black] (t-text) {$T(t)$};
            \draw[-] (comp-8.north east) -- ++(3mm, 3mm) node[right] (cmp-text) {\it Comparators};
    \end{scope}
    \begin{scope}[on background layer]
        \foreach \pos in {0.125, 0.25, 0.375, 0.5, 0.625, 0.75, 0.875} {
            \draw[opacity=0.3] ($(bitmap.north west)!\pos!(bitmap.north east)$) -- ($(bitmap.south west)!\pos!(bitmap.south east)$);
        }
        \foreach \pos/\val in {0.0625/1, 0.1875/1, 0.3125/0, 0.6875/0, 0.8125/1, 0.9375/0} {
            \coordinate (bitmap-start) at ($(bitmap.north west)!\pos!(bitmap.north east)$);
            \node[opacity=0.2] at ($(bitmap.west)!\pos!(bitmap.east)$) {\val};
        }
    \end{scope}
    \node[arbiter, below=6mm of bitmap] (arbiter) {Round-robin Arbiter};
    \draw[->, line width=2pt] (arbiter.south) -- +(0, -6mm)
        node[below] {Head-drop Queue Index};
    % \path (t-text.south) -- (bitmap.north west) node[pos=0.5, left=1mm] (headdrop-maintain-text) {\numcircled{1}};
    \path (bitmap.north east) -- (bitmap.north east|-comp-8.south)
        node[pos=0.6, right=1mm, numcircle] (headdrop-maintain-text) {1};
    \node[numcircle] at (arbiter-|headdrop-maintain-text) {2};
\end{tikzpicture}}
    \vspace{-0.1in}
    \caption{Head-drop selector}
    \label{fig:headdrop-queue-select}
    \vspace{-0.2in}
\end{figure}

\mypara{Selecting a head-drop queue}
As analyzed in~\textsection\ref{sec:motv:problem},
a BM can work anomalously
when the queue length is higher than the threshold (\ie~$T(t)$).
Thus, \sysname{} considers a queue over-allocated
if and only if its queue length is higher than $T(t)$.
Moreover,
rather than maintaining the longest queue
that can incur non-trivial costs,
\sysname{} maintains all over-allocated queues
and selects a queue in a round-robin manner.
The above functions are achieved in the head-drop selector module,
whose circuit diagram is shown in \figurename~\ref{fig:headdrop-queue-select}.
The head-drop selector module contains two parts:
maintaining the over-allocated queues (\fillcircle{1})
and iterating over them (\fillcircle{2}).

% Thus, \sysname{} considers a queue as over-allocated
% if and only if its length is larger than the threshold given by DT (\ie~$T(t)$).
\fillcircle{1}
To maintain the over-allocated queues,
\sysname{} uses a bitmap,
whose length is equal to the number of queues,
to indicate whether a queue is over-allocated.
Each bit of the bitmap represents whether the queue length exceeds $T(t)$.
For example, the bitmap \texttt{00000101} denotes that queue 1 and queue 3 are over-allocated.
Each bit is set by comparing the queue length to $T(t)$,
using some simple combinatorial logic that consists of some cheap comparators.

\fillcircle{2}
With multiple over-allocated queues,
\sysname{} iterates over them with a round-robin arbiter,
which is a common hardware component in
high-speed crossbar switches~\cite{
    Micro99PPE, NoCRouterArch, ISSS02Arbiter,
}.
The round-robin arbiter takes the bitmap as input
and gives the index of an over-allocated queue.

% \reffig{fig:headdrop-queue-select}
% shows the circuit diagram of iterating the head-drop queues.
% \sysname{} finds the next queue for head drop with the PPE (\fillcircle{1}).
% The PPE takes the bitmap and the index of the last head-dropped queue as inputs
% and determines the index of the next non-zero bit in the bitmap,
% which is the index of the next head-drop queue.
% % The inputs of the PPE are the bitmap and the index of the previous head-drop queue.
% % PPE will consider the bit of previous queue index as the lowest priority,
% % and find the index of the first non-zero bit in priority order.
% % The output of the priority encoder is the index of the first significant bit of the masked bitmap,
% % which is the index of the next head-drop queue.
% Finding the target head-drop queue,
% \sysname{} can move on to the next queue
% by setting the current queue index to the lowest priority (\fillcircle{2}).
% This is achieved by storing the current queue index to some flip-flops,
% which are connected to the PPE's lowest-priority input signals.
% % For example, when the bitmap is \texttt{11001} and the target queue index is 1,
% % the bitmask in the next round will be \texttt{11110}.
% % The masked bitmap will be \texttt{11000},
% % which discards the head-drop queue in the current round.

\mypara{Resolving the conflicts between the output scheduler and head-drop selector}
The head-drop operation also needs to consume the read bandwidth of PD memory and cell pointer memory.
Thus, the read conflict arises
when both output scheduler and head-drop selector fetch packets simultaneously.
% The read operation from the scheduler module can conflict with that from \sysname{}.
In such cases,
the head-drop selector should give way to the output scheduler.
Otherwise, \sysname{} may not guarantee line-rate forwarding,
which is unacceptable.
Thus, \sysname{} uses a fixed-priority arbiter
to resolve the conflict.
The read requests from the head-drop selector are blocked
whenever the output scheduler needs to fetch a packet.

\mypara{Performing the head-drop action}
If a head-drop request is granted,
\sysname{} extracts a packet at the head of the queue and drops it.
% As mentioned in \textsection\ref{sec:back:buffer},
% a queue is maintained as a linked list of Packet Descriptors (PDs)
% that lies in the PD memory,
% and the occupied buffer is maintained through cell
As explained in \textsection\ref{sec:motv:opportunities},
the head-drop operation
does not need to access the cell data memory.
% and clear the physical memory area.
Instead, \sysname{} simply dequeues the corresponding PDs from the PD linked list,
and returns the cell pointers of the packet to the free cell pointer list.
% Thus, the head-drop operations do not
% consume the bandwidth of the packet buffer.
Moreover,
as the head-drop action shares lots of operations with the normal dequeue action,
it can be synthesized into the existing packet dequeue pipelines.

\begin{figure}
    \centering
    \includegraphics[width=.9\linewidth]{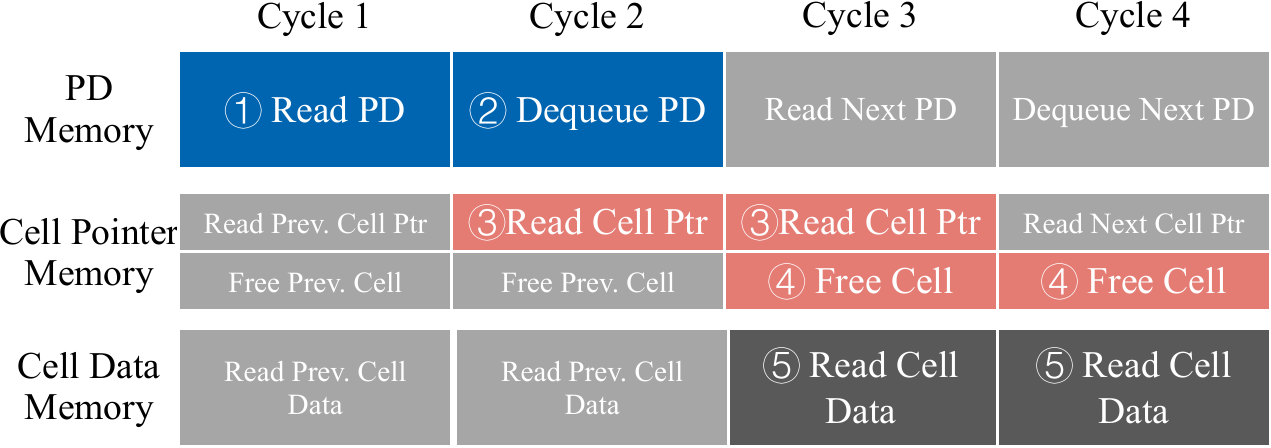}
    \vspace{-.1in}
    \caption{Packet dequeue pipeline}\label{fig:cycles}
    \vspace{-.2in}
\end{figure}
Specifically,
as shown in \figurename~\ref{fig:cycles},
a dequeue action mainly consists of 5 operations:
\numcircled{1} Read a PD from PD memory;
\numcircled{2} Dequeue the PD (\ie~advance the head of the PD linked lists);
\numcircled{3} Read cell pointer;
\numcircled{4} Free cell (\ie~move the cell pointer to the free cell pointer list);
\numcircled{5} Read cell data.
With a packet partitioned into multiple cells,
operation \numcircled{3}, \numcircled{4}, and \numcircled{5} can be executed multiple times.
Moreover, as PD memory, cell pointer memory, and cell data memory are physically separate,
accesses to them can be parallelized.
In this pipeline,
the only difference between the head-drop action and the normal dequeue action is
that the head-drop action does not need to read the cell data (\ie~operation \numcircled{5}).
Thus, with \sysname{}, these operations can be recomposed to
% The existing switch performs $\numcircled{1} \to \numcircled{2} \to \numcircled{3} \to \numcircled{3} \to \numcircled{4}$ in order.
% With \sysname{}, the operations can be recomposed to
\tikzset{
    opnode/.style={
        draw, circle, inner sep=1pt,
        line width=1pt,
    },
    headdrop/.style={
        opnode,
        draw=none, fill=dadiannaored, text=white,
    },
    demux/.style={
        draw=dadiannaored, trapezium, trapezium angle=135,
        anchor=center, rotate=-90, line width=1pt,
        inner sep=3pt,
    },
}
\begin{center}
\begin{tikzpicture}
    \tikzset{
        every node/.style = {opnode},
    }
    \node[] at (0, 0) (op1) {1};
    \node[right=0.07\linewidth of op1] (op2) {2};
    \node[right=0.07\linewidth of op2] (op3) {3};
    \node[right=0.07\linewidth of op3] (op4) {4};
    \node[right=0.07\linewidth of op4] (op5) {5};
    \node[right=0.07\linewidth of op5, dadiannaolinegray] (op6) {1};
    \node[right=0.07\linewidth of op6, dadiannaolinegray] (op7) {2};
    \node[draw=none, right=0.08\linewidth of op7, dadiannaolinegray] (op8) {$\cdots$};
    % \node[right=0.08\linewidth of op4, demux] (demux) {};
    % \node[right=0.08\linewidth of demux.west] (op5) {5};
    % \node[right=0.08\linewidth of demux.east, draw=none, text=dadiannaored] (op6) {NOP};
    % \node[right=0.12\linewidth of op3] (op4) {4};
    \begin{scope}[->, line width=1pt]
        \draw[] (op1) -- (op2);
        \draw[] (op2) -- (op3);
        \draw[] (op3) -- (op4);
        \draw[] (op4) -- (op5);
        \draw[] (op5) -- (op6);

        \draw[dadiannaolinegray] (op6) -- (op7);
        \draw[dadiannaolinegray] (op7) -- (op8);
        \draw[dadiannaored]
            ($(op4)!0.5!(op5)$) -- ++(0, -3mm) -| ($(op5)!0.5!(op6)$);
    \end{scope}
    % \draw[->, line width=1pt] (op4) -- (demux);
    % \draw[->, line width=1pt] (demux.north|-op5) -- (op5);
    % \draw[->, line width=1pt] (demux.north|-op5) -- (op5);
    % \draw[->, line width=1pt, dadiannaored] (demux.north|-op6) -- (op6);
    % \draw[->, line width=1pt] (op3) -- (op4);
    % \node at (3, 0) (op3) {\numcircled{3}};
    % \node at (4, 0) (op4) {\numcircled{4}};
\end{tikzpicture}
\end{center}
where
\tikz[baseline=(char.base)]{\draw[dadiannaored, ->, line width=1pt] (0, 0) -- +(5mm, 0)}
denotes the (extra) execution path introduced by \sysname{}.
% where \tikz[baseline=(char.base)]{\node[demux, inner sep=2pt] (op5) {}} denotes a demultiplexer.
% and \tikz[baseline=(char.base)]{\node[headdrop, inner sep=0.5pt] (op5) {\small 5}} denotes
% the operation of dropping the PD.

\subsection{Parameter Settings}\label{sec:params}
\sysname{} includes a parameter $\alpha$,
which influences the buffer efficiency and fairness of \sysname{}.
In this part, we analyze and discuss the settings of $\alpha$.

Since \sysname{} directly uses DT for admission control,
we can follow the method and conclusion in~\cite{TON98DT}
to analyze \sysname{}'s efficiency and fairness.
According to~\cite{TON98DT},
the amount of reserved free buffer in the steady state can be given by
\begin{equation}
    F = \frac{B}{1 + \alpha N}
    \label{eq:free-buffer-size}
\end{equation}
where $B$ is the buffer size,
$N$ is the number of congested queues whose length is equal to the threshold.
\refeq{eq:free-buffer-size} denotes that
\sysname{} reserves less free buffer with a larger $\alpha$, resulting in higher efficiency.
Nevertheless, the efficiency improvement is undermined as the $\alpha$ becomes larger.
For example,
when $N=1$,
the free buffer reservation is $B/9$ with $\alpha = 8$ and $B/17$ with $\alpha=16$,
resulting in only a 5.2\% increase in buffer utilization
(note that $\alpha$ is typically a factor of two to facilitate the threshold calculation).
Thus, it is not necessary to set a very high value for $\alpha$.

On the other hand, according to~\cite{TON98DT},
when bursty traffic arrives at inactive queues,
\sysname{} can fairly allocate buffer only if
\begin{equation}
    R \leqslant V\left(1 + \frac{1+\alpha N}{\alpha M}\right)
    \label{eq:healthy-condition}
\end{equation}
where $R$ is the arrival rate of traffic bursts,
$V$ is the expulsion rate,
$M$ is the number of queues receiving traffic bursts,
and $N$ is the number of over-allocated queues.
Inequality~\eqref{eq:healthy-condition} can be rewritten as
\begin{equation}
    \frac{1}{\alpha} \geqslant \left(\frac{R}{V} - 1\right) M - N
    \label{eq:healthy-condition-tmp1}
\end{equation}
Inequality~\eqref{eq:healthy-condition-tmp1} implies that
$\alpha$ can be set to a larger value with a higher packet expulsion rate.
For example, with one over-allocated queue (\ie~$N=1$) and one burst arrival (\ie~$M = 1$),
Inequality~\eqref{eq:healthy-condition-tmp1} can be rewritten
as $\frac{1}{\alpha} \geqslant \frac{R}{V} - 2$.
When $V \geqslant R/2$, $\alpha$ can be arbitrarily high in theory.
% When $V \leqslant R/3$, $\alpha$ should be lower than 1.

In sum, setting $\alpha$ involves a tradeoff between efficiency and fairness,
and this tradeoff is less tense with a higher expulsion rate.
Our experiments (in \textsection\ref{sec:eval:params}) show that $\alpha = 8$ is an appropriate choice.
Increasing $\alpha$ further does not result in notable improvement in efficiency,
but can result in unfairness.

\subsection{Discussions}
\mypara{Impact of preemption}
One might wonder whether the preemption (\ie~expelling packets by head drop)
introduced by \sysname{} could impact normal packet processing (\eg~scheduling) at switches.
Our design ensures that the preemption does not cause any impacts
for the following two reasons.
\numcircled{1}~With the fixed-priority arbiter,
the preemption only occurs when the output scheduler is not fetching packets from the queue,
avoiding interference with the ongoing packet processing.
\numcircled{2}~If the output scheduler needs to fetch packets from the queue while \sysname{} is performing preemption,
the preemption process can be interrupted at any time.
Specifically, as shown in Figure 9,
interruption occurs at the beginning of either Cycle 1/2 or Cycle 3/4.
In the former case (at the beginning of Cycle 1 or 2),
\sysname{} has not yet modified to the queue (\ie~PD linked list),
allowing the output scheduler to dequeue packets as if \sysname{} were not present.
% and thus the output scheduler can dequeue packets as if \sysname{} was not present.
In the latter case (at the beginning of Cycle 3 or 4),
the PD to be expelled has already been removed from the queue by the end of Cycle 2,
so the output scheduler will believe that the corresponding packet has been dequeued,
thereby starting to dequeue a new packet.

\mypara{What if there is no redundant bandwidth?}
Even without redundant memory bandwidth,
\sysname{}'s performance remains stable.
Without redundant memory bandwidth (\ie~the memory bandwidth is saturated),
\sysname{} operates similarly to Dynamic Threshold (DT),
which is sufficient because there is no need to agilely adjust buffer allocations.
Specifically, memory bandwidth saturation occurs
only when all ports are transmitting and receiving packets at full rate.
In this case, the overall ingress traffic rate is nearly equal to the egress traffic rate.
From the buffer’s perspective,
the traffic arrival rate matches the departure rate,
leading to relatively stable buffer occupancy in each queue,
thereby reducing the need for agile buffer management.

\section{Implementation}\label{sec:impl}
In this section,
we first analyze the hardware costs of \sysname{}'s components.
Then we describe the implementations of a proof-of-concept hardware prototype based on P4
and a full-featured software prototype based on DPDK.
% Next, we describe the details of each prototype.
% \subsection{FPGA-based Hardware Prototype}

\subsection{Hardware Cost}\label{sec:impl:asic}
To evaluate the hardware cost of \sysname{},
we implement the head-drop selector (with a 64-bit bitmap), arbiter,
and head-drop executor
with 215, 11, and 60 lines of Verilog code, respectively.
To analyze the overhead of each hardware component,
we use Vivado Design Suite~\cite{Vivado}
to evaluate the FPGA resource consumption.
Furthermore,
we use Design Compiler~\cite{DesignCompiler} to synthesize each component
on an open-source 45nm ASIC technology library~\cite{45nmASICLib}
to evaluate the timing, chip area, and power consumption.

% \begin{table}
%     \small
%     \color{blue}
%     \caption{\small \revised{FPGA Resource Consumption}}\label{tab:fpga-cost}
%     \begin{center}
%         \begin{tabularx}{\linewidth}[c]{Y|Y|Y|Y}\toprule
%           \bf Module & \bf LUTs & \bf Flip Flops & \bf BRAM \\\midrule
%           Selector   & 218      & 46             & 0 \\
%           Arbiter    & 3        & 0              & 0 \\
%           Executor   & 47       & 7              & 0 \\
%           \bottomrule
%         \end{tabularx}
%     \end{center}
% \end{table}
% \begin{table}
%     \small
%     \color{blue}
%     \caption{\small \revised{ASIC Cost}}\label{tab:asic-cost}
%     \begin{center}
%         \begin{tabularx}{\linewidth}[c]{Y|Y|c|c}\toprule
%           \bf Module & \bf Timing ($ns$) & \bf Area ($\mu m^2$) & \bf Power ($mW$) \\\midrule
%           Selector   & 1.44              & 3,425.42             & 0.162 \\
%           Arbiter    & 0.17              & 23.93                & 0.003 \\
%           Executor   & 0.38              & 730.70               & 0.044 \\
%           \bottomrule
%         \end{tabularx}
%     \end{center}
% \end{table}
\begin{table}
    \small
    \caption{\small Hardware Cost}\label{tab:hardware-cost}
    \begin{center}
        \begin{tabularx}{\linewidth}[c]{cYYYYY}\toprule
                        & \multicolumn{2}{c}{\bf FPGA Cost} & \multicolumn{3}{c}{\bf ASIC Cost} \\ \cmidrule(r){2-3} \cmidrule(l){4-6}
          \bf Module  & \bf \multirow{2}{*}{LUTs} & \bf Flip Flops & \bf Timing ($ns$)  & \bf Area ($mm^2$)  & \bf Power ($mW$)\\\midrule
          Selector    & 1262                      & 47             & 1.49               & 0.023              & 0.895 \\
          Arbiter     & 3                         & 0              & 0.17               & 2.3e-5             & 0.003 \\
          Executor    & 47                        & 7              & 0.38               & 7.3e-4             & 0.044 \\
          \bottomrule
        \end{tabularx}
    \end{center}
\end{table}

\tablename~\ref{tab:hardware-cost}
shows the FPGA resource consumption of each component reported by Vivado
and the ASIC cost reported by Design Compiler.
Note that the numbers are per switch chip,
as all queues share the same head-drop selector, arbiter, and executor.
The majority of hardware cost comes from the head-drop selector.
Specifically,
the head-drop selector requires $\sim$1200 LUTs and dozens of Flip Flops,
incurring little resource consumption on FPGA
(note that an FPGA chip typically
contains hundreds of thousands of LUTs and Flip Flops~\cite{AlveoU50}).
Besides, the timing reports show that
the head-drop selector has a delay of less than 1.5ns,
indicating that \sysname{} can expel a packet every 2 cycles with a 1GHz clock,
which is fast enough, as expelling a packet usually requires several cycles to dequeue multiple cell pointers.
% The timing reports show that
% the delay of every component is less than 1ns,
% indicating that \sysname{} can reach a clock frequency of 1GHz.
Furthermore, \sysname{} introduces less than 0.03$mm^2$ ASIC area and 1$mW$ power,
which are negligible for a commodity switch ASIC.

\subsection{P4-based Hardware Prototype}\label{sec:impl:p4}
We implement \sysname{} with 616 lines of code in P4,
which can run on our Intel Tofino switch.
As the TM cannot be programmed,
we cannot implement the head-drop selector and fixed-priority arbiter.
Consequently,
we only implement the packet admission module and head-drop operations
to examine the benefits brought by actively expelling packets for over-allocated queues.

\mypara{Admission}
The admission control is implemented in the ingress pipeline.
The admission module requires queue lengths and threshold.
In our switch,
queue lengths can be directly obtained at the egress pipeline.
We follow the method in \cite{FB, SIGCOMM21AIFO} to make them available at the ingress pipeline.
Specifically,
we generate special packets called SYNC packets.
These packets read the queue lengths from the egress pipeline and return to the ingress pipeline via recirculation,
where the queue lengths are copied to the ingress register array.
The threshold is also calculated as SYNC packets traverse through the egress pipeline
and is synchronized to the ingress pipeline in the same manner.
Whenever a packet arrives at the ingress pipeline,
we compare the length of the corresponding queue to the threshold given by DT,
deciding whether to accept the packet.

\mypara{Head drop}
The head drop is implemented in the egress pipeline.
Whenever a packet arrives,
we compare the threshold to the length of the queue it comes from.
The head drop operation is conducted by setting the metadata \texttt{drop\_ctl}.
Specifically, we set \texttt{drop\_ctl=1} if the queue length is higher than $T(t)$,
which tells the switch to drop the packet.
\begin{figure*}[!t]
    \centering
    \begin{subfigure}[]{.23\linewidth}
        \centering
        \includegraphics[width=\linewidth]{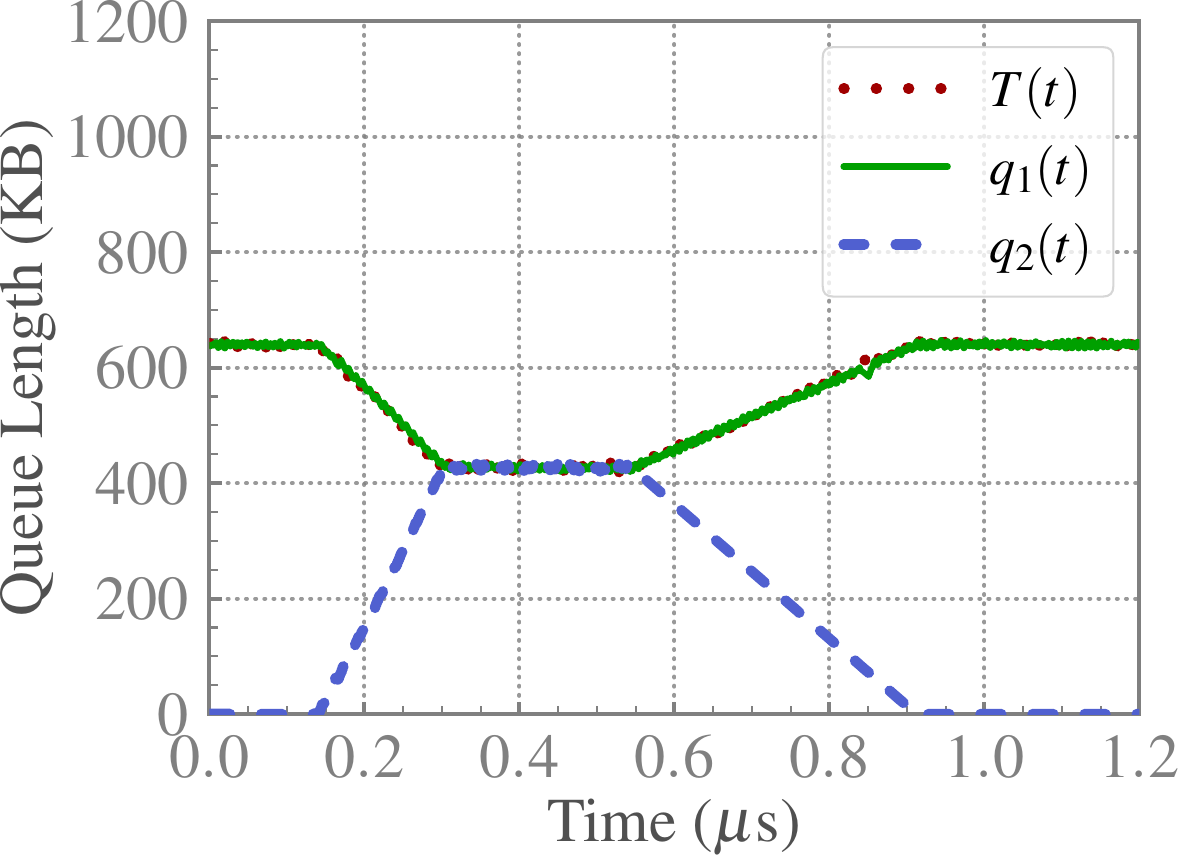}
        \caption{\sysname{}, $\alpha=1$}\label{fig:tofino:qevo:pbuffer-1}
    \end{subfigure}
    \hfil
    \begin{subfigure}[]{.23\linewidth}
        \centering
        \includegraphics[width=\linewidth]{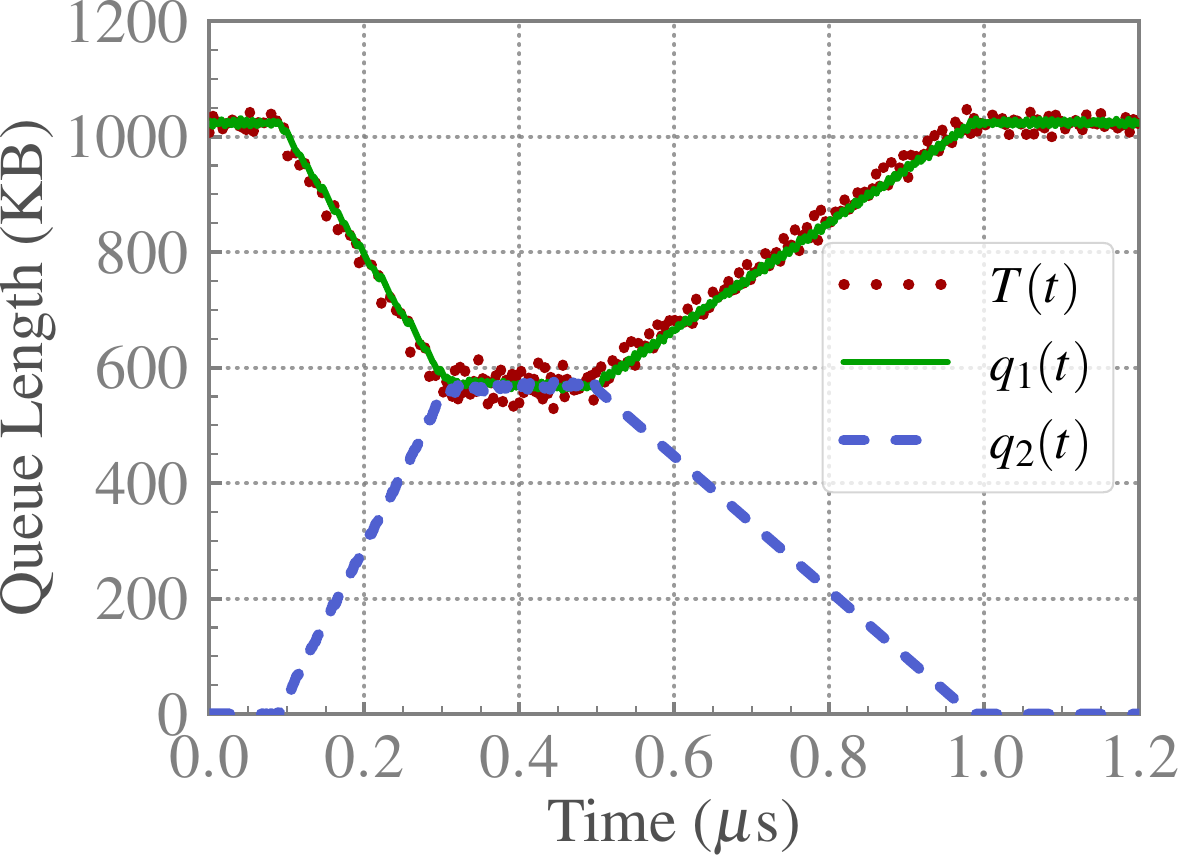}
        \caption{\sysname{}, $\alpha=4$}\label{fig:tofino:qevo:pbuffer-4}
    \end{subfigure}
    \hfil
    \begin{subfigure}[]{.23\linewidth}
        \centering
        \includegraphics[width=\linewidth]{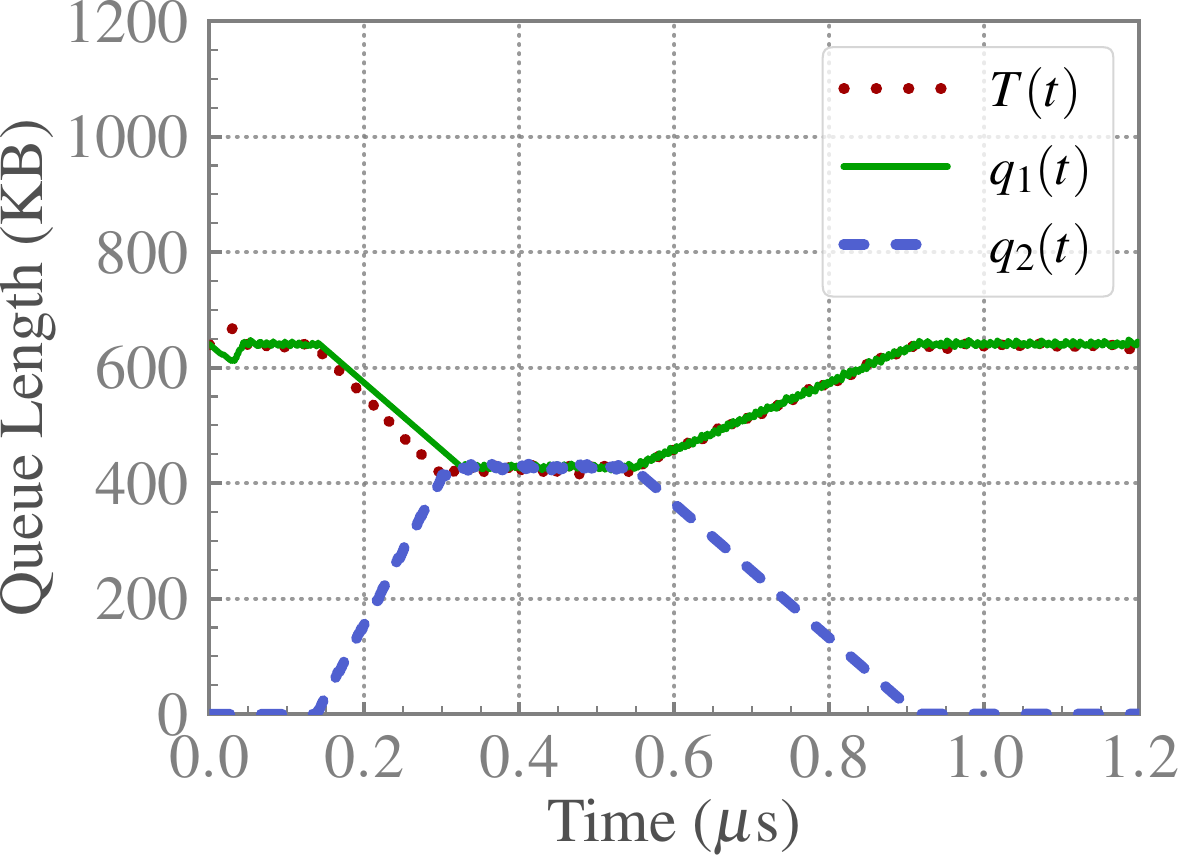}
        \caption{DT, $\alpha=1$}\label{fig:tofino:qevo:dt-1}
    \end{subfigure}
    \hfil
    \begin{subfigure}[]{.23\linewidth}
        \centering
        \includegraphics[width=\linewidth]{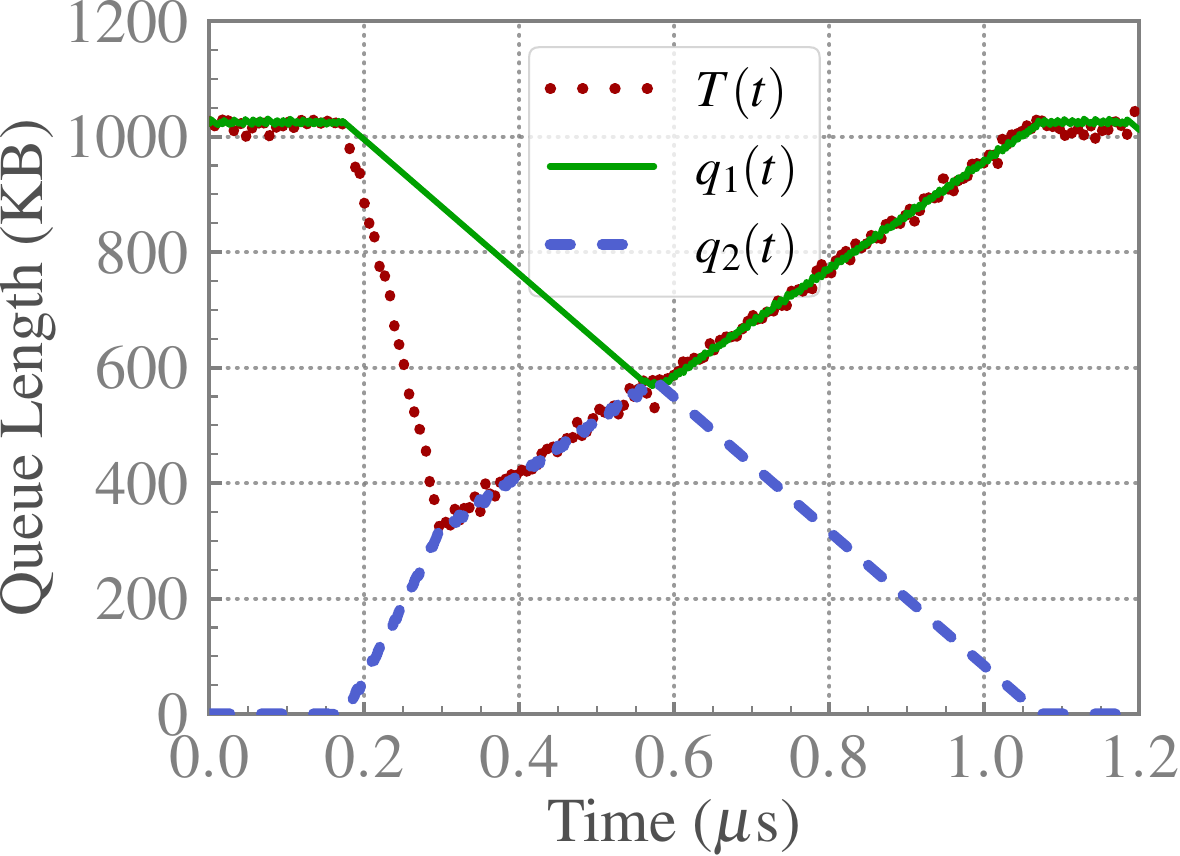}
        \caption{DT, $\alpha=4$}\label{fig:tofino:qevo:dt-4}
    \end{subfigure}
    \caption{[P4] Queue length evolution}\label{fig:tofino:qevo}
\end{figure*}
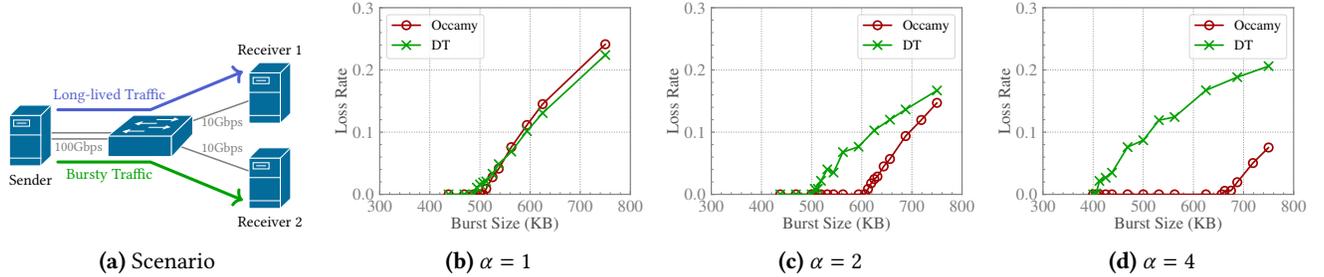
\begin{figure*}[!t]
    \centering
    \begin{subfigure}[b]{.23\linewidth}
        \resizebox{\linewidth}{!}{\begin{tikzpicture}[]
    \node[switch] at (0, 0) (sw) {};
    \node[server, left=1 of sw] (sender) {};
    \node[server, above right=-0.5 and 1 of sw] (recver1) {};
    \node[server, below right=-0.5 and 1 of sw] (recver2) {};
    \begin{scope}[on background layer, every path/.style={gray, line width=1}]
        \draw[double, double distance=3pt]
            (sender.center) -- (sw.center)
            node[pos=0.4, below] {\small 100Gbps};
        \draw (sw.center) -- (recver1.center)
            node[pos=0.6, below] {\small 10Gbps};
        \draw (sw.center) -- (recver2.center)
            node[pos=0.6, above] {\small 10Gbps};
    \end{scope}
    \begin{scope}[every path/.style={line width=2}]
        \draw[->, darkblue]
            ([yshift=6mm] sender.east) to
            ([yshift=6mm] sw.center) to
            ([yshift=5mm] recver1.west)
            node[pos=0, above left=6mm and -5mm] {Long-lived Traffic};
        \draw[->, darkgreen]
            ([yshift=-6mm] sender.east) to
            ([yshift=-6mm] sw.center) to
            ([yshift=-5mm] recver2.west)
            node[pos=0, below left=6mm and -2mm] {Bursty Traffic};
    \end{scope}
    \node[below of=sender] {Sender};
    \node[above of=recver1] {Receiver 1};
    \node[below of=recver2] {Receiver 2};
\end{tikzpicture}}
        \caption{Scenario}\label{fig:tofino:burst:scenario}
    \end{subfigure}
    \hfil
    \begin{subfigure}[b]{.23\linewidth}
        \centering
        \includegraphics[width=\linewidth]{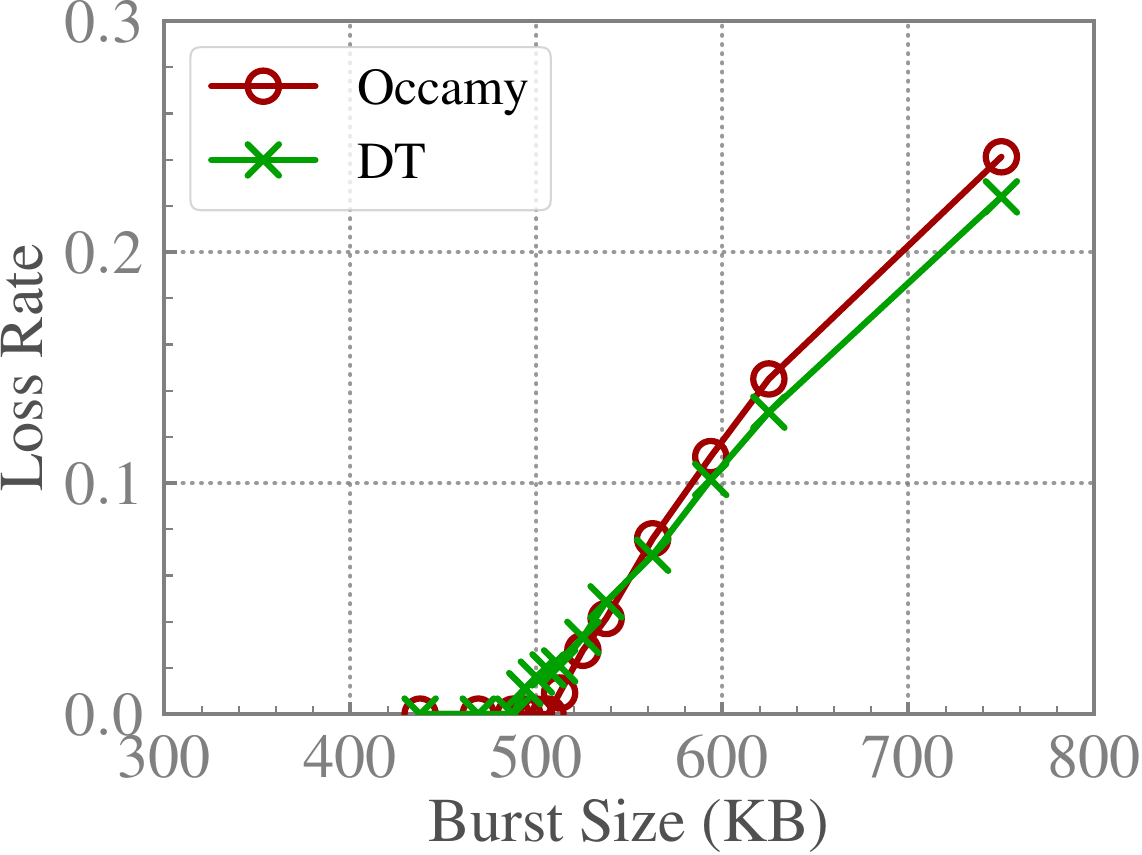}
        \caption{$\alpha=1$}\label{fig:tofino:burst:alpha-1}
    \end{subfigure}
    \hfil
    \begin{subfigure}[b]{.23\linewidth}
        \centering
        \includegraphics[width=\linewidth]{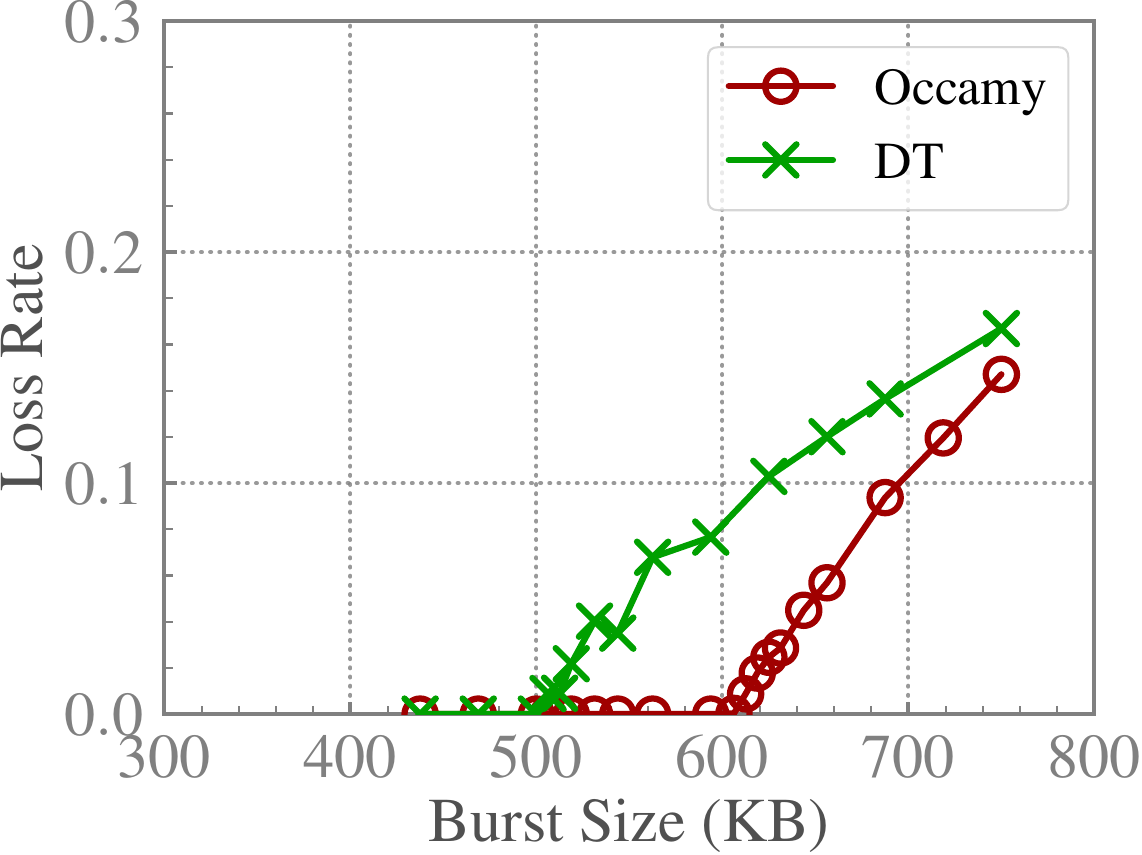}
        \caption{$\alpha=2$}\label{fig:tofino:burst:alpha-2}
    \end{subfigure}
    \hfil
    \begin{subfigure}[b]{.23\linewidth}
        \centering
        \includegraphics[width=\linewidth]{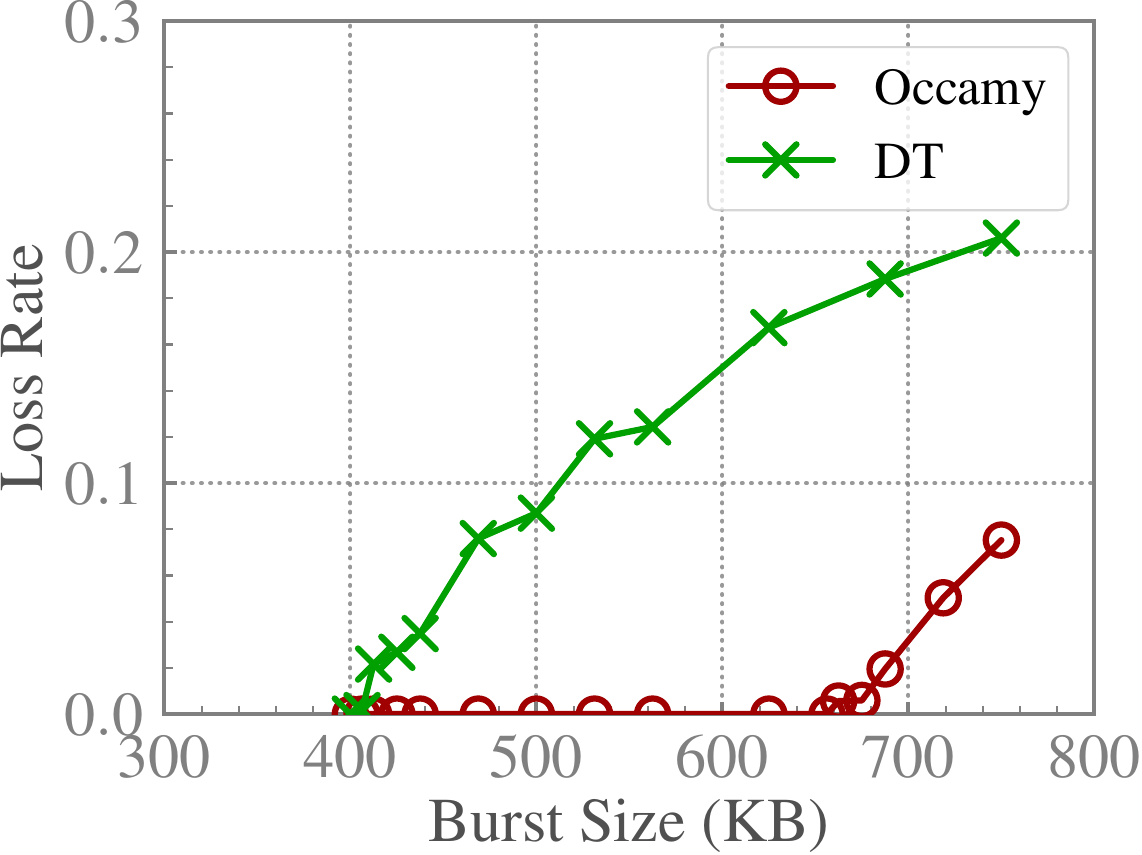}
        \caption{$\alpha=4$}\label{fig:tofino:burst:alpha-4}
    \end{subfigure}
    \caption{[P4] Ability of absorbing traffic bursts with different $\alpha$s}\label{fig:tofino:burst}
\end{figure*}

\subsection{DPDK-based Software Prototype}\label{sec:impl:dpdk}
% To compare \sysname{} with Pushout\footnotemark,
We also implement a software prototype of \sysname{} with DPDK.
% \footnotetext{
%     Pushout cannot be implemented with P4.
%     This is because Pushout needs to push out packets from other queues
%     when a packet arrives at the ingress pipeline.
%     However, in P4, the ingress pipeline and egress pipeline are separated,
%     and the dequeue actions cannot be performed at the ingress pipeline.
% }
We use a server with eight 10Gbps NICs to emulate the switch
and implement \sysname{} on top of it.
The prototype consists of four modules:
an RX module, a forwarding module, a TX module, and an expulsion module.
% and a statistics maintenance module.
Each module runs on separate CPU cores.
\sysname{} is implemented on the forwarding module and expulsion module.

\mypara{RX, forwarding, and TX module}
The RX module polls each NIC port for packets
and delivers them to the forwarding module through an RX queue.
The forwarding module fetches packets from the RX queue,
and determines the destination queue of each packet.
Then the forwarding module executes \sysname{}'s admission control component,
\ie~it decides whether to admit the packet based on DT.
Passing the BM, the packet is put into the corresponding output queue.
The TX module fetches packets from the output queues and delivers them to the NIC.
Furthermore, to accelerate the TX module,
we assign a dedicated CPU core to send packets for each NIC port.

% \mypara{Forwarding module}
% The forwarding module maintains a forwarding table and determines the destination queue of each packet.
% Then it decides whether a packet is allowed to be enqueued based on DT.
% Passing the BM, the packet is put into the corresponding output queue.

\mypara{Expulsion module}
The expulsion module has two functions.
\numcircled{1}
It ensures that packet expulsion is performed
only with redundant memory bandwidth.
To achieve this goal,
we maintain a token bucket,
and generate tokens based on the switching capacity.
Specifically,
our software switch has eight 10Gbps ports
and thus can forward packets at a rate of 80Gbps.
We assume that each packet is partitioned into 200B cells
(Note that this is only an assumption, \ie~actually packets are not divided into cells),
which means that the switch can dequeue a cell every 20ns.
Consequently, we generate a token every 20ns and put it into a bucket\footnotemark{}.
When a packet is dequeued or dropped,
the corresponding number of tokens is removed from the bucket.
To guarantee line-rate forwarding,
the TX module is always allowed to dequeue packets and remove tokens from the bucket
even without enough available tokens
(in this case the number of tokens will become negative).
% We allow the number of tokens to be negative so that the tx module is always granted with enough tokens.
In comparison,
only with enough available tokens
can the expulsion module fetch a packet.
In this way, the expulsion module
can only utilize the redundant memory bandwidth for packet expulsion.
\numcircled{2} The expulsion module selects an over-allocated queue and drops the head packet.
To achieve this,
we maintain a bitmap to track which queues are over-allocated.
Then the expulsion module iterates over the over-allocated queues
and drops a packet at the head of the queue
(if there are enough tokens).

\footnotetext{
  In practice, it is difficult to achieve such a fine granularity.
  Instead, we generate $d/20$ tokens every time,
  where $d$ is the interval between token generations.
}

\section{Evaluation}\label{sec:eval}
In this section, we evaluate \sysname{}
with both testbed experiments
(\textsection\ref{sec:eval:p4}, \textsection\ref{sec:eval:dpdk},
and \textsection\ref{sec:eval:params})
and ns-3 simulations (\textsection\ref{sec:eval:sim}).

\begin{figure*}[!t]
    \centering
    \begin{subfigure}[b]{.23\linewidth}
        \includegraphics[width=\linewidth]{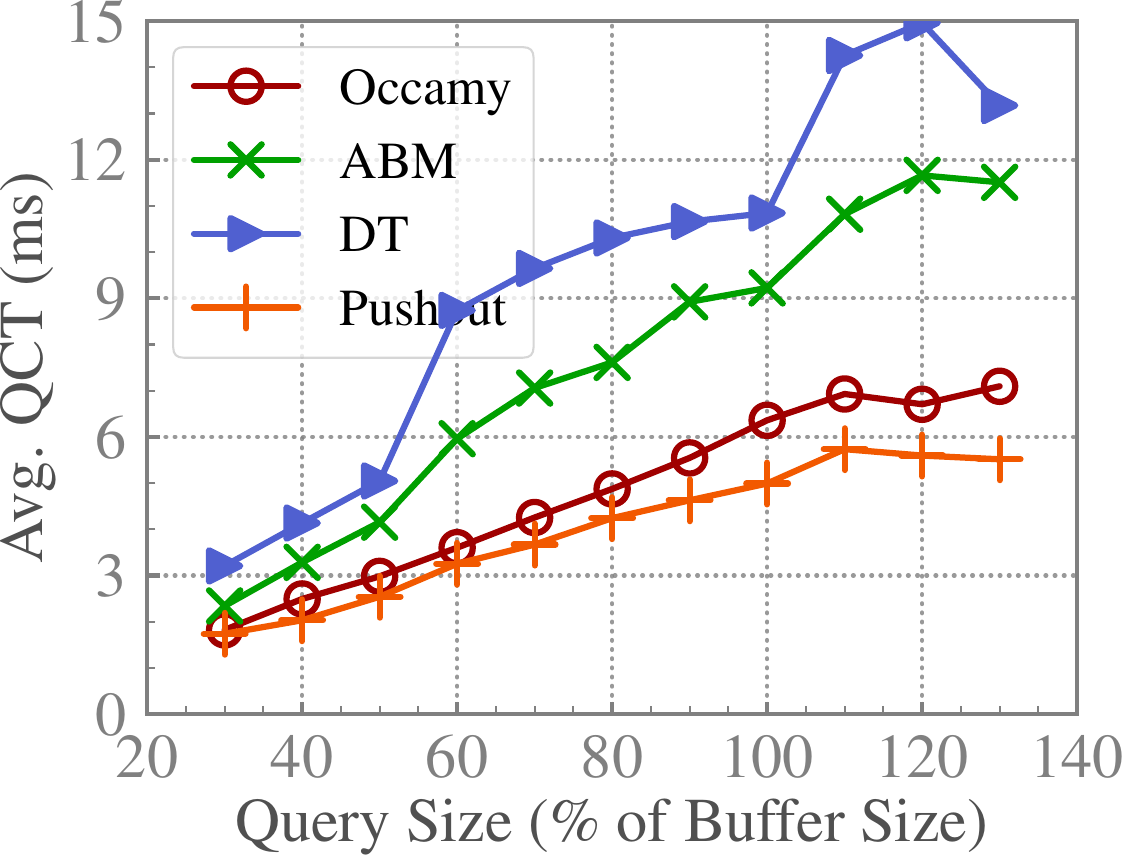}
        \caption{Query: Average QCT}\label{fig:dpdk:burst:avg-qct}
    \end{subfigure}
    \hfil
    \begin{subfigure}[b]{.23\linewidth}
        \centering
        \includegraphics[width=\linewidth]{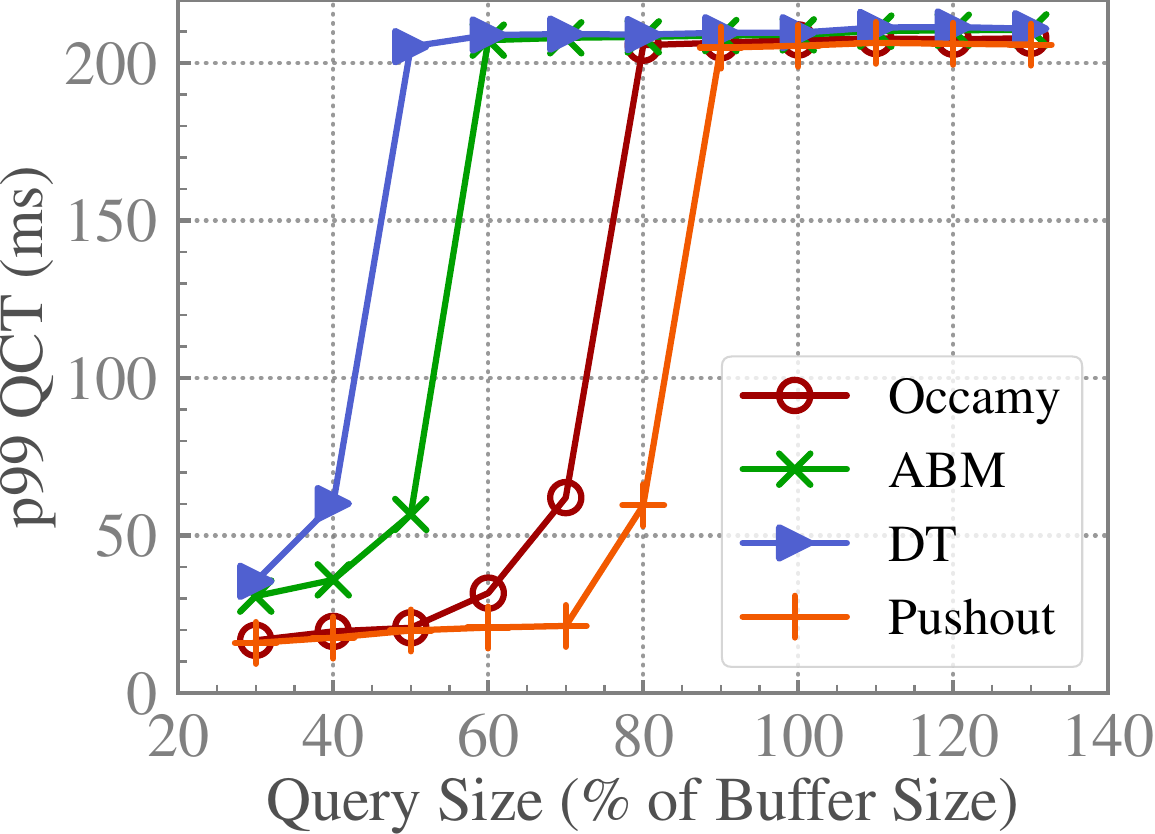}
        \caption{Query: 99th QCT}\label{fig:dpdk:burst:tail-qct}
    \end{subfigure}
    \hfil
    \begin{subfigure}[b]{.23\linewidth}
        \centering
        \includegraphics[width=\linewidth]{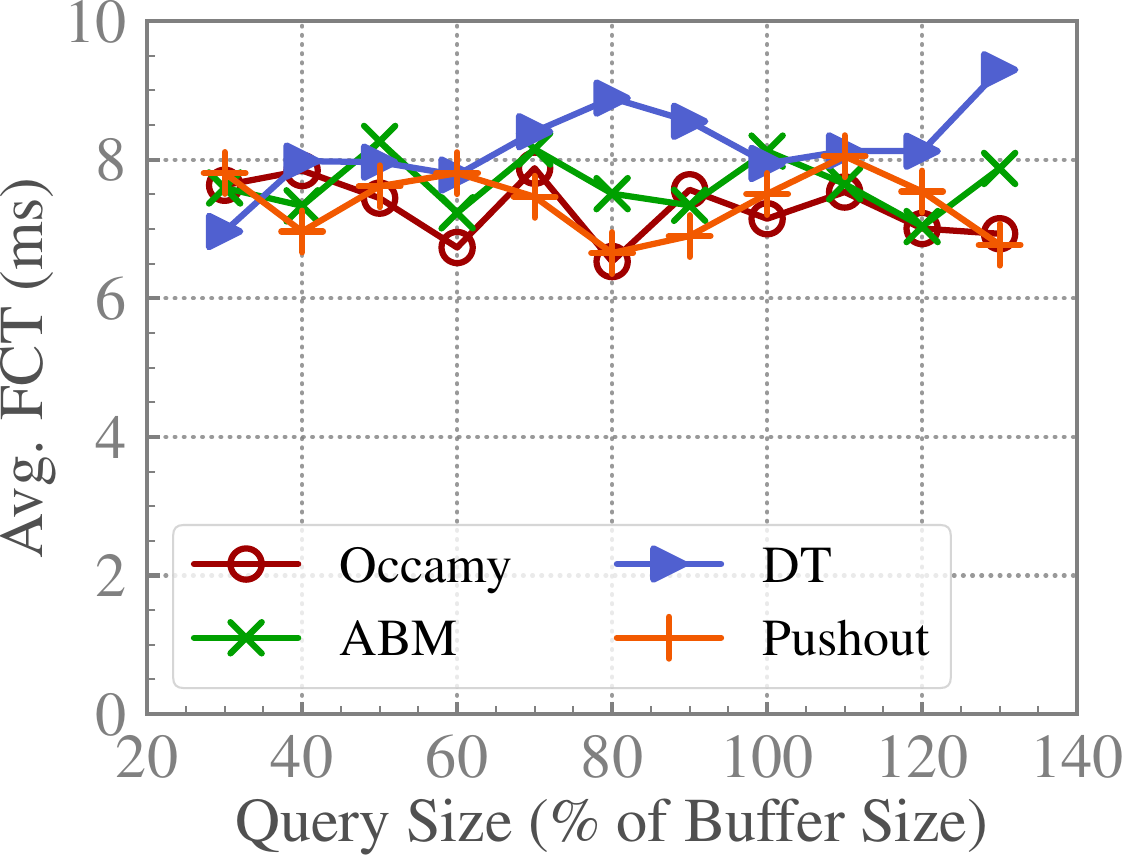}
        \caption{Overall Bg: Average FCT}\label{fig:dpdk:burst:avg-fct}
    \end{subfigure}
    \hfil
    \begin{subfigure}[b]{.23\linewidth}
        \centering
        \includegraphics[width=\linewidth]{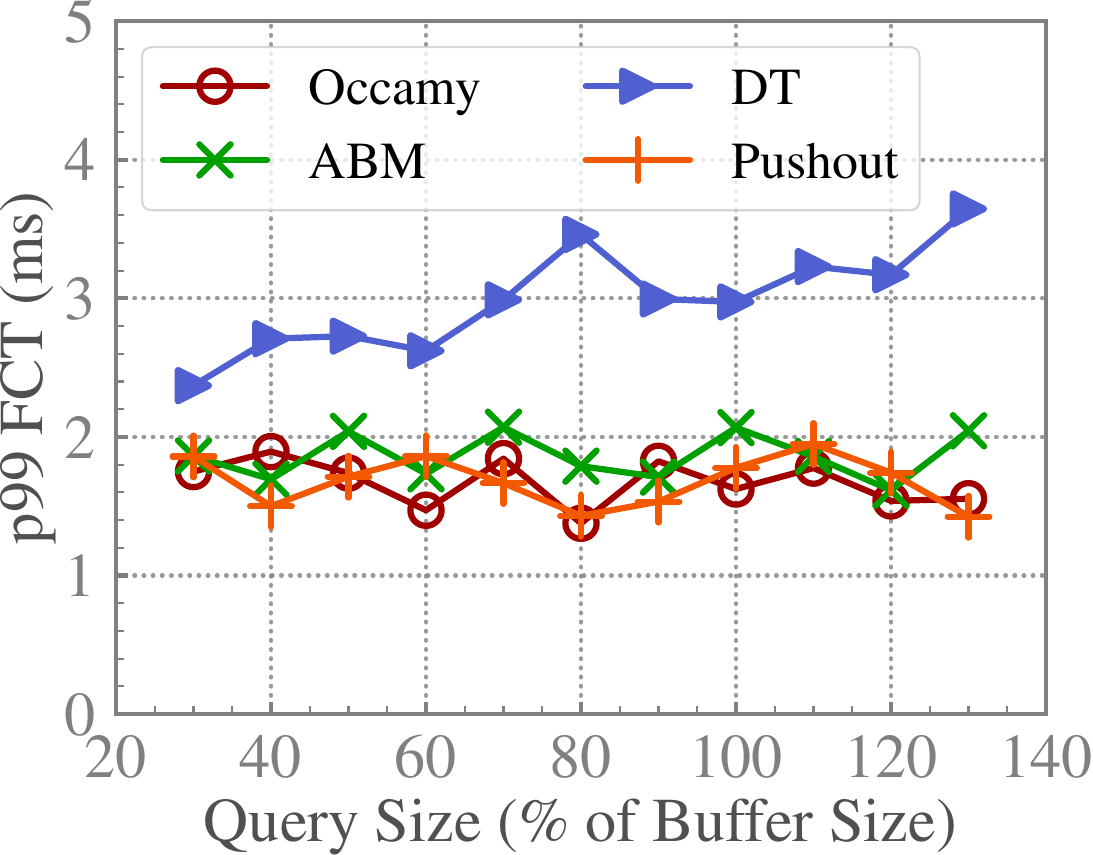}
        \caption{Small Bg: 99th FCT}\label{fig:dpdk:burst:tail-fct}
    \end{subfigure}
    \caption{Ability of absorbing traffic bursts}\label{fig:dpdk:burst}
\end{figure*}

\subsection{P4-based Hardware Prototype}\label{sec:eval:p4}
\mypara{Testbed Setup}
We build a small testbed with one sender and two receivers.
The topology is shown in \reffig{fig:tofino:burst:scenario}.
Each server has a 16-core 2.10GHz x86 CPU and 16GB memory.
The sender is equipped with a dualport 100GbE NIC,
connected to the switch with two 100Gbps links.
Each receiver is equipped with an 10GbE NIC,
connected to the switch with a 10Gbps link.
The switch has a 6.5Tbps Intel Tofino switch chip.
The traffic are generated by \texttt{Pktgen-DPDK}~\cite{pktgen-dpdk}.

\mypara{\sysname{} is agile and can quickly adjust buffer allocation}
For this experiment, we consider a scenario shown in \reffig{fig:tofino:burst:scenario}.
At the beginning,
we generate long-lived traffic from the sender to receiver 1.
After a while, we generate bursty traffic from the sender to receiver 2.
The long-lived traffic is ongoing throughout the experiment,
while the bursty traffic only lasts for $\sim$0.8$\mu$s.
The long-lived traffic and bursty traffic are sent from different NIC ports
and heading for different receivers
so as not to affect each other beyond the switch.

\figurename~\ref{fig:tofino:qevo} shows the queue length evolution of \sysname{} and DT.
\figurename~\ref{fig:tofino:qevo:pbuffer-1} and \reffig{fig:tofino:qevo:pbuffer-4}
show that \sysname{} can quickly adjust buffer allocation as the bursty traffic arrives,
and the bursty traffic does not drop packets
until it is allocated with the fair-share amount of buffer.
In comparison,
\figurename~\ref{fig:tofino:qevo:dt-1} and \figurename~\ref{fig:tofino:qevo:dt-4}
show that only with enough free buffer reservation
can DT adjust buffer allocation in time (\ie~$\alpha=1$).
\figurename~\ref{fig:tofino:qevo:dt-4} shows that,
without enough buffer reservation (\ie~$\alpha=4$),
DT is not able to quickly release the over-allocated buffer for the long-lived traffic.
As a result, the packets of the newly arrived bursty traffic are dropped
before obtaining the deserved buffer allocation.

% \begin{figure}[!t]
%     \centering
%     \begin{subfigure}[]{.46\linewidth}
%         \centering
%         \includegraphics[width=\linewidth]{burst-absorption-tofino-pbuffer.pdf}
%         \caption{pBuffer}\label{fig:eval:burst:tofino:pbuffer}
%     \end{subfigure}
%     \hfil
%     \begin{subfigure}[]{.46\linewidth}
%         \centering
%         \includegraphics[width=\linewidth]{burst-absorption-tofino-pbuffer.pdf}
%         \caption{DT}\label{fig:eval:burst:tofino:dt}
%     \end{subfigure}
%     \caption{[Tofino] Burst absorption}\label{fig:eval:burst:tofino}
% \end{figure}
%
\mypara{\sysname{} can absorb more traffic bursts}
We evaluate the ability of burst absorption
with both hardware prototype and software prototype.
For the hardware prototype, we use the same experiment setup as above.
\reffig{fig:tofino:burst} shows the loss rate of bursty traffic
with different burst sizes.
We make two observations:

\emph{(1)~\sysname{} can absorb more bursty traffic
by avoiding anomalous behavior.}
With the same amount of free buffer reservation,
\sysname{} can absorb more bursty traffic than DT.
For example,
\figurename~\ref{fig:tofino:burst:alpha-4} shows that
\sysname{} can absorb 57\% more bursty traffic
than DT with an $\alpha$ of 4.
This is attributed to the agility of \sysname{},
\ie~it can quickly make buffer for newly arrived bursty traffic.
In comparison, DT is not agile enough,
thereby reluctantly dropping packets of bursty traffic
before getting fair buffer share.

\emph{(2)~\sysname{} can absorb more bursty traffic by higher buffer efficiency.}
\figurename~\ref{fig:tofino:burst:alpha-1} and
\figurename~\ref{fig:tofino:burst:alpha-4} show that,
with $\alpha=4$,
\sysname{} can absorb 29\% more bursty traffic than that with $\alpha=1$.
In comparison, DT absorbs 12\% less bursty traffic
with $\alpha=4$ than that with $\alpha=1$.
This is because \sysname{} is agile even with a little free buffer reservation,
and thus can utilize more buffer for burst absorption.
In comparison, DT is not agile enough without enough free buffer reservation,
and thus cannot improve burst absorption by higher buffer efficiency.

\subsection{DPDK-based Software Prototype}\label{sec:eval:dpdk}
\mypara{Testbed Setup}\label{sec:eval:setup}
% \mypara{Software-switch-based testbed}
We build a small testbed with eight hosts.
Each host is a Dell PowerEdge R730 server with a 16-core Intel Xeon E5-2620 2.1GHz CPU
and 16GB DDR4 memory.
Each server is equipped with an Intel 82599 10GbE NIC,
connected to a DPDK-based software switch with a 10Gbps link.
The software switch is emulated by a server
with two Intel XL710 Quad Port 10GbE NICs.
Following the features of Broadcom Tomahawk switch chip~\cite{BroadcomTomahawkBuffer},
the switch contains 5.12KB buffer-per-port-per-Gbps (\ie~410KB in total).
Different from the testbed in P4-based evaluations,
the traffic is sent through Linux kernel's network stack,
which allow us to use DCTCP as the congestion control algorithm.
The ECN threshold is set to 65 packets as suggested by~\cite{SIGCOMM10DCTCP}.
% Specifically,
% for the hardware switch, the total buffer size is set to 1.25MB.
% For the software switch, the total buffer size is set to to 256KB.
We compare \sysname{} with DT, ABM~\cite{SIGCOMM22ABM}, and Pushout,
where ABM is a recently proposed non-preemptive BM that can ensure performance isolation.
Unless otherwise specified,
we set $\alpha=1$ for DT (as suggested by~\cite{TON98DT}),
$\alpha=2$ for ABM to achieve better performance\footnotemark,
and $\alpha=8$ for \sysname{}.
\footnotetext{We do not set a larger $\alpha$ for unscheduled packets
because we assume that the end hosts remain untouched.
}

% \subsection{\sysname{} Benefits}\label{sec:eval:testbed}

\mypara{\sysname{} can absorb more traffic bursts}
% For this experiment, we connect four servers to a four-port server-emulated switch with 1Gbps links.
We use the open-source traffic generator~\cite{NSDI16MQECN} to generate two kinds of traffic.
\numcircled{1} Bursty query traffic:
A client on each host periodically sends queries to 16 servers on other hosts
(each host runs 2 servers).
Receiving the query,
each server will generate a response to the client.
The total amount of response data (termed as query size)
is varied to generate different burst sizes.
The queries are generated according to a Poisson process,
with a network load of 1\%.
% The load of query traffic is 1\%.
\numcircled{2} Background traffic:
This kind of traffic follows a 1-to-1 pattern.
We generate background flows according to a Poisson process.
The sender and receiver are randomly chosen,
and the flow size follows the web-search distribution~\cite{SIGCOMM10DCTCP}.
The load of background traffic is 50\%.

\reffig{fig:dpdk:burst} shows the query completion time (QCT) of the query traffic
and the flow completion time (FCT) of the background traffic.
% Here, slowdown is the ratio between the actual value and the ideal value without other traffic.
We make two observations.

(1)~\sysname{} can significantly improve the performance of query traffic.
% For the average QCT slowdown (\figurename~\ref{fig:dpdk:burst:avg-qct}),
% \sysname{}'s performance is within $\sim$\inlinetodo{}\% of Pushout,
% which is considered as the optimal BM.
% \sysname{} can achieve comparable performance to Pushout.
% \sysname{}'s average QCT is within $\sim$\inlinetodo{}\% of Pushout.
\figurename~\ref{fig:dpdk:burst:avg-qct} shows that,
compared to DT and ABM,
\sysname{} can reduce the average QCT by up to $\sim$55\% and $\sim$42\%,
respectively.
% For the 99th percentile QCT slowdown (\reffig{fig:dpdk:burst:tail-qct}),
% \sysname{}'s performance is within 6.3-10.1\% of Pushout.
\figurename~\ref{fig:dpdk:burst:tail-qct} shows that
\sysname{} can avoid retransmission timeout (RTO)
with the burst size below 80\% of the buffer size,
which is 33\% and 60\% higher than DT and ABM, respectively.
(2)~\sysname{}'s performance gain of burst absorption does not come at the cost of hurting background flows.
\figurename~\ref{fig:dpdk:burst:avg-fct} shows that
the average FCT of \sysname{} is comparable to that of DT and ABM.
\figurename~\ref{fig:dpdk:burst:tail-fct} shows that the 99th percentile FCT for small flows
(\ie~flow size < 100KB)
is comparable to ABM and up to $\sim$57\% shorter than DT.
This is because \sysname{} only drops the over-allocated buffer of background flows,
rather than seizing their deserved buffer.

\begin{figure*}
\begin{minipage}[b]{.49\linewidth}
    \centering
    \begin{subfigure}[b]{0.46\linewidth}
        \centering
        \includegraphics[width=\linewidth]{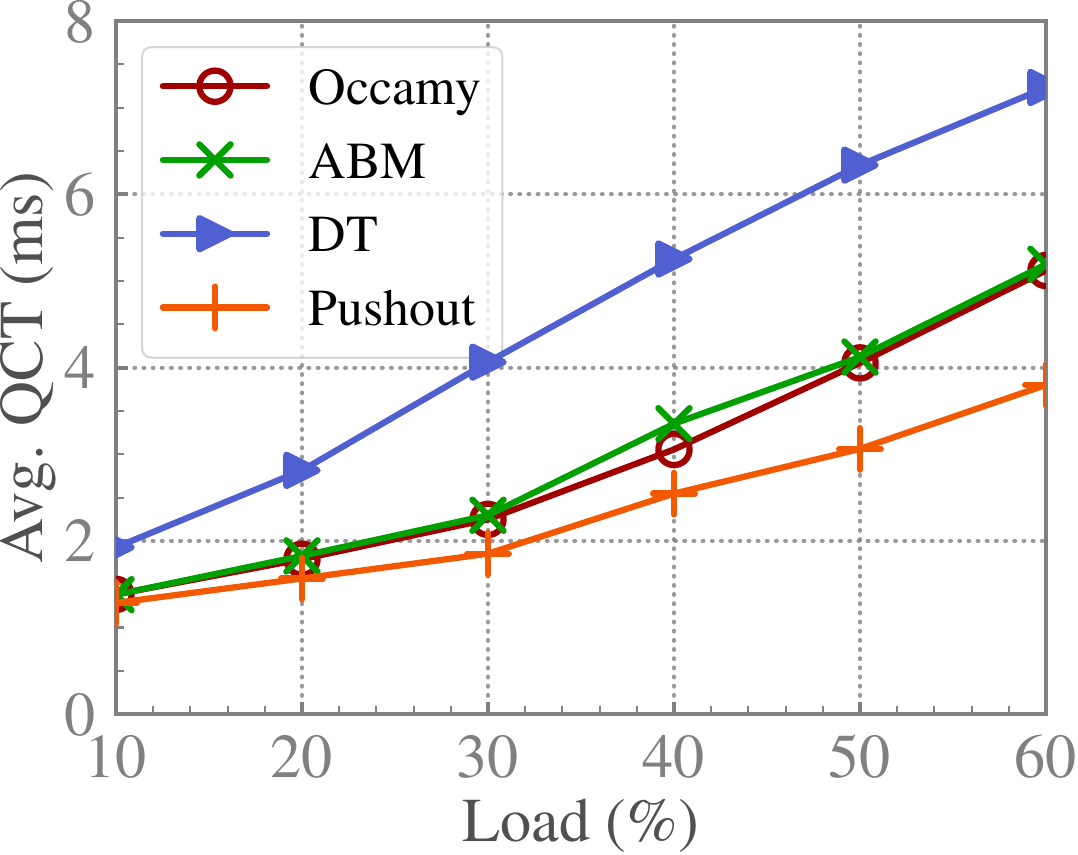}
        \caption{Average QCT}\label{fig:dpdk:isolation:avg-qct}
    \end{subfigure}
    \hfil
    \begin{subfigure}[b]{0.48\linewidth}
        \centering
        \includegraphics[width=\linewidth]{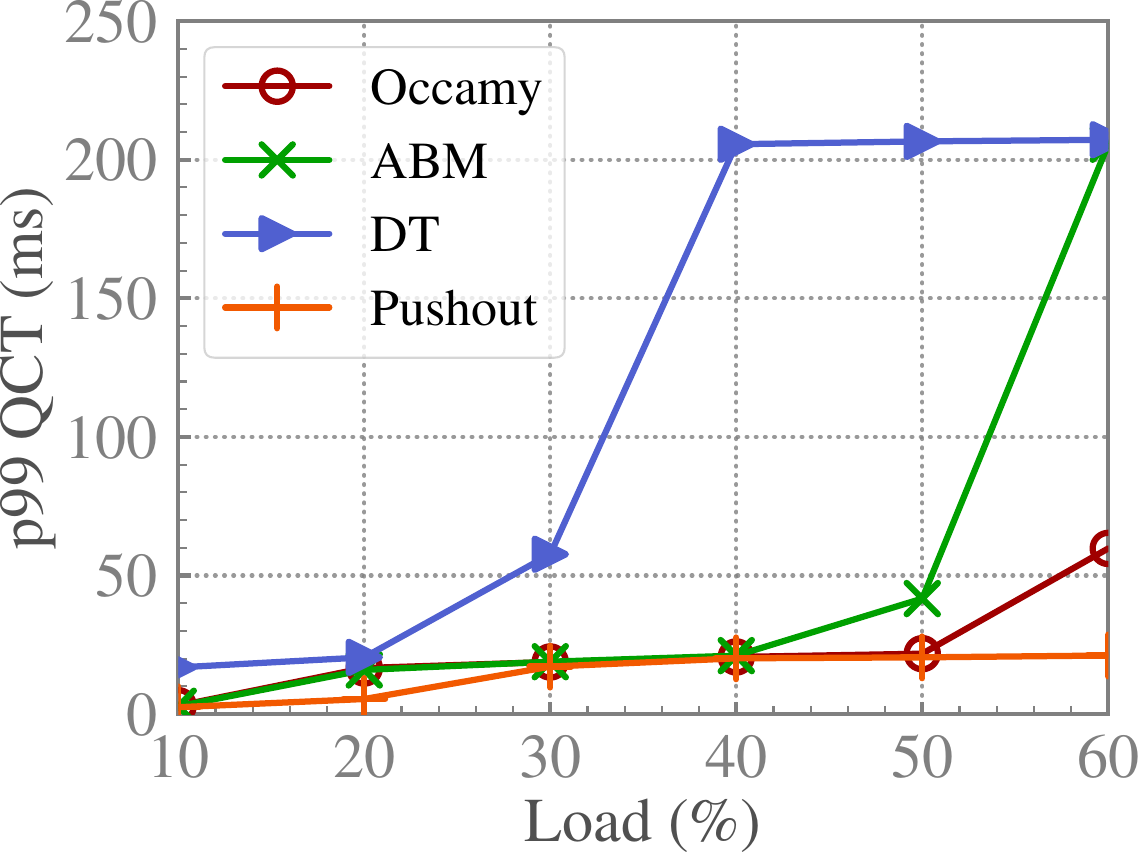}
        \caption{99th QCT}\label{fig:dpdk:isolation:tail-qct}
    \end{subfigure}
    % \hfil
    % \begin{subfigure}[]{0.46\linewidth}
    %     \centering
    %     \includegraphics[width=\linewidth]{isolation-avg-fct-dpdk.pdf}
    %     \caption{Average FCT}\label{fig:dpdk:isolation:avg-qct}
    % \end{subfigure}
    % \hfil
    % \begin{subfigure}[]{0.46\linewidth}
    %     \centering
    %     \includegraphics[width=\linewidth]{isolation-small-tail-fct-dpdk.pdf}
    %     \caption{Tail FCT of Small Flows}\label{fig:dpdk:isolation:tail-qct}
    % \end{subfigure}
    \caption{Performance isolation}
    \label{fig:dpdk:isolation}
\end{minipage}
\hfil
\begin{minipage}[b]{.49\linewidth}
    \begin{subfigure}[b]{.46\linewidth}
        \centering
        \includegraphics[width=\linewidth]{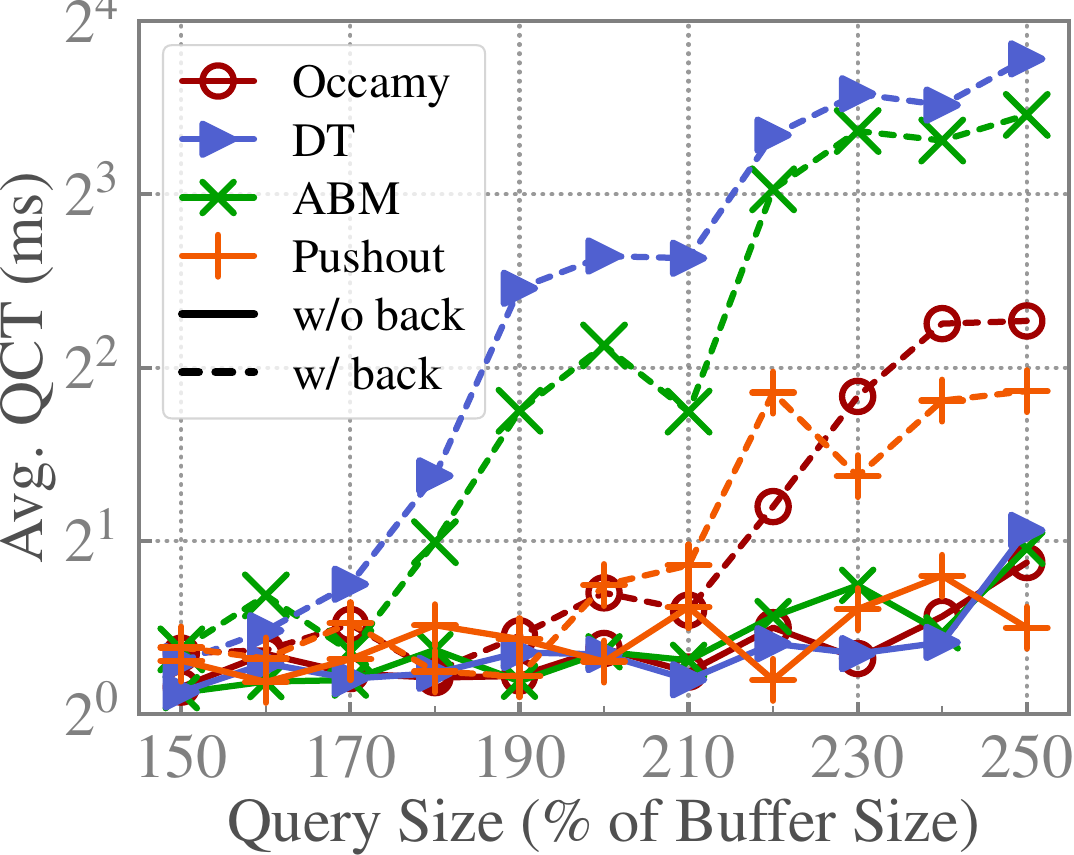}
        \caption{Average QCT}\label{fig:dpdk:buffer-choke:avg-qct}
    \end{subfigure}
    \hfil
    \begin{subfigure}[b]{.46\linewidth}
        \centering
        \includegraphics[width=\linewidth]{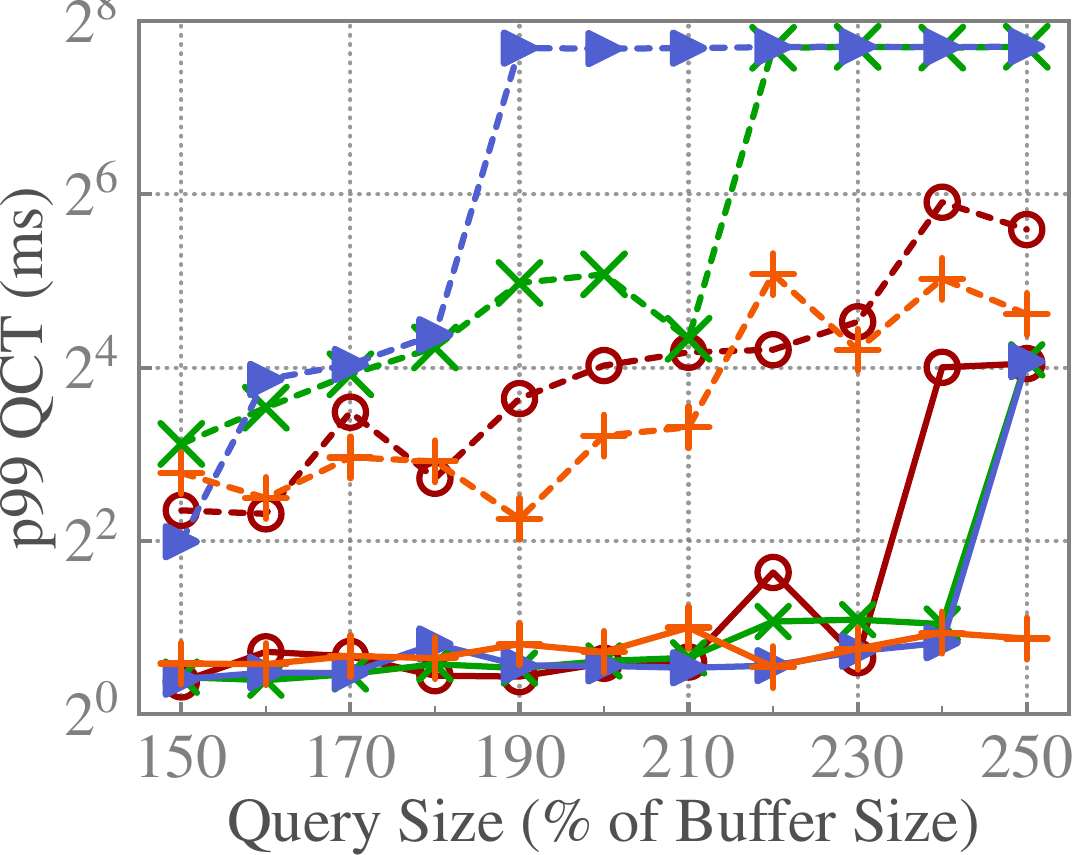}
        \caption{99th QCT}\label{fig:dpdk:buffer-choke:tail-qct}
    \end{subfigure}
    \caption{Mitigation of buffer choking}
    \label{fig:dpdk:buffer-choke}
\end{minipage}
\end{figure*}

\mypara{\sysname{} can ensure performance isolation}
% For this experiment, we set up our host and switch with two priorities.
% The background flows are assigned with the lower priority,
% while the query flows are assigned with the higher priority.
% For the higher-priority queue, we set the $\alpha$ of DT, ABM, and \sysname{} to 8.
% For the lower-priority queue, we set the $\alpha$ to 1.
% Other settings are kept unchanged.
% We let a host receive query traffic
% and let another host receive background traffic from the other three hosts.
% The low-priority traffic and high-priority traffic will be congested at different ports,
% and thus it is expected that the low-priority traffic should not affect the performance of high-priority traffic.
%
For this experiment,
we set up our switch with two service queues per port,
which are fairly scheduled with Deficit Round Robin~(DRR).
The background traffic and query traffic
are assigned to different queues for performance isolation.
The background flows use CUBIC as their congestion control algorithm.
Other settings remain unchanged.
We vary the load of background traffic to
examine its impact on the performance of query traffic.
% We let a host query for both query traffic
% and background traffic from other three hosts
% so that two queues in the same port will be both congested.
% We consider two scenarios.
% (1) Scenario 1: we start 1 background flow,
% and the background flow will not occupy buffer.
% (2) Scenario 2: we start 3 background flows,
% and the background flows will occupy some buffer.
% In both scenarios, query traffic should receive 50\% bandwidth.
% Thus, hopefully, the performance of query traffic should be the same.

\figurename~\ref{fig:dpdk:isolation:avg-qct} and \figurename~\ref{fig:dpdk:isolation:tail-qct}
show the average QCT and 99th percentile QCT of the query traffic, respectively.
\figurename~\ref{fig:dpdk:isolation:tail-qct} shows that,
as the load of background traffic increases,
DT and ABM can result in RTOs for query traffic,
which significantly extends the QCT.
Since the background traffic and the query traffic are put into different queues,
the RTOs mainly result from the inability to quickly adjust the buffer allocation.
% Specifically, \figurename~\ref{fig:dpdk:isolation:tail-qct}
% shows that DT and ABM result in RTOs when the network load reaches 40\% and 60\%, respectively.
In comparison, \sysname{} can quickly adjust the buffer allocation
by actively expelling the packets for over-allocated queues,
achieving significantly lower 99th percentile QCT.

\mypara{\sysname{} can effectively mitigate the buffer choking problem}
For this experiment, we set up our host and switch with two priority queues.
The query flows are assigned with the high priority (HP),
whereas the background flows are assigned with the low priority (LP).
For the HP queue, we set the $\alpha$ of DT, ABM, and \sysname{} to 8
so that more buffer is allocated to HP traffic.
For the LP queue, we set the $\alpha$ to 1.
Other settings remain unchanged.
% The experiment settings are the same as above,
We let a host receive both query flows and background flows from the other hosts,
so that two priority queues are congested at the same port simultaneously.
Hopefully, the LP background traffic should not affect the HP query traffic.

\figurename~\ref{fig:dpdk:buffer-choke:avg-qct} and \figurename~\ref{fig:dpdk:buffer-choke:tail-qct}
show the average and 99th percentile QCT, respectively.
The solid line depicts the QCT without background traffic,
whereas the dashed line depicts the QCT with background traffic.
We can observe that \sysname{} achieves similar performance to Pushout,
\ie~the background traffic does not affect the performance of query traffic much
even though they share the same buffer.
In contrast, the background traffic significantly extends the QCT of DT and ABM.
\figurename~\ref{fig:dpdk:buffer-choke:avg-qct} shows that,
the background traffic can extend the average QCT of DT by up to $\sim$6.6$\times$.
\figurename~\ref{fig:dpdk:buffer-choke:tail-qct} shows that
the background traffic can extend the 99th percentile QCT of DT by up to $\sim$60$\times$.
ABM achieves better QCT than DT, which is expected
since ABM restricts the queue length of LP queues.
However, it cannot radically address the buffer choking problem.
\figurename~\ref{fig:dpdk:buffer-choke:avg-qct} shows that
the background traffic can extend the average QCT of ABM by up to $\sim$5.7$\times$.
This is because ABM is also non-preemptive
and relies on natural queue drain to adjust buffer allocation.
% However, with both HP and LP queues congested,
% the LP queue drains slowly.

% This is because
% the buffer occupied by the low-priority background traffic
% is draining slowly,
% while the high-priority query traffic is thirsting for buffer.
% the buffer occupancy of DT and ABM can only be drained by sending out the traffic.
% With the port busy sending high-priority query traffic,
% In comparison, \sysname{} can drain the buffer of the low-priority traffic by head drop,
% effectively mitigating this problem.

\begin{figure}[!t]
    \centering
    % \begin{subfigure}[b]{.23\linewidth}
    %     \centering
    %     \includegraphics[width=\linewidth]{alpha-dt-avg-qct-dpdk.pdf}
    %     \caption{DT: Average QCT}\label{fig:dpdk:alpha:dt:avg-qct}
    % \end{subfigure}
    % \hfil
    \begin{subfigure}[]{.46\linewidth}
        \centering
        \includegraphics[width=\linewidth]{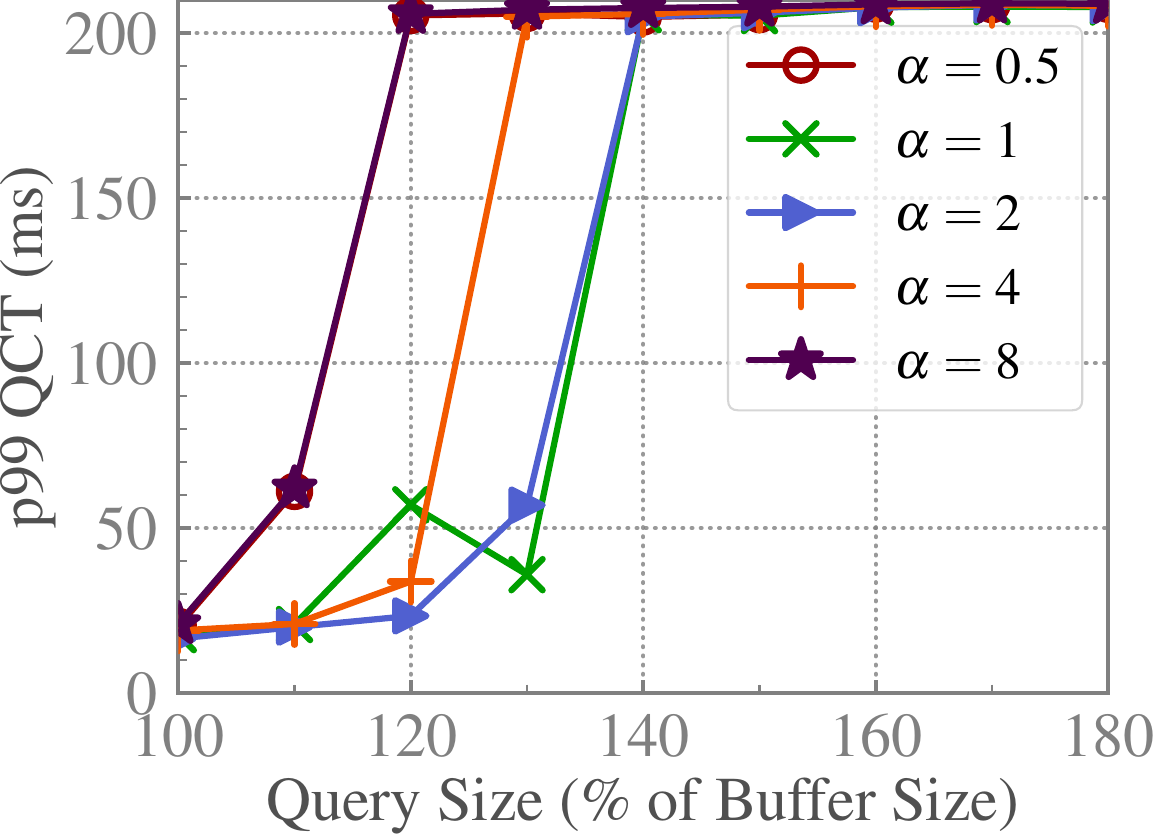}
        \caption{DT: 99th QCT}\label{fig:dpdk:alpha:dt:tail-qct}
    \end{subfigure}
    % \hfil
    % \begin{subfigure}[b]{.23\linewidth}
    %     \centering
    %     \includegraphics[width=\linewidth]{alpha-pBuffer-avg-qct-dpdk.pdf}
    %     \caption{\sysname{}: Average QCT}\label{fig:dpdk:alpha:pbuffer:avg-qct}
    % \end{subfigure}
    \hfil
    \begin{subfigure}[]{.46\linewidth}
        \centering
        \includegraphics[width=\linewidth]{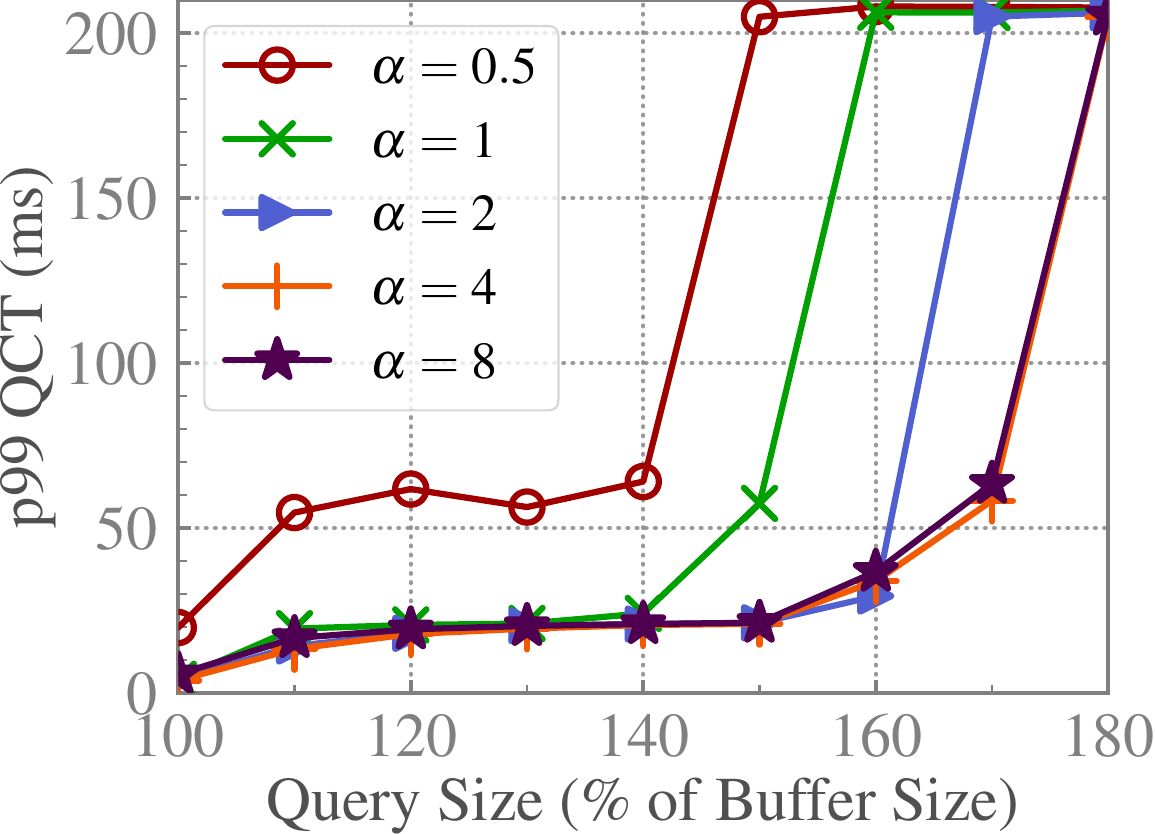}
        \caption{\sysname{}: 99th QCT}\label{fig:dpdk:alpha:pbuffer:tail-qct}
    \end{subfigure}
    \caption{Impact of $\alpha$}
    \label{fig:dpdk:alpha}
\end{figure}
\begin{figure*}[!t]
    \centering
    % \begin{subfigure}[b]{.23\linewidth}
    %     \centering
    %     \includegraphics[width=\linewidth]{sim/burst-absorb-avg-qct-load-reno-search.pdf}
    %     \caption{TCP}\label{fig:sim:burst-absorb:qct:dctcp}
    % \end{subfigure}
    % \hfil
    \begin{subfigure}[b]{.24\linewidth}
        \centering
        \includegraphics[width=\linewidth]{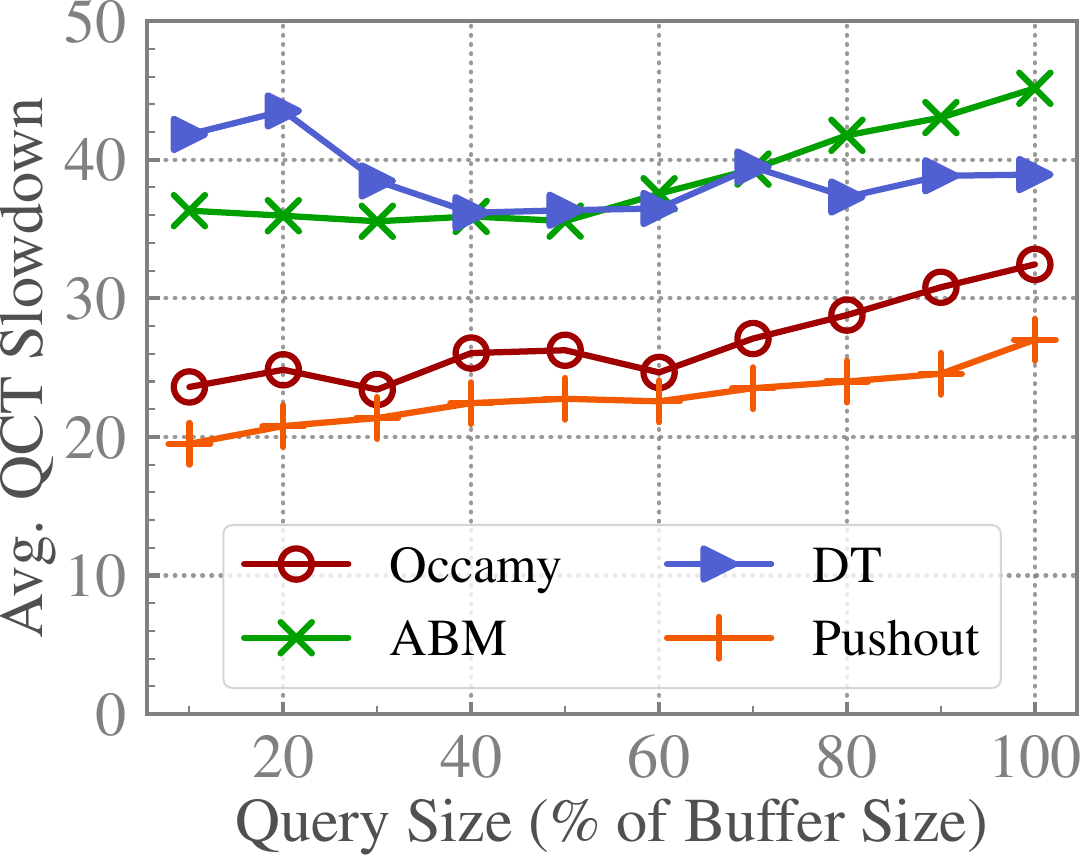}
        \caption{Query: Average QCT}\label{fig:sim:burst-absorb:avg-qct}
    \end{subfigure}
    \hfil
    \begin{subfigure}[b]{.24\linewidth}
        \centering
        \includegraphics[width=\linewidth]{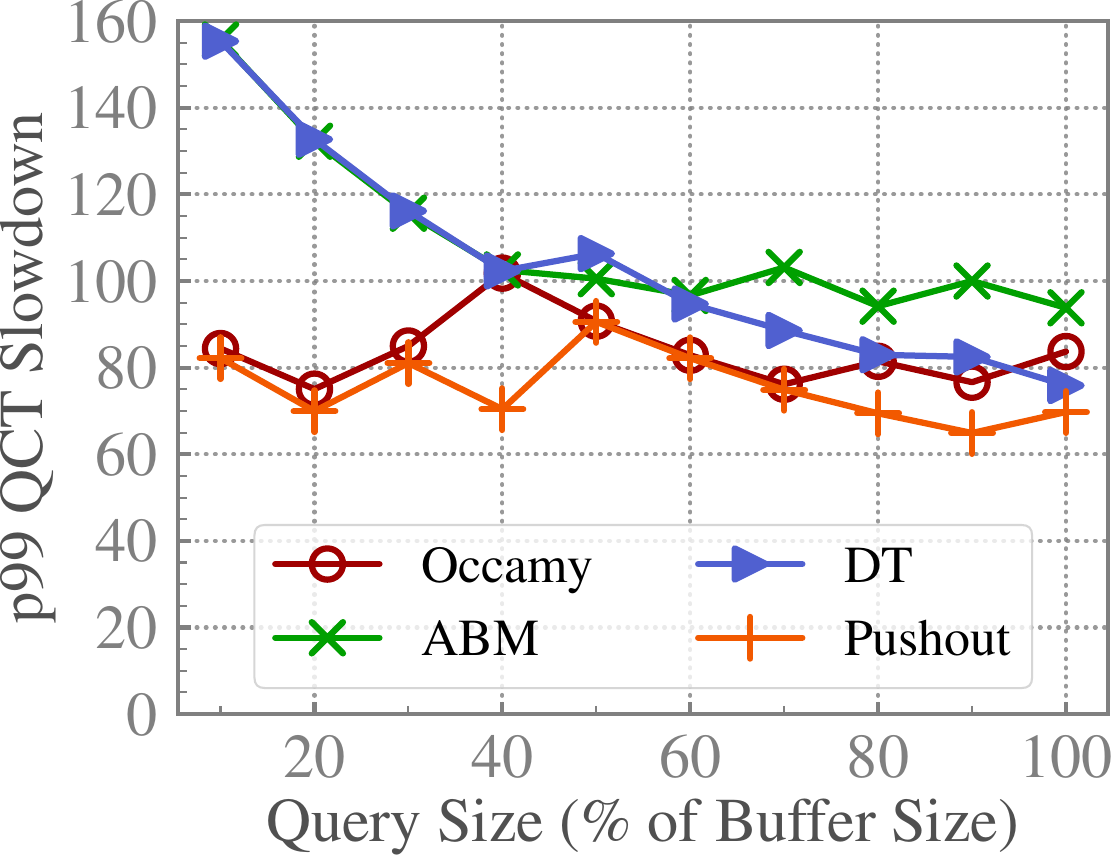}
        \caption{Query: 99th QCT}\label{fig:sim:burst-absorb:tail-qct}
    \end{subfigure}
    \hfil
    \begin{subfigure}[b]{.24\linewidth}
        \centering
        \includegraphics[width=\linewidth]{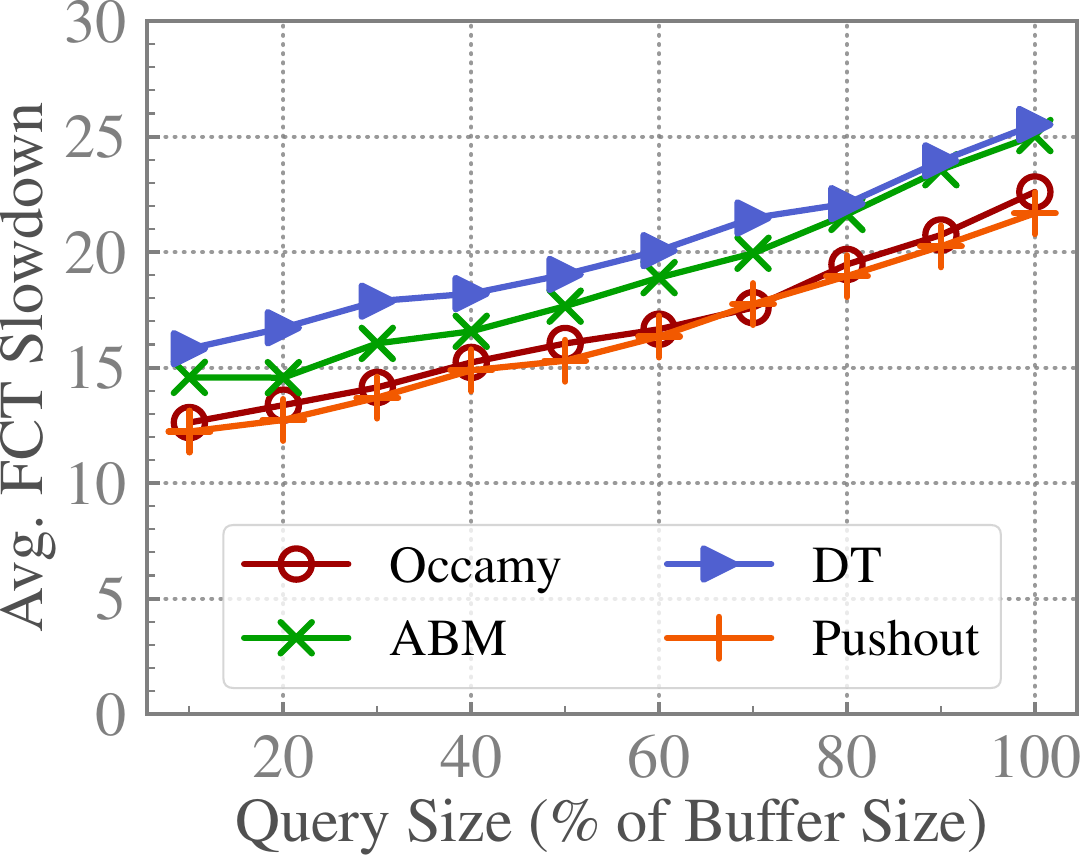}
        \caption{Overall Bg: Average FCT}\label{fig:sim:burst-absorb:avg-fct}
    \end{subfigure}
    \hfil
    \begin{subfigure}[b]{.24\linewidth}
        \centering
        \includegraphics[width=\linewidth]{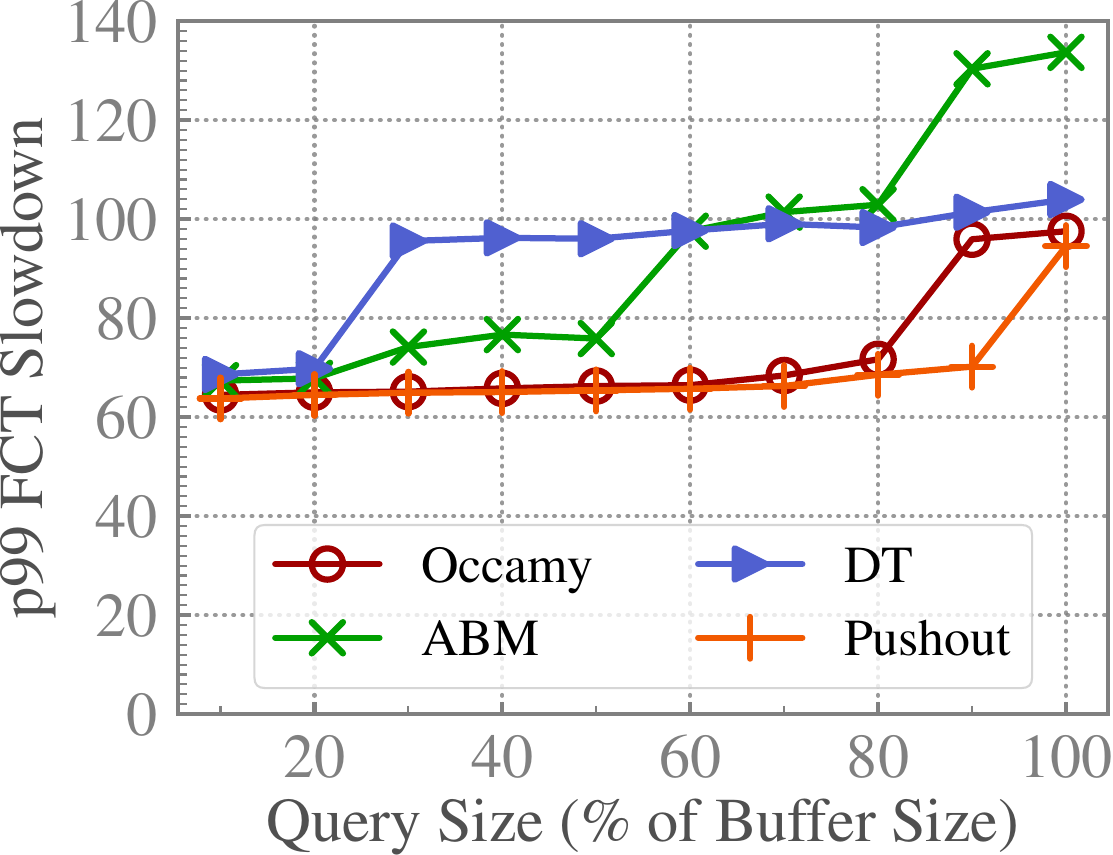}
        \caption{Small Bg: 99th FCT}\label{fig:sim:burst-absorb:small-tail-fct}
    \end{subfigure}
    % \hfil
    % \begin{subfigure}[b]{.23\linewidth}
    %     \centering
    %     \includegraphics[width=\linewidth]{sim/burst-absorb-avg-qct-load-hpcc-search.pdf}
    %     \caption{HPCC: Average QCT}\label{fig:sim:burst-absorb:qct:hpcc}
    % \end{subfigure}
    % \hfil
    % \begin{subfigure}[b]{.24\linewidth}
    %     \centering
    %     \includegraphics[width=\linewidth]{sim/burst-absorb-avg-fct-flowsize-dctcp.pdf}
    %     \caption{FCT}\label{fig:sim:search:powertcp:qct}
    % \end{subfigure}
    \caption{Query Completion Time (QCT) and Flow Completion Time (FCT)}
    \label{fig:sim:burst-absorb}
\end{figure*}
\begin{figure*}
    \begin{minipage}[b]{.49\linewidth}
        \begin{subfigure}[b]{.49\linewidth}
            \centering
            \includegraphics[width=\linewidth]{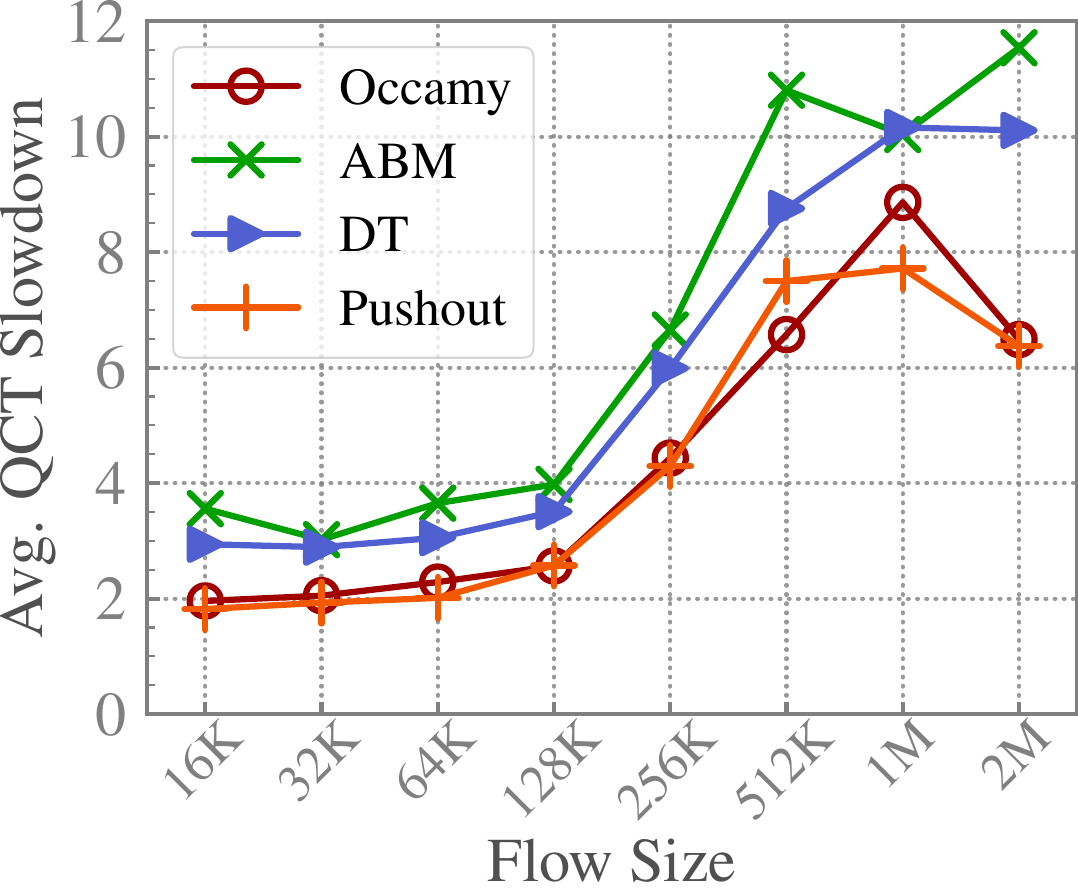}
            \caption{Query: Average QCT}
            \label{fig:sim:alltoall:avg-qct}
        \end{subfigure}
        \hfil
        \begin{subfigure}[b]{.49\linewidth}
            \centering
            \includegraphics[width=\linewidth]{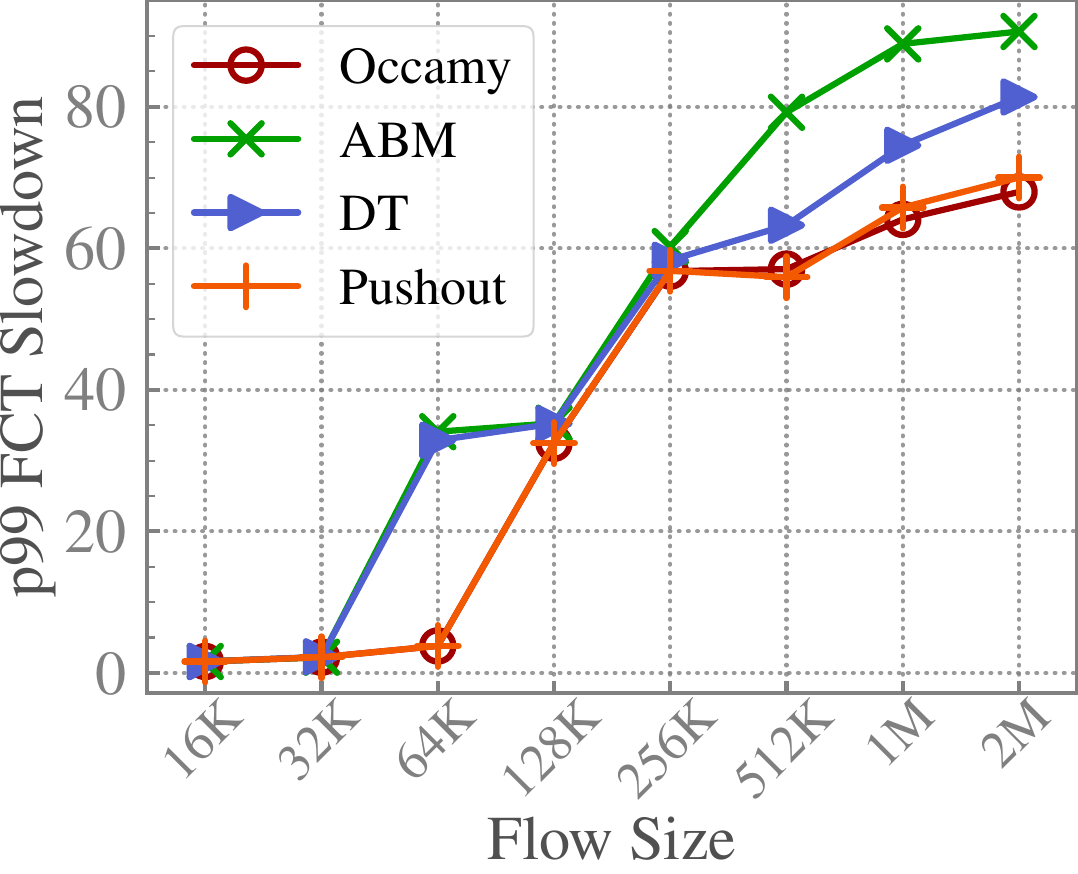}
            \caption{Overall Bg: 99th FCT}
            \label{fig:sim:alltoall:tail-fct}
        \end{subfigure}
        \caption{Performance with all-to-all traffic}
        \label{fig:sim:alltoall}
    \end{minipage}
    \hfil
    \begin{minipage}[b]{.49\linewidth}
        \begin{subfigure}[b]{.49\linewidth}
            \centering
            \includegraphics[width=\linewidth]{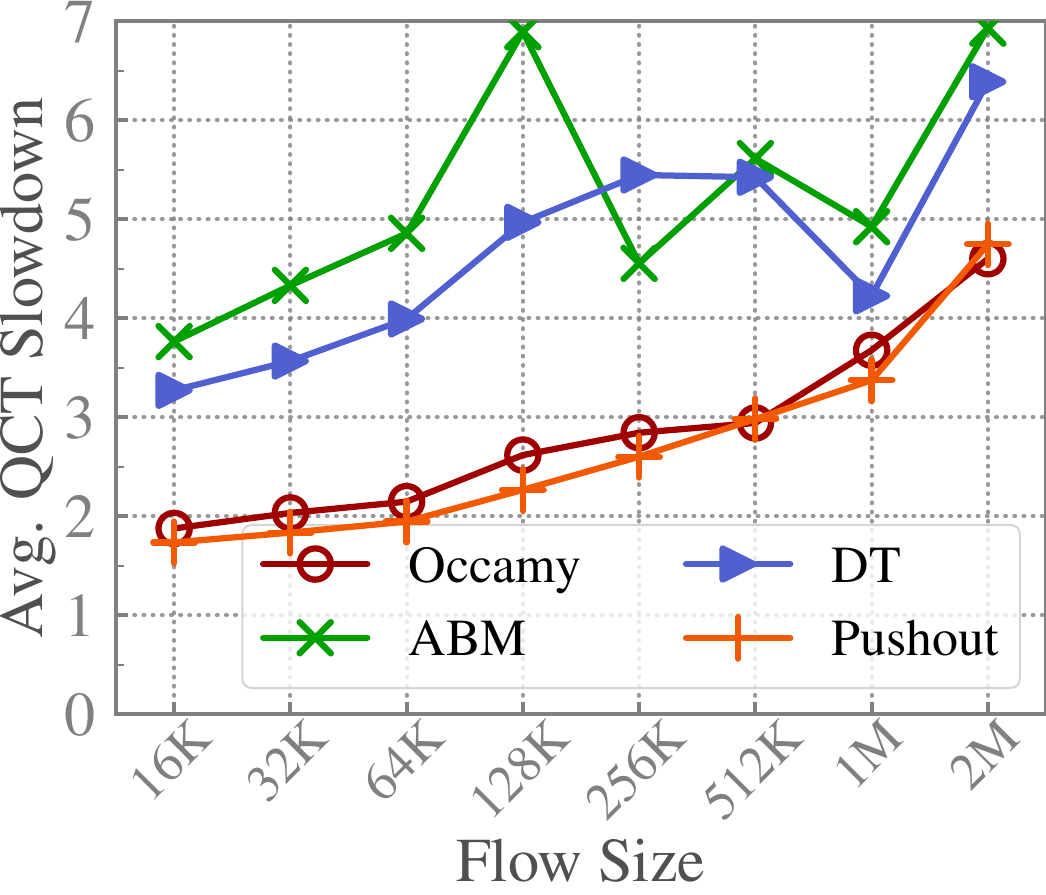}
            \caption{Query: Average QCT}
            \label{fig:sim:allreduce:avg-qct}
        \end{subfigure}
        \hfil
        \begin{subfigure}[b]{.49\linewidth}
            \centering
            \includegraphics[width=\linewidth]{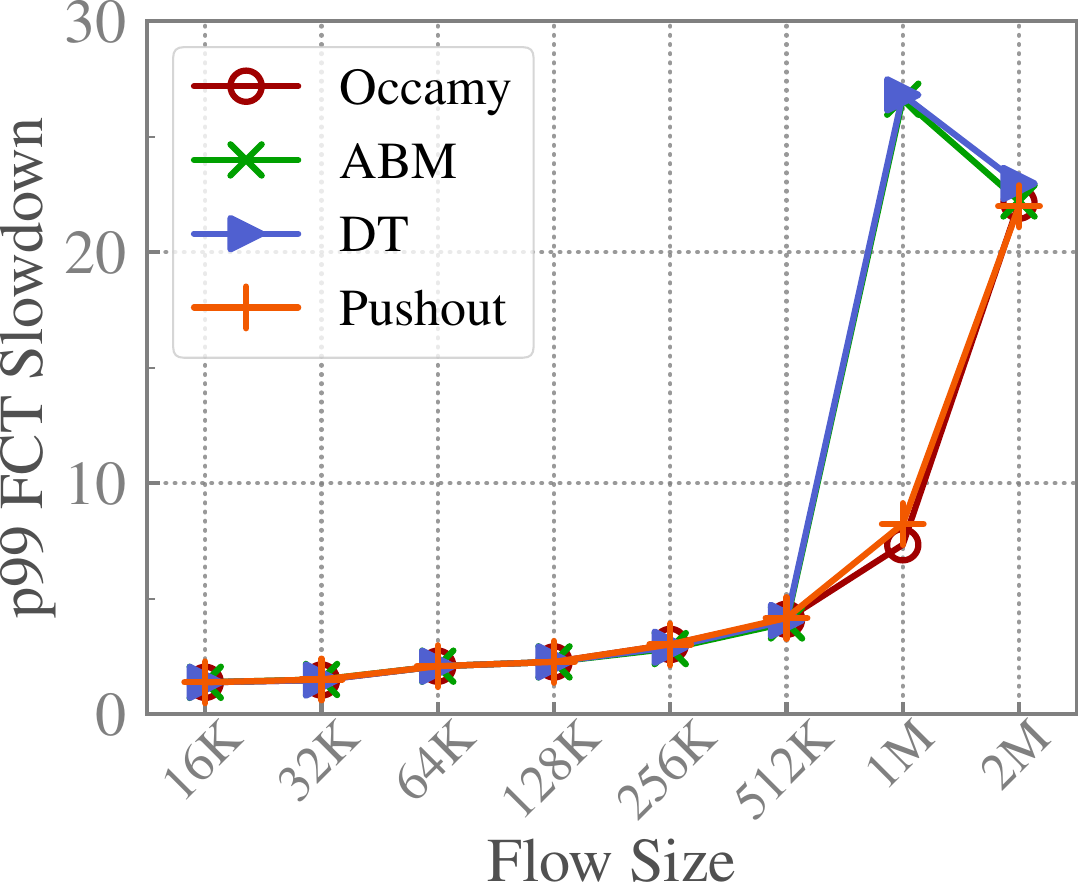}
            \caption{Overall Bg: 99th FCT}
            \label{fig:sim:allreduce:tail-fct}
        \end{subfigure}
        \caption{Performance with all-reduce traffic}
        \label{fig:sim:allreduce}
    \end{minipage}
\end{figure*}

\subsection{Parameter Settings}\label{sec:eval:params}
We use our DPDK-based testbed to explore the impact of parameter $\alpha$.
% To evaluate the performance isolation with different $\alpha$s,
The same as previous experiments,
we put background traffic and query traffic into two queues,
which are fairly scheduled by DRR.
Other settings are kept unchanged.
\figurename~\ref{fig:dpdk:alpha} shows the 99th percentile QCT
of DT and \sysname{} with different $\alpha$s.
\figurename~\ref{fig:dpdk:alpha:dt:tail-qct}
shows that DT achieves better performance with $\alpha = 1$ or $\alpha = 2$.
The performance is degraded with either smaller $\alpha$ or larger $\alpha$.
This is because DT is more inefficient with small $\alpha$,
and is prone to anomalous behavior with large $\alpha$.
In comparison,
\figurename~\ref{fig:dpdk:alpha:pbuffer:tail-qct}
shows that \sysname{} achieves better performance with larger $\alpha$.
This is because \sysname{} can avoid anomalous behavior under the support of fast buffer expulsion,
and thus can use a large $\alpha$ to improve efficiency.
Nonetheless,
we also observe that the performance improvement is undermined
with $\alpha$ larger than 4.
Specifically,
\figurename~\ref{fig:dpdk:alpha:pbuffer:tail-qct} shows that
the performance with $\alpha=8$ is very close to that with $\alpha=4$.
This is because the improvement of buffer utilization
is small with large $\alpha$
(as analyzed in \textsection\ref{sec:params}).
% For example,
% increasing $\alpha$ from 4 to 8
% only improves the buffer utilization by $\sim$8.9\%
% according to \refeq{eq:free-buffer-size}
% (note that $\alpha$ is a factor of two to facilitate the calculation of threshold).
Thus, we recommend setting $\alpha$ to 8.

\subsection{Large-scale Simulations}\label{sec:eval:sim}
\mypara{Topology}
We simulate a 128-host leaf-spine topology
with 8 spine switches and 8 leaf switches.
Each leaf switch is connected to 16 hosts with 100Gbps down links
and 8 spine switches with 100Gbps up links.
The base RTT across spine is 80$\mu$s.
We employ ECMP for multi-path load balancing.
% Each switch has 2MB buffer shared among all queues,
% which emulates the Broadcom Tomahawk switch chip
To emulate the Broadcom Tomahawk switch chip,
we make every 8 ports share 4MB buffer~\cite{BroadcomTomahawkBuffer, TON21BCC}.
As a result, the leaf switches contain 12MB buffer in total,
whereas the spine switches contain 8MB buffer.

\mypara{Workload}
We generate query traffic and background traffic
similarly to the previous DPDK-based experiments.
Each server generates 200 queries per second.
% Each host queries for some data from other 32 hosts,
% which are randomly chosen from other racks.
% The total query size is 80\% of the buffer size.
% The queries are generated based on a Poisson process
% wit an average rate of 200 queries per second per server.
% The total query size is varied across 10\%-100\% of the buffer.
% (2) The background traffic follows a 1-to-1 pattern.
% We generate flows based on web-search~\cite{SIGCOMM10DCTCP}
% flow size distributions.
% Flow arrivals follow a Poisson process.
% and we vary the average arrival rate to change the traffic load.
% The source and destination for each flow are randomly chosen.
The load of background traffic is 90\%.
We use DCTCP~\cite{SIGCOMM10DCTCP} as the congestion control algorithm,
with ECN threshold set to 720KB (\ie~0.72BDP) as suggested by~\cite{TON21BCC}.
The minimum RTO is set to 5ms.

% \mypara{Transports}
% We evaluate \sysname{} with different kinds of congestion control algorithms:
% TCP (loss-based),
% We use DCTCP~\cite{SIGCOMM10DCTCP} 
% TIMELY (RTT-based)~\cite{SIGCOMM15TIMELY},
% HPCC (INT-based)~\cite{SIGCOMM19HPCC}, PowerTCP (power-based)~\cite{NSDI22PowerTCP}.
% For TCP, we use NewReno as the congestion control algorithm.
% For DCTCP, we set $K=720$KB (\ie~0.72BDP) according to~\cite{TON21BCC}.
% For DCTCP, we set $K=65$ packets according to~\cite{SIGCOMM10DCTCP}.
% \footnotetext{
%     We do not use CUBIC because we find that ns-3's CUBIC does not contain the TCP-friendly region,
%     and thus may increase its congestion window too slow.
%     Furthermore, with short RTT,
%     the CUBIC's congestion window evolution is similarly as NewReno.
% }

% \mypara{Compared schemes}
% We compare \sysname{} with three BM schemes:
% DT~\cite{TON98DT}, ABM~\cite{SIGCOMM22ABM}, Pushout~\cite{GLOBECOM91Pushout}.
% We have tuned their parameters to achieve their best performance.
% For DT, we set $\alpha = 1$ as suggested in \cite{TON98DT}.
% For ABM, we set $\alpha = 4$, which can achieve better performance than that of $\alpha=1$.
% For \sysname{}, we set $\alpha = 8$ as it is more adaptive.
% The frequency of head-drop operation is limited to 50Mhz (\ie~drop a packet every 20ns).
% Other settings are kept unchanged.
% as analyzed in \textsection\ref{sec:desigin:vacation}.
% For all BM schemes,
% we do not set a larger $\alpha$ for unscheduled packets as~\cite{SIGCOMM22ABM}
% as we assume that the end hosts are unmodified.

\mypara{Performance}
\figurename~\ref{fig:sim:burst-absorb}
shows the performance of query traffic (in terms of QCT slowdown) and background traffic (in terms of FCT slowdown).
Here, slowdown is the ratio between the actual value and the ideal value without other traffic.
% The results are consistent with our testbed experiments.
We make two observations.
(1) \sysname{} can improve the performance of bursty query traffic.
% \figurename~\ref{fig:sim:burst-absorb:avg-qct} and \figurename~\ref{fig:sim:burst-absorb:tail-qct}
% show the average and 99th percentile QCT slowdown, respectively.
% The QCT of \sysname{} is close to that of Pushout
% and is significantly better than DT and ABM.
% \sysname{}'s average QCT and 99th percentile QCT are within $\sim$\inlinetodo{} of Pushout.
\figurename~\ref{fig:sim:burst-absorb:avg-qct} shows that
\sysname{} reduces the average QCT slowdown of DT and ABM
by up to $\sim$44\% and $\sim$36\%, respectively.
\figurename~\ref{fig:sim:burst-absorb:tail-qct} shows that
\sysname{} reduces the 99th percentile QCT slowdown of DT and ABM
by up to $\sim$46\%.
(2) \sysname{} is also beneficial to the background flows.
\figurename~\ref{fig:sim:burst-absorb:avg-qct} shows that
\sysname{} reduces the average FCT slowdown of DT and ABM
by up to $\sim$20\% and $\sim$13\%, respectively.
\figurename~\ref{fig:sim:burst-absorb:tail-qct} shows that,
for small flows,
\sysname{} reduces the 99th percentile FCT slowdown of DT and ABM
by up to $\sim$32\%.
% Actually, \reffig{fig:sim:search:dctcp:fct} shows that
% \sysname{} can improve the average FCT by up to $\sim$38.6\% and $\sim$16.2\%
% compared with DT and ABM, respectively.
% Similar results can be obtained with data-mining workload
% and other CCs (see Appendix~\ref{sec:appendix:mining}).

% \mypara{Impact of head-drop frequency}
%
% \mypara{Impact of $\alpha$}
%
% \mypara{Impact of buffer size}

% \mypara{Performance of performance isolation}
% For this simulation, we enable two service queues for each port in the switch.
% The queues are fairly served by the DRR algorithm.
% The query traffic and the background traffic are put into different queues.
% Other settings are kept unchanged.
%
% \mypara{Performance of buffer lock problem}
% For this simulation, we enable two priority queues for each port in the switch.
% The query traffic is put into the higher-priority queue,
% and the background traffic id put into lower-priority queue.
% Other settings are keep unchanged.

\mypara{Performance with all-to-all/all-reduce traffic}
To examine \sysname{}'s performance in prevalent AI scenarios,
we generate all-to-all and all-reduce background traffic.
For all-to-all traffic,
every host sends the same amount of data to all other hosts.
For all-reduce traffic,
we generate flows based on
the prevailing double binary tree algorithm~\cite{PC09DoubleBinaryTree},
with all flows having an identical size.

\figurename~\ref{fig:sim:alltoall}
and \figurename~\ref{fig:sim:allreduce}
show that
\sysname{} can improve both the QCT for query traffic
and FCT for background traffic.
Specifically,
in the all-to-all scenario,
\sysname{} can improve the average QCT and 99th percentile FCT of DT
by up to $\sim$33\% and $\sim$88\%, respectively.
In the all-reduce scenario,
\sysname{} can improve the average QCT and 99th percentile FCT of DT
by up to $\sim$48\% and $\sim$73\%, respectively.

\mypara{Performance with higher query traffic rate}
In previous scenarios, the load of query traffic is not very high.
In this part, we examine \sysname{}'s performance
with higher query traffic rate.
We vary the load of query traffic
from 10\% (\ie~186 queries per second per server) to 80\%
(\ie~1,490 queries per second per server)\footnotemark.
\footnotetext{
Load = \# queries\_per\_second $\times$ query\_size $\times$ oversubscription\_ratio / link\_capacity,
where query\_size=3.2MB, oversubscription\_ratio=2, and link\_capacity=100Gbps.
}
The query size is 80\% of the buffer size (\ie~3.2MB).
The load of background traffic is 10\%.

\begin{figure}
    \begin{subfigure}[b]{.49\linewidth}
        \centering
        \includegraphics[width=\linewidth]{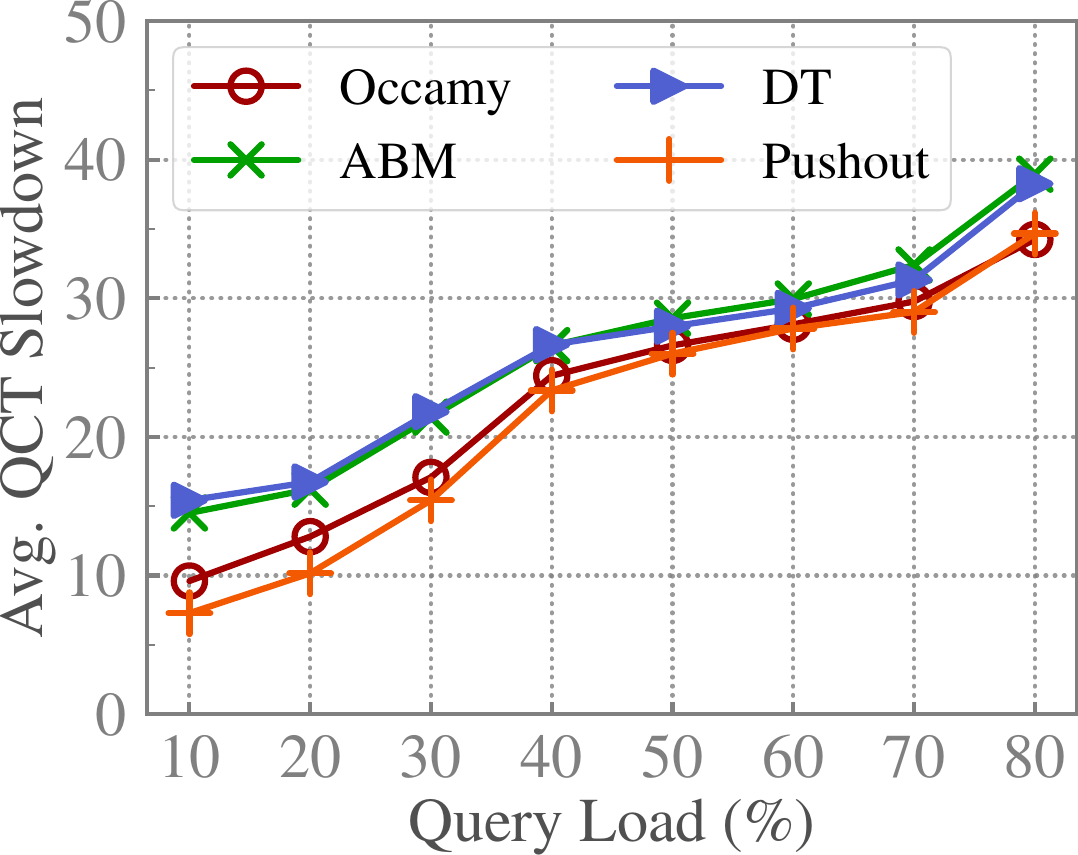}
        \caption{Query: Average QCT}
        \label{fig:sim:query-load:avg-qct}
    \end{subfigure}
    \hfil
    \begin{subfigure}[b]{.49\linewidth}
        \centering
        \includegraphics[width=\linewidth]{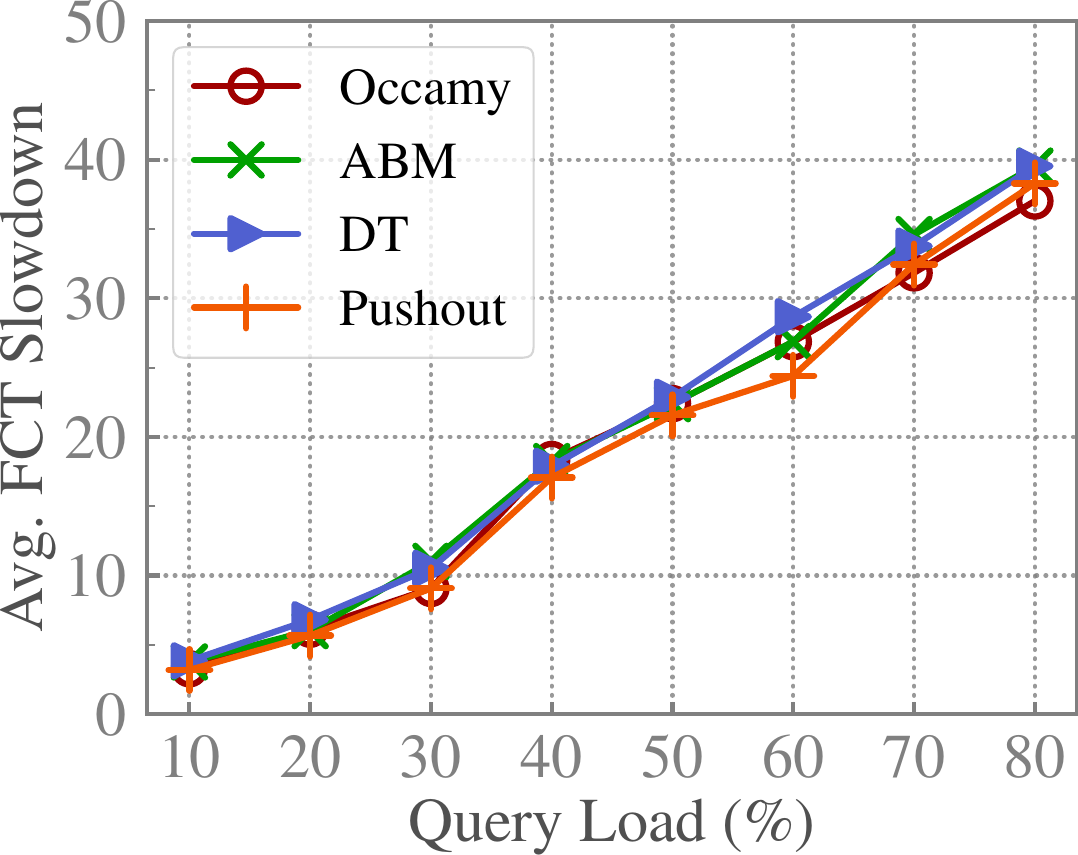}
        \caption{Overall Bg: Average FCT}
        \label{fig:sim:query-load:avg-fct}
    \end{subfigure}
    \caption{Performance with higher query traffic load}
    \label{fig:sim:query-load}
\end{figure}

\figurename~\ref{fig:sim:query-load:avg-qct} shows that
\sysname{} can improve the average QCT of DT and ABM
by up to $\sim$38\% and $\sim$34\%, respectively.
Besides, the performance improvements of both \sysname{} and Pushout
are more notable at lower query load.
This is because DT's inefficiency
is more pronounced with fewer active ports~\cite{TON98DT},
and thus \sysname{} has more space to improve performance.
\figurename~\ref{fig:sim:query-load:avg-fct} shows that
BM has little influence on the performance of background traffic.
This is because the background traffic is light-loaded,
and thus has minimal requirement on buffer.

\begin{figure*}[!t]
    \centering
    \begin{subfigure}[b]{.24\linewidth}
        \centering
        \includegraphics[width=\linewidth]{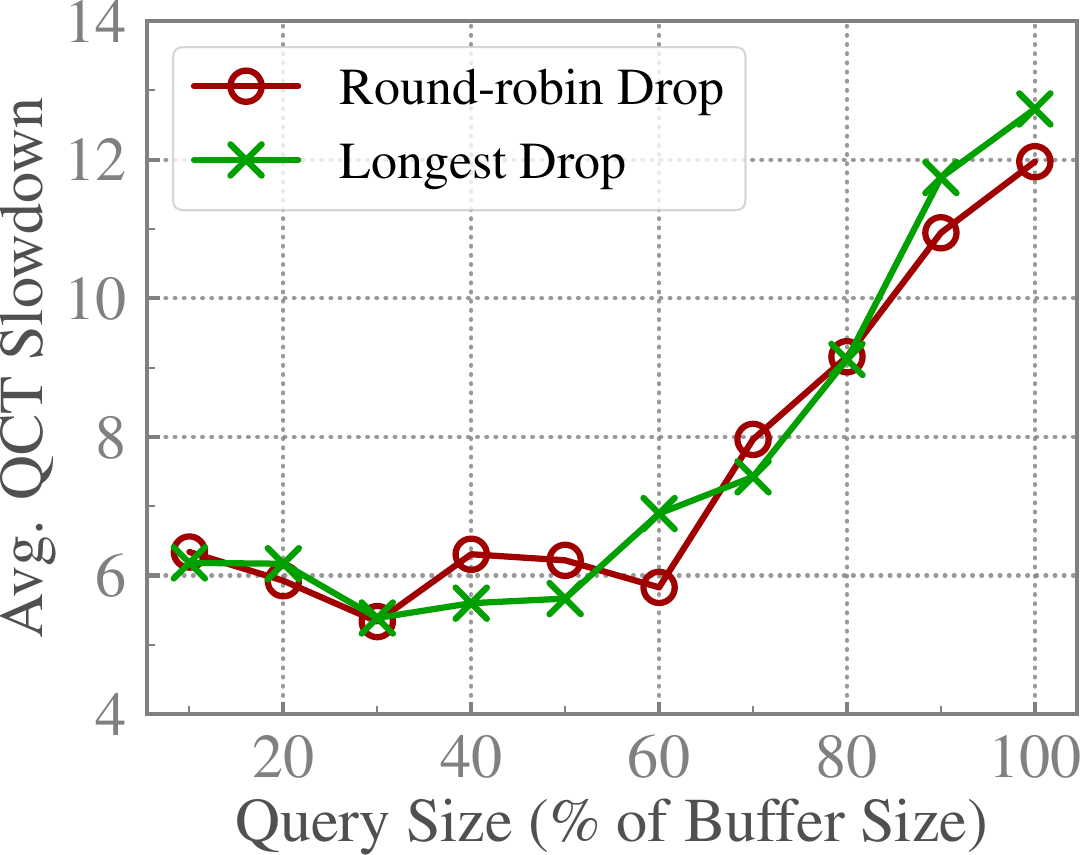}
        \caption{Query: Average QCT}\label{fig:sim:longest-drop:avg-qct}
    \end{subfigure}
    \hfil
    \begin{subfigure}[b]{.24\linewidth}
        \centering
        \includegraphics[width=\linewidth]{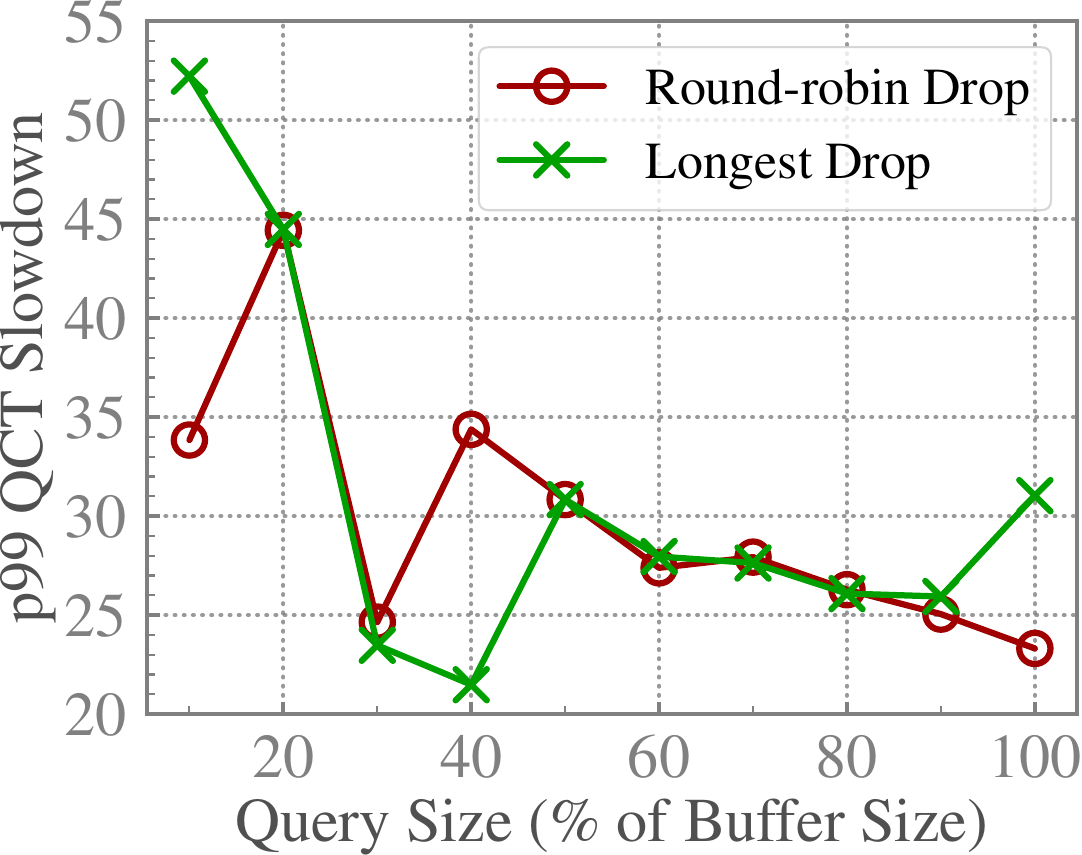}
        \caption{Query: 99th QCT}\label{fig:sim:longest-drop:tail-qct}
    \end{subfigure}
    \hfil
    \begin{subfigure}[b]{.24\linewidth}
        \centering
        \includegraphics[width=\linewidth]{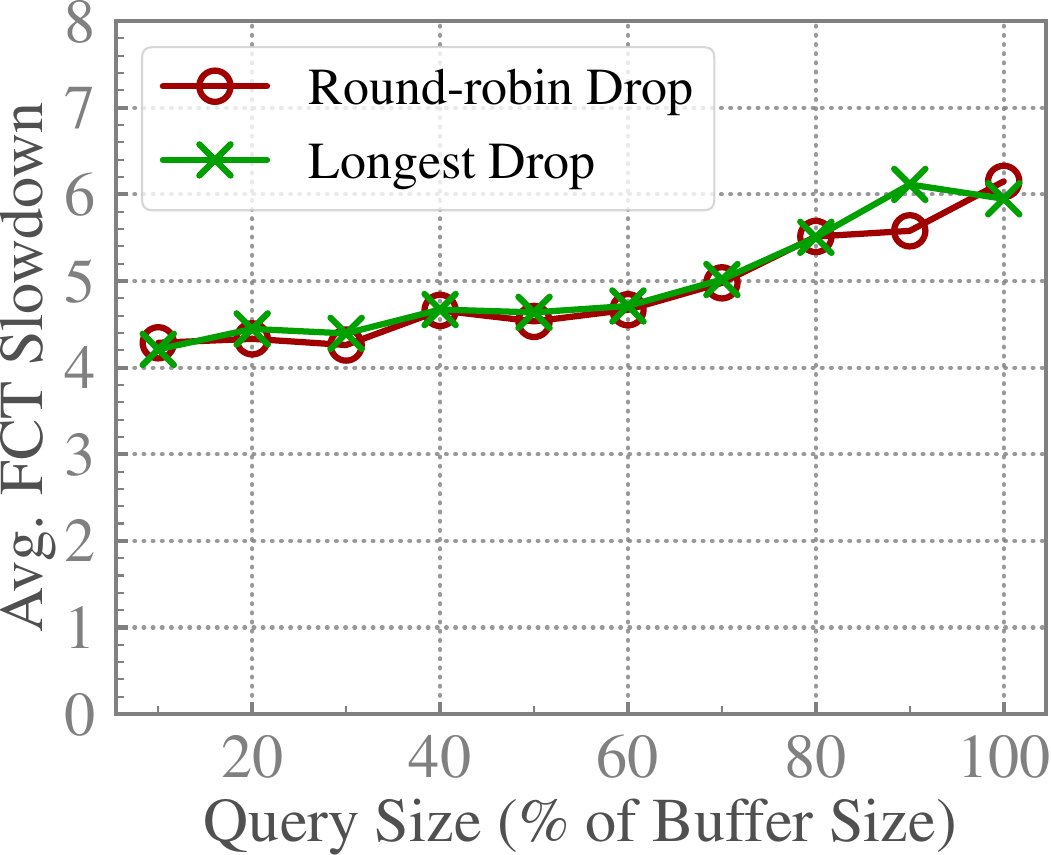}
        \caption{Overall Bg: Average FCT}\label{fig:sim:longest-drop:avg-fct}
    \end{subfigure}
    \hfil
    \begin{subfigure}[b]{.24\linewidth}
        \centering
        \includegraphics[width=\linewidth]{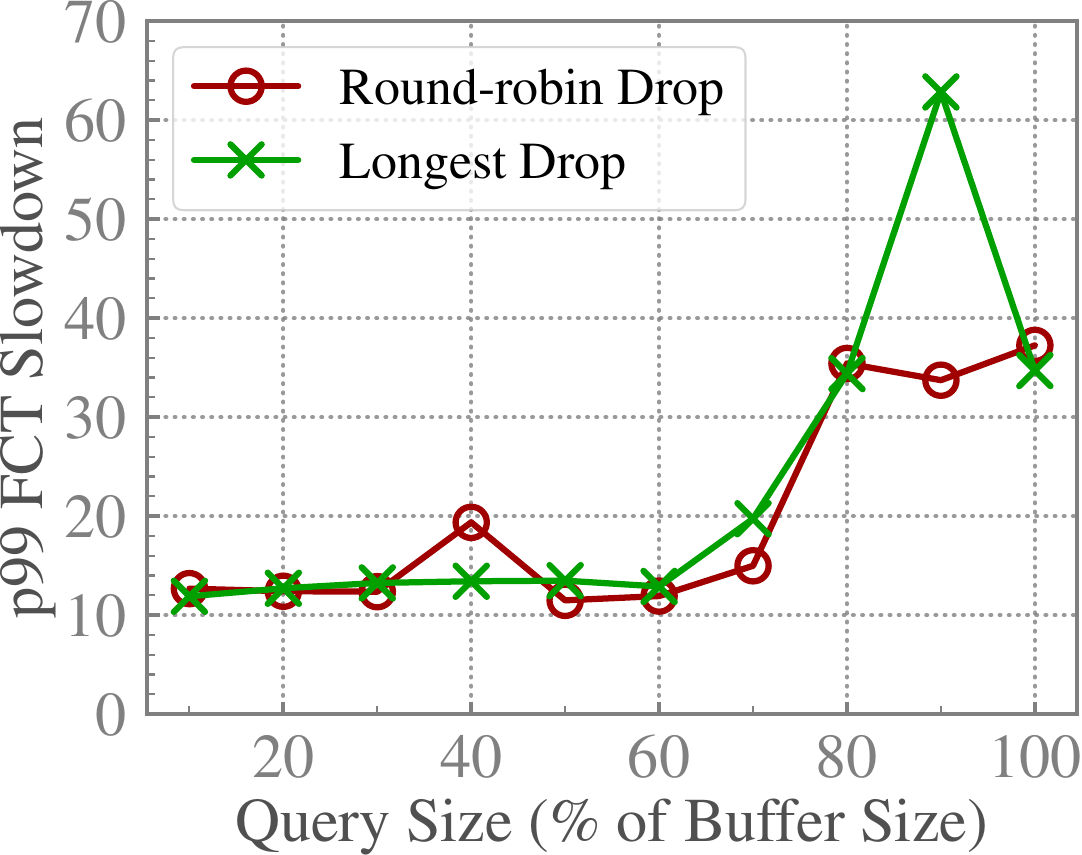}
        \caption{Small Bg: 99th FCT}\label{fig:sim:longest-drop:small-tail-fct}
    \end{subfigure}
    \caption{[Simulation] The effectiveness of round-robin drop}
    \label{fig:sim:longest-drop}
\end{figure*}
\begin{figure*}[!t]
    \centering
    \begin{subfigure}[b]{.24\linewidth}
        \centering
        \includegraphics[width=\linewidth]{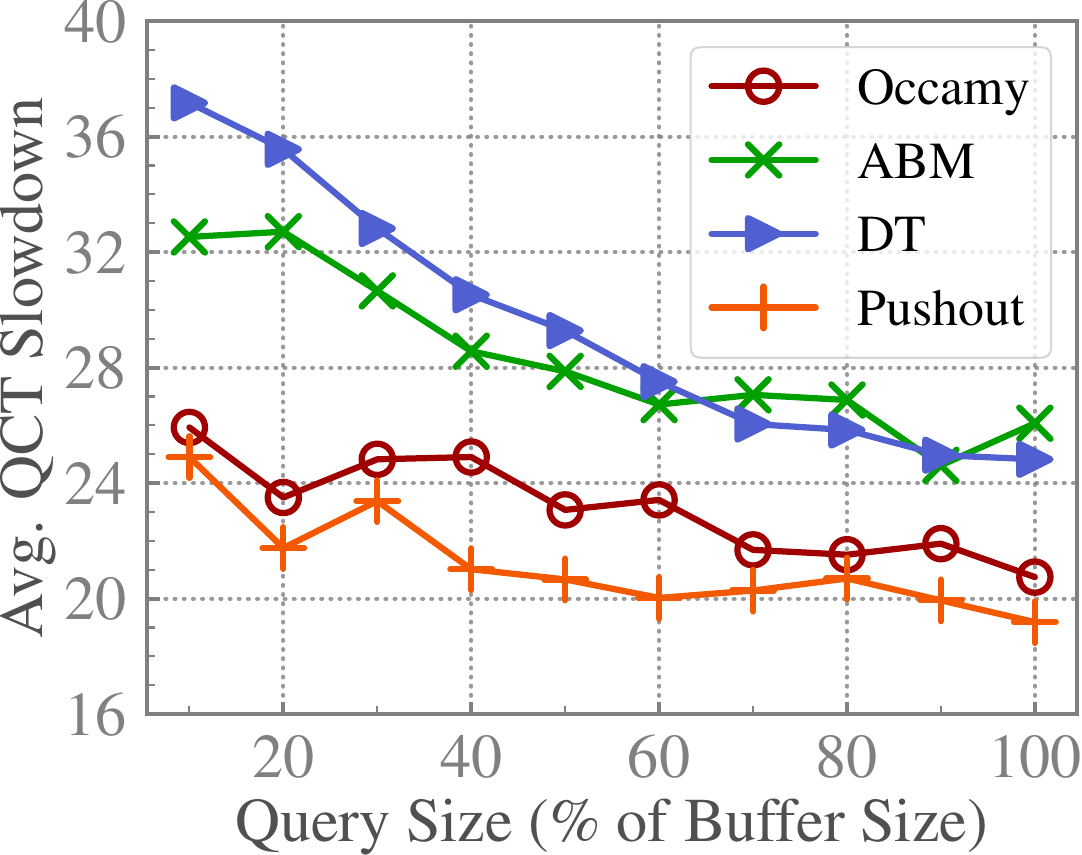}
        \caption{Query: Average QCT}\label{fig:sim:heavy-load:avg-qct}
    \end{subfigure}
    \hfil
    \begin{subfigure}[b]{.24\linewidth}
        \centering
        \includegraphics[width=\linewidth]{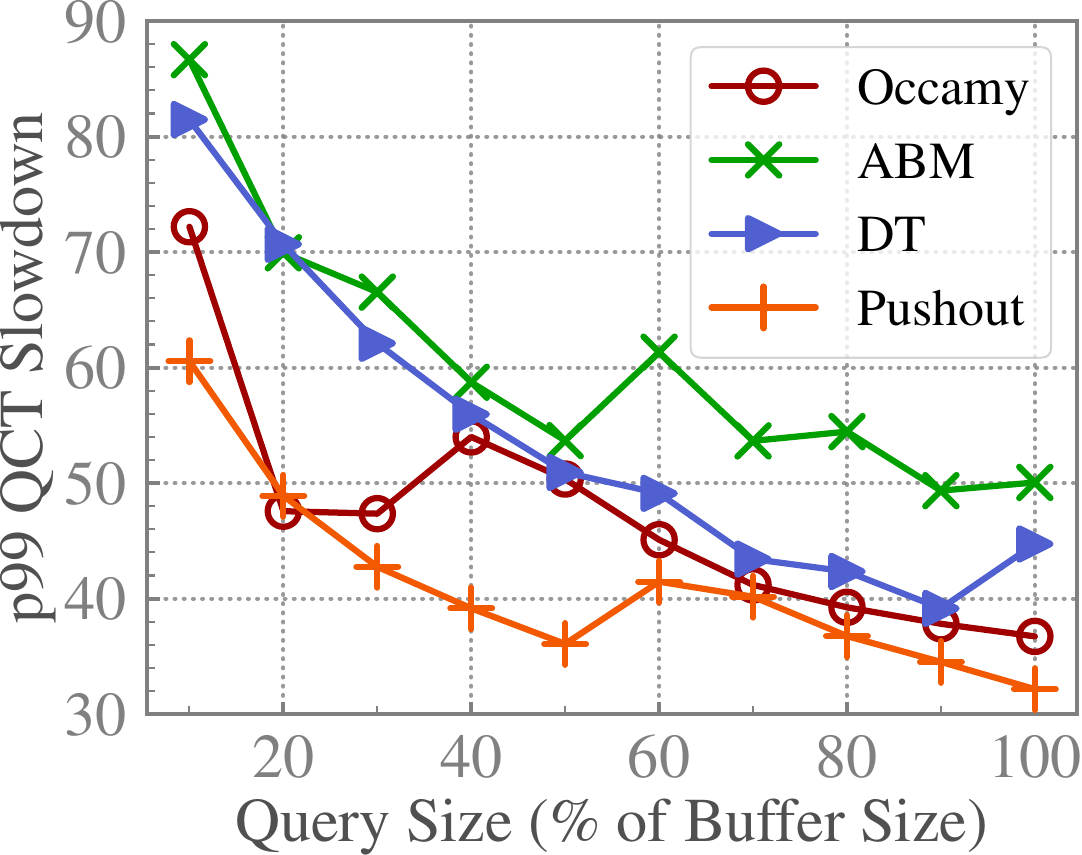}
        \caption{Query: 99th QCT}\label{fig:sim:heavy-load:tail-qct}
    \end{subfigure}
    \hfil
    \begin{subfigure}[b]{.24\linewidth}
        \centering
        \includegraphics[width=\linewidth]{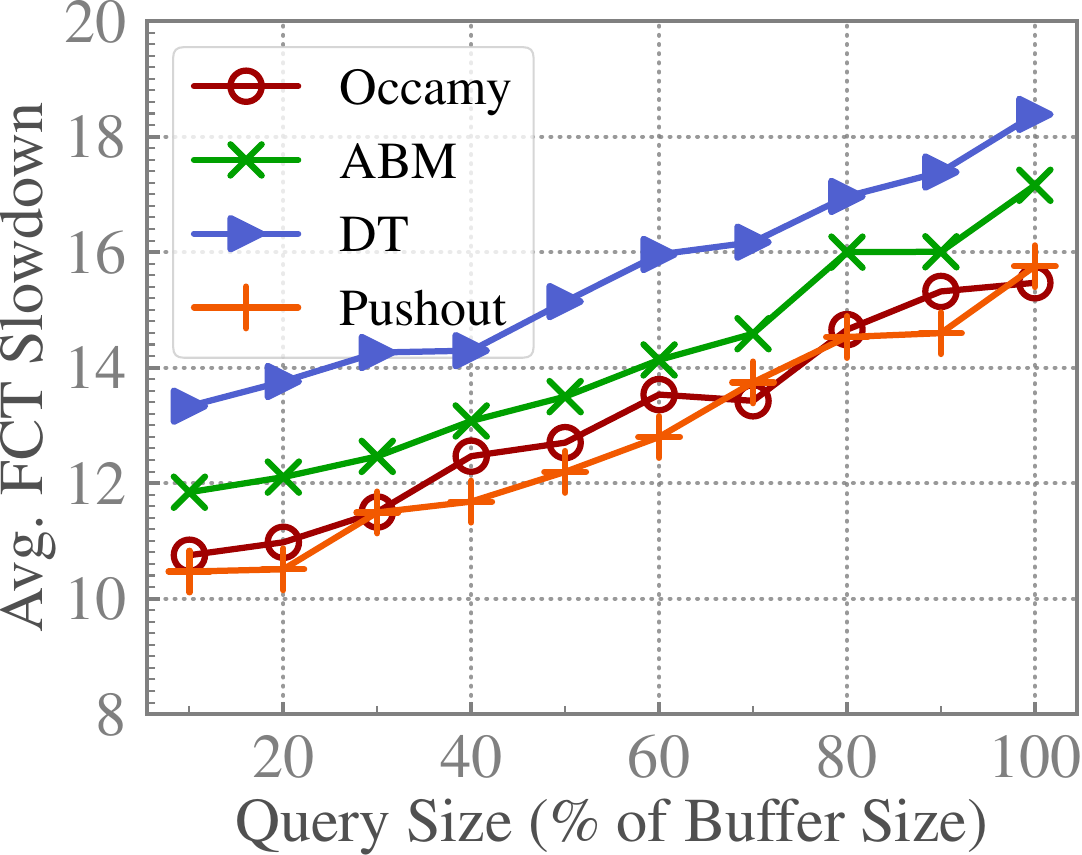}
        \caption{Overall Bg: Average FCT}\label{fig:sim:heavy-load:avg-fct}
    \end{subfigure}
    \hfil
    \begin{subfigure}[b]{.24\linewidth}
        \centering
        \includegraphics[width=\linewidth]{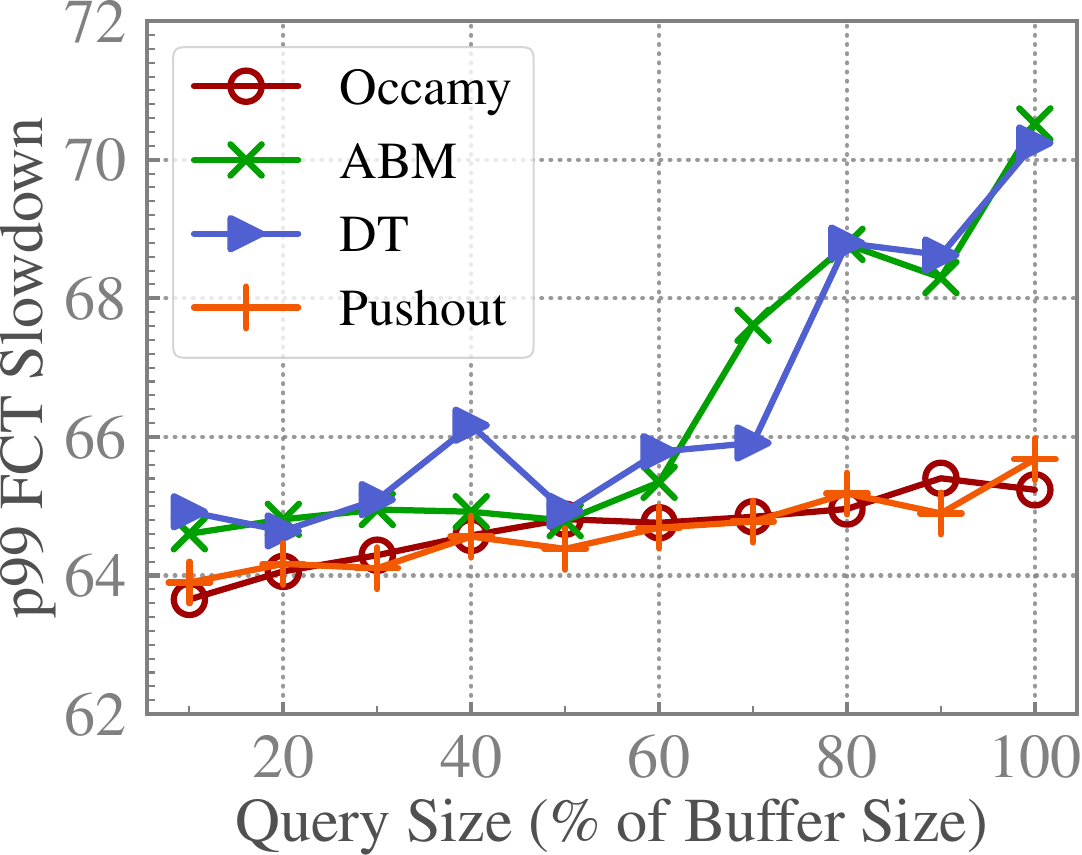}
        \caption{Small Bg: 99th FCT}\label{fig:sim:heavy-load:small-tail-fct}
    \end{subfigure}
    \caption{[Simulation] Performance with 120\% load}
    \label{fig:sim:heavy-load}
\end{figure*}

%
% \begin{figure*}[!t]
%     \centering
%     \begin{subfigure}[b]{.24\linewidth}
%         \centering
%         \includegraphics[width=\linewidth]{sim/cc-avg-qct-load-dctcp.pdf}
%         \caption{CUBIC}\label{fig:sim:cc:cubic}
%     \end{subfigure}
%     \hfil
%     \begin{subfigure}[b]{.24\linewidth}
%         \centering
%         \includegraphics[width=\linewidth]{sim/cc-avg-qct-load-dctcp.pdf}
%         \caption{DCTCP}\label{fig:sim:cc:dctcp}
%     \end{subfigure}
%     \hfil
%     \begin{subfigure}[b]{.24\linewidth}
%         \centering
%         \includegraphics[width=\linewidth]{sim/cc-avg-qct-load-timely.pdf}
%         \caption{TIMELY}\label{fig:sim:cc:timely}
%     \end{subfigure}
%     \hfil
%     \begin{subfigure}[b]{.24\linewidth}
%         \centering
%         \includegraphics[width=\linewidth]{sim/cc-avg-qct-load-hpcc.pdf}
%         \caption{HPCC}\label{fig:sim:cc:hpcc}
%     \end{subfigure}
%     \caption{[Simulation] Average QCT with different CCs}
%     \label{fig:sim:cc}
% \end{figure*}
\begin{figure*}[!t]
    \centering
    \begin{subfigure}[b]{.24\linewidth}
        \centering
        \includegraphics[width=\linewidth]{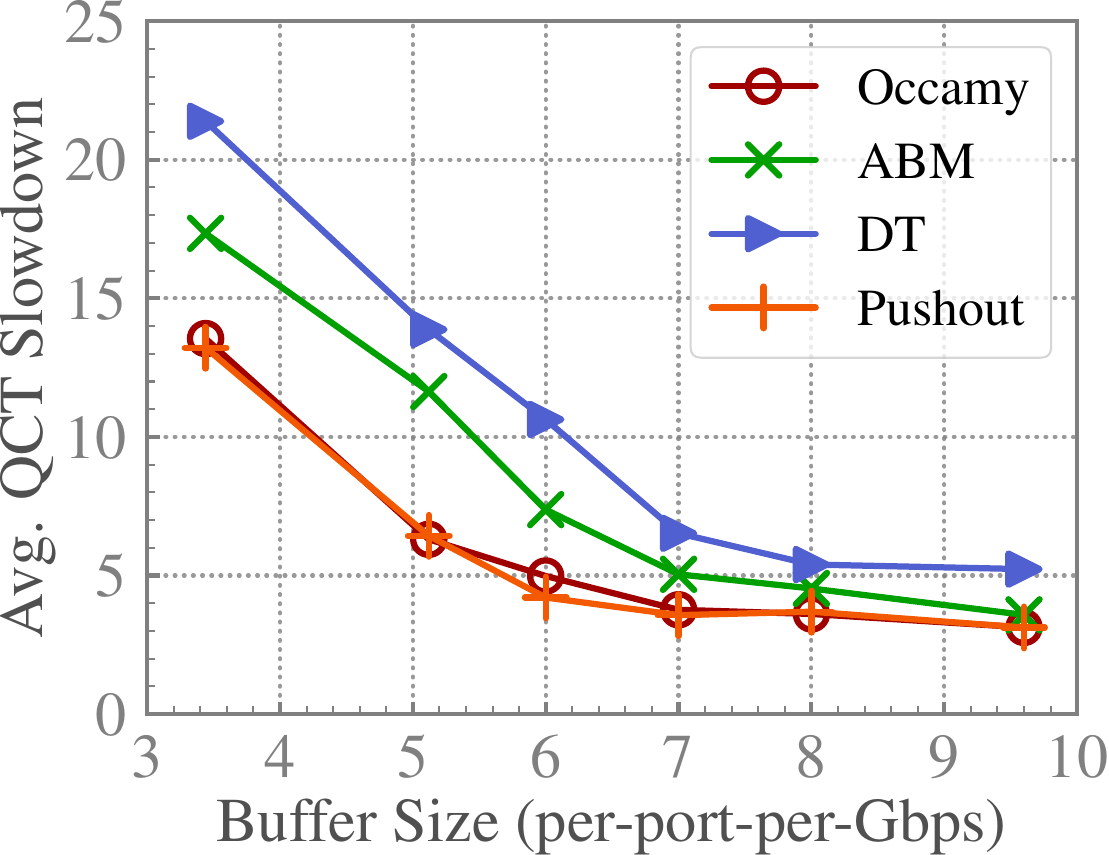}
        \caption{Query: Average QCT}\label{fig:sim:buffer-size:avg-qct}
    \end{subfigure}
    \hfil
    \begin{subfigure}[b]{.24\linewidth}
        \centering
        \includegraphics[width=\linewidth]{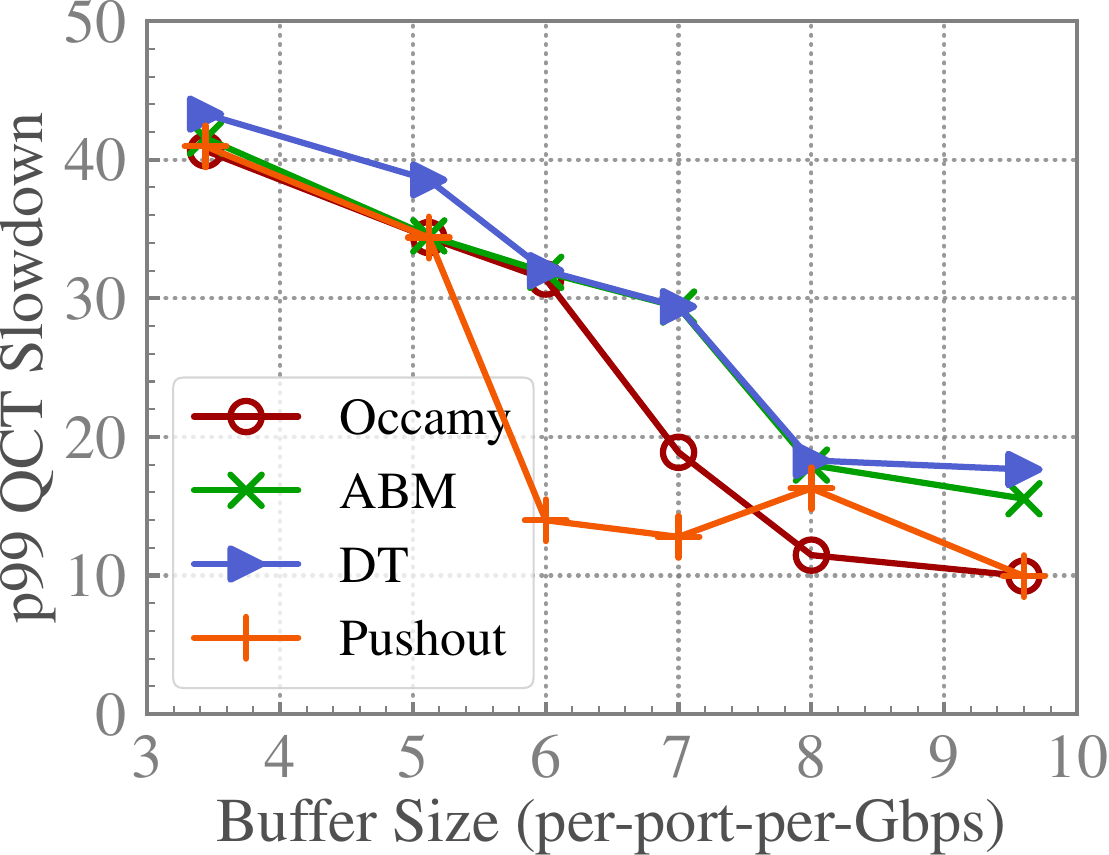}
        \caption{Query: 99th QCT}\label{fig:sim:buffer-size:tail-qct}
    \end{subfigure}
    \hfil
    \begin{subfigure}[b]{.24\linewidth}
        \centering
        \includegraphics[width=\linewidth]{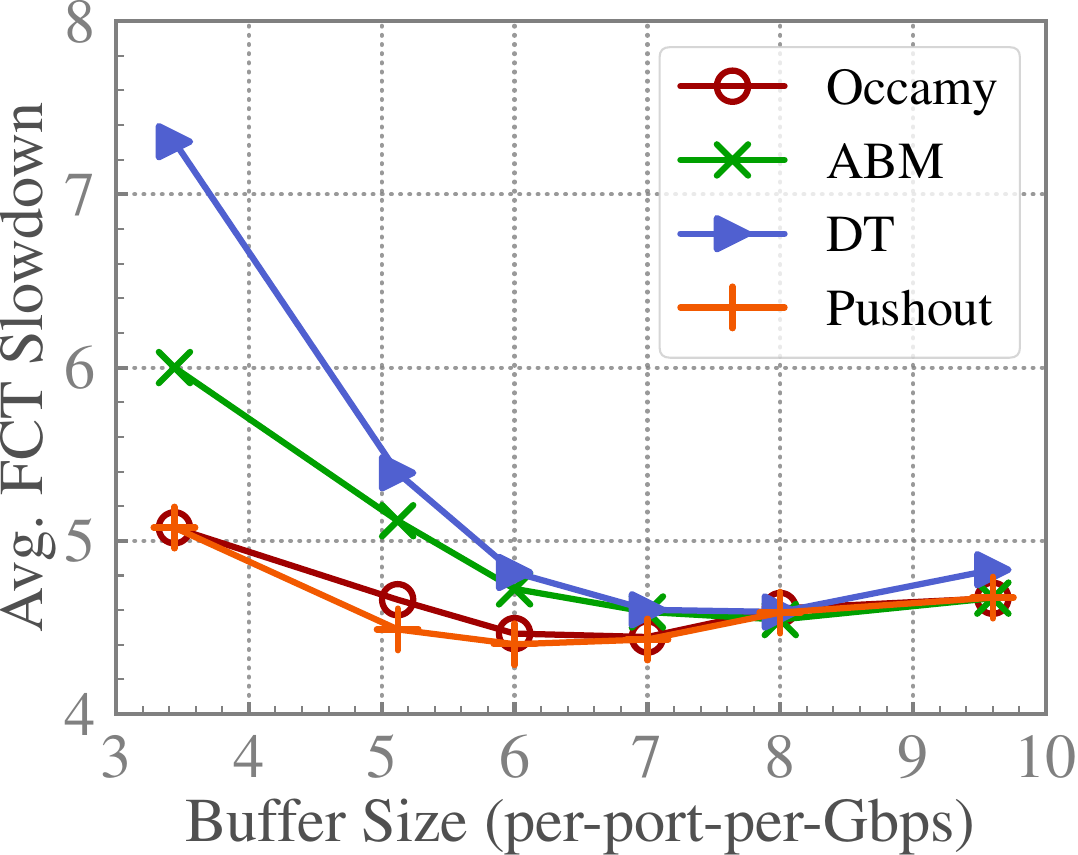}
        \caption{Overall Bg: Average FCT}\label{fig:sim:buffer-size:avg-fct}
    \end{subfigure}
    \hfil
    \begin{subfigure}[b]{.24\linewidth}
        \centering
        \includegraphics[width=\linewidth]{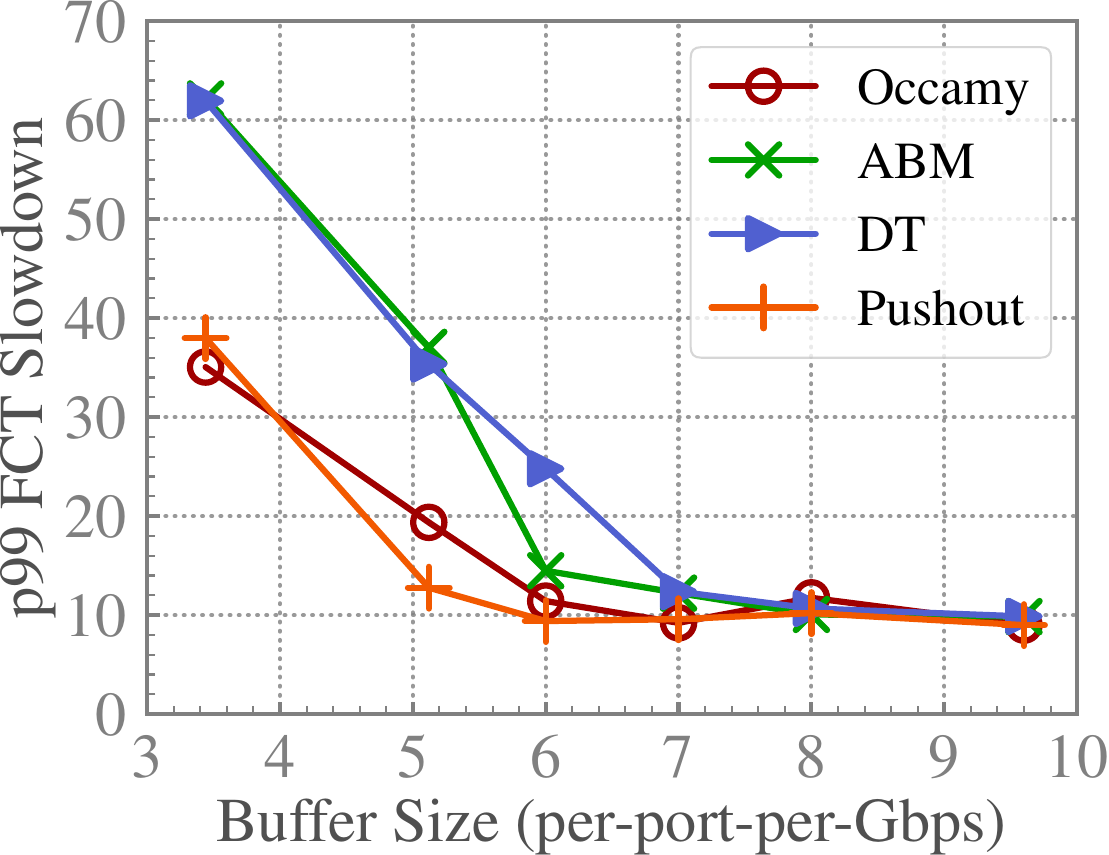}
        \caption{Small Bg: 99th FCT}\label{fig:sim:buffer-size:small-tail-fct}
    \end{subfigure}
    \caption{[Simulation] Performance with different buffer sizes}
    \label{fig:sim:buffer-size}
\end{figure*}

\mypara{Effectiveness of round-robin drop}
For simplicity,
\sysname{} drops packets in a round-robin manner,
avoiding maintaining the longest queue.
In this section, we examine how such simplicity affects the performance.
We use large-scale simulations to compare \sysname{} to its variant
that always drops the longest queue.
The load of background traffic is 40\%.
Other settings are the same as those in \textsection\ref{sec:eval:sim}.

\figurename~\ref{fig:sim:longest-drop} shows the performance
of \sysname{} with round-robin drop and \sysname{} with longest queue drop.
We can observe that
\sysname{}'s performance is close to that with longest queue drop.
For query traffic,
\figurename~\ref{fig:sim:longest-drop:avg-qct} shows that
the difference in average QCT is within $\sim$15\%.
% \figurename~\ref{fig:sim:longest-drop:tail-qct} shows that
% the difference of 99th percentile QCT is within \inlinetodo{$\sim$16\%}.
For background traffic,
\figurename~\ref{fig:sim:longest-drop:avg-fct} shows that
the difference in average FCT (\figurename~\ref{fig:sim:longest-drop:avg-fct})
is within $\sim$8.8\%.
% \figurename~\ref{fig:sim:longest-drop:small-tail-fct} shows that
% the difference of 99th percentile FCT for small flows is within \inlinetodo{7.8\%}
% (except for the case with a query size of 80\%).

\mypara{Performance with heavy network load}
\sysname{} relies on redundant memory bandwidth for expelling packets from the over-allocated queues.
Thus, one may wonder whether \sysname{} can still bring benefits under a very high load.
In this part, we evaluate the performance of \sysname{} with heavy load.
The simulation settings are the same as those in \textsection\ref{sec:eval:sim}
except that the load of background traffic is 120\%.
\figurename~\ref{fig:sim:heavy-load} shows the performance of query traffic and background traffic.
With heavy network load,
\sysname{} can still achieve high performance
for both query traffic and background traffic.
This is because the congestion inside the network is usually unbalanced.
For example, with incast traffic pattern,
the congestion is mainly on down links,
and the up link may be idle.
As a result, there is still some redundant bandwidth for dropping over-allocated buffer.

\mypara{Impact of the buffer size}
% With smaller buffer, a BM should be faster to adapt the buffer allocations.
The buffer size can be smaller with higher-speed switch chips in the future.
In this part, we examine how the buffer size impacts the performance of \sysname{}.
The simulation settings are the same as those in \textsection\ref{sec:eval:sim}.
We vary the buffer size from 3.44KB per-port-per-Gbps (Intel Tofino)
to 9.6KB per-port-per-Gbps (Broadcom Trident2).
The load of background traffic is 40\% and the query size is 40\% of the buffer size.
\figurename~\ref{fig:sim:buffer-size} shows the performance with different buffer sizes.
We can observe that
\sysname{} can always bring some benefit with different buffer sizes.
For example,
\figurename~\ref{fig:sim:buffer-size:avg-qct}
shows that,
compared with DT,
\sysname{} can improve the average QCT by $\sim$36.7\% and $\sim$40.3\%
with 3.44KB and 9.6KB buffer-per-port-per-Gbps, respectively.

% \subsection{\sysname{} Deep Dive}
%
% \mypara{Impact of Transport Protocols}
%
% \mypara{Parameter Sensitivity}
%
% \mypara{Hardware Overhead}
%
% \mypara{Extreme Shallow Buffer}
%
% \mypara{Impact of HeadDrop Frequency}

\section{Related Work}\label{sec:rlwk}
% There is a long history on BM schemes.
% Generally, these BM schemes can be divided into two categories:
% preemptive schemes and non-preemptive schemes.
% Next, we will summarize them respectively.

\mypara{Preemptive BM}
Preemptive schemes (which are also called Pushout)
can arbitrarily overwrite or evict packets residing in the buffer.
They had been proved to be optimal, but were considered to be hard to implement.
SDR~\cite{TCOM84Pushout} accepts a packet whenever there is free buffer space
and either drops the arriving packet or swaps out an accepted packet when the buffer becomes full.
% The study of preemptive BM schemes can be originated to 1981,
% Thareja and Agrawala proposed Stationary Delayed Resolution (SDR) policy~\cite{1981Pushout, TCOM84Pushout}.
% SDR drops a packet whenever there is free buffer space.
% When the buffer becomes full, SDR can either drop the newly-arrived packet
% or swap out a previously-accepted packet.
% SDR assumes that traffic arrival follows a Markov process,
% and uses a search technique to find an optimal policy on deciding which packet to swap out/drop.
DoD~\cite{GLOBECOM91Pushout} proposed to purge the longest queue when the buffer becomes full.
% It had been shown to be optimal in terms of throughput.
% Wei \etal~\cite{GLOBECOM91Pushout} proposed Drop on Demand (DoD) policy,
% which purges the longest queue when buffer becomes full.
% This improves the practicality of the preemptive BM.
% They also showed that the policy is optimal
% in terms of switch throughput in many cases.
POT~\cite{JSAC95Pushout} pushes out an accepted packet
only when the length of corresponding queue for the newly arrived packet is below a certain threshold.
% Cidon \etal~\cite{JSAC95Pushout} showed that DoD is optimal for symmetric system.
% They proved that the optimal policy is push-out with threshold type (POT),
% where the action of pushing out another packet is performed
% only if the length of queue for newly-arrived packet is below some threshold.
LossPass~\cite{TCC22LossPass} proposes to evict packets of large flows
to make room for newly arrived small flows.
Some other work~\cite{CC96Pushout, CL97Pushout} studied the implementation of Pushout.
Choudhury and Hahne~\cite{CC96Pushout} proposed an implementation of Pushout
that can reduce the required memory footprint for maintaining queues.
% As finding the longest queue is difficult,
QPO~\cite{CL97Pushout} proposed to discard packets
in the near-longest queue, which is easy to be maintained.
However, it still entails complex enqueue operations.
Several studies~\cite{JSAC91Pushout, GLOBECOM93Pushout, TON94Pushout, JHSN94Pushout}
focused on Pushout schemes with multiple (loss) priorities.
% Their basic idea is to discard an ordinary/low-priority cell
% when a vital/high-priority cell arrives and seeing the queue saturated.
However, as analyzed in \textsection\ref{sec:back:bm},
these schemes can introduce unacceptable implementation overhead.
% Recently, LossPass~\cite{TCC22LossPass} proposes to evict packets of large flows
% to vacate buffer for newly arrived small flows.
% However, it requires the switch to differentiate between short and large flows
% and put them into different queues.

\mypara{Non-preemptive BM}
Non-preemptive schemes only drop packets before they enter into the buffer.
They are simple to implement.
% The related work mainly focuses on the condition of packet drop,
% aiming at achieving both fairness and efficiency.
Typically, they use threshold(s) to restrict the length of each queue.
SMXQ~\cite{TCOM78ST}, SMA~\cite{TCOM77ST}, SMQMA~\cite{TCOM80ST}, and Harmonic~\cite{INFOCOM02HarmonicBM}
use static threshold(s) to provide a minimum buffer guarantee and restrict the maximum queue length for fairness.
To adapt BM to varying traffic loads,
DT~\cite{INFOCOM96DT, TON98DT, TON02DT, GLOBECOM99DT}, PSPP~\cite{HSNMC02PSPP}, and DFT~\cite{IJCS07DFT}
dynamically adjust the threshold(s) based on the buffer occupancy and queue length.
Among them, DT~\cite{INFOCOM96DT, TON98DT, TON02DT} has become the de facto BM in commodity switch chips due to its simplicity.
% Dynamic threshold schemes~\cite{INFOCOM84BM, INFOCOM88BM, TON98DT, TON02DT, GLOBECOM99DT, INFOCOM02GA-FT, TON05GA-FT, IJCS07DFT}
% allow the threshold to be dynamically changed.
Other studies~\cite{INFOCOM84BM, INFOCOM88BM, INFOCOM02GA-FT, TON05GA-FT}
proposed to leverage adaptive control to dynamically adjust the thresholds based on the traffic load.
% DT~\cite{TON98DT, TON02DT} adapts the threshold based on the amount of free buffer size.
% Due to its simplicity and adaptivity, it has become the de-facto BM in commodity switch chips.
% Fan \etal~\cite{GLOBECOM99DT} proposed to adjust threshold based on the total buffer occupancy.
% Genetic Algorithm-Fuzzy Threshold (GA-FT)~\cite{INFOCOM02GA-FT, TON05GA-FT}
% leverages evolutionary computing and fuzzy logic to adapt the threshold to the traffic conditions.
% DFT~\cite{IJCS07DFT} employs a separate threshold for each queue
% and dynamically adjust the threshold based on both the free buffer size and the queue length.
In recent years,
as the buffer becomes increasingly insufficient,
the importance of BM has been brought to attention again.
Many BM schemes have been proposed to improve burst absorption and adaptivity.
% The majority of BMs focus on non-preemptive approaches
% as preemptive ones are considered to be hard to realize.
EDT~\cite{INFOCOM15Burst}, FAB~\cite{BS19FAB}, TDT~\cite{INFOCOM21TDT},
Smartbuf~\cite{IWQoS21Smartbuf}, and Protean~\cite{INFOCOM23Protean}
improve the ability of burst absorption by allocating more buffer to bursty traffic.
NDT~\cite{INFOCOM22NDT} leverages deep reinforcement learning to dynamically adjust the parameters of DT as the traffic pattern changes.
ABM~\cite{SIGCOMM22ABM} adapts the threshold based on both total buffer occupancy and queue drain time to achieve predictable burst tolerance,
performance isolation, and bounded buffer drain time simultaneously.
Reverie~\cite{NSDI24Reverie} uses the moving average queue length to improve burst absorption.
However, these non-preemptive schemes cannot quickly adjust the buffer allocation
as the over-allocated buffer can only be released by naturally sending out the traffic.
Credence~\cite{NSDI24Credence} makes packet drop decisions based on machine-learning-based predictions of future packet arrivals.
However, it requires ML algorithms in the dataplane, which can incur high overhead.
% Although not purposely designed for shared buffer management,
% AIFO~\cite{SIGCOMM21AIFO} also contains some BM mechanisms.
% AIFO \emph{proactively} allocates a fraction of buffer as headroom for absorbing bursts.
% In comparison,
% \sysname{} mainly relies on \emph{reactively} vacates packets
% when the buffer becomes over-allocated.

\section{Conclusion}\label{conc}
In this paper, we argue that today's BM requires agile adjustment of buffer allocation facing dynamic traffic,
while the agility of the de facto BM is limited by its non-preemptive nature.
% while the de facto BM, designed over two decades ago,
% now fails to meet the requirement
% due to the trends in switch chips and traffic characteristics.
% We argue that the underlying factor lies in its non-preemptive nature.
On the other hand, although preemptive BMs were considered as difficult to implement in history,
we find that the new advances in switch chips have
opened the door to realizing preemptive mechanisms.
We propose \sysname{},
which utilizes the redundant memory bandwidth to
quickly expel the packets for the over-allocated queues.
Experiment and simulation results
show that \sysname{} significantly outperforms non-preemptive BMs
in terms of burst absorption, performance isolation, and buffer choking mitigation.

\begin{acks}
    % We would like to thank our shepherd, Gaël Thomas,
    % and the anonymous reviewers for their helpful feedback.
    We would like to thank Yuxuan Li for the help for the evaluations on Design Compiler.
    This work is supported in part by
    \grantsponsor{nsfc-2023}{National Natural Science Foundation of China}{}
    under Grant No. \grantnum{nsfc-2023}{62372363}.
\end{acks}

%-------------------------------------------------------------------------------
\printbibliography{}

@book{SwitchBook,
    title={{The All-New Switch Book: The Complete Guide to LAN Switching Technology}},
    author={Seifert, Rich and Edwards, Jim},
    year={2008},
    publisher={Wiley},
    edition={2},
}

@book{NoCRouterArch,
    title={{Microarchitecture of Network-on-chip Routers}},
    subtitle={{A Designer's Perspective}},
    author={Dimitrakopoulos, Giorgos and Psarras, Anastasios and Seitanidis, Ioannis},
    volume={1025},
    year={2015},
    publisher={Springer}
}

@inproceedings{IMC17Burst,
    author = {Zhang, Qiao and Liu, Vincent and Zeng, Hongyi and Krishnamurthy, Arvind},
    title = {{High-resolution Measurement of Data Center Microbursts}},
    booktitle = {ACM IMC},
    year = {2017},
}

@inproceedings{IMC22Burst,
    author = {Ghabashneh, Ehab and Zhao, Yimeng and Lumezanu, Cristian and Spring, Neil and Sundaresan, Srikanth and Rao, Sanjay},
    title = {{A Microscopic View of Bursts, Buffer Contention, and Loss in Data Centers}},
    year = {2022},
    booktitle = {ACM IMC},
}

@inproceedings {OSDI16DDC,
    author = {Peter X. Gao and Akshay Narayan and Sagar Karandikar and Joao Carreira and Sangjin Han and Rachit Agarwal and Sylvia Ratnasamy and Scott Shenker},
    title = {{Network Requirements for Resource Disaggregation}},
    booktitle = {USENIX OSDI},
}

@inproceedings{CoNEXT21Floodgate,
    author = {Liu, Kexin and Tian, Chen and Wang, Qingyue and Zheng, Hao and Yu, Peiwen and Sun, Wenhao and Xu, Yonghui and Meng, Ke and Han, Lei and Fu, Jie and Dou, Wanchun and Chen, Guihai},
    title = {{Floodgate: Taming Incast in Datacenter Networks}},
    booktitle = {ACM CoNEXT},
    year = {2021},
}

@INPROCEEDINGS{HPSR01SharedMemory,
    author={Sundar Iyer and Ramana Rao Kompella and Nick McKeown},
    booktitle={IEEE HPSR},
    title={{Analysis of a Memory Architecture for Fast Packet Buffers}},
    year={2001},
}

@inproceedings{BS19Yahoo,
    author={Yihua He and Nitin Batta and Igor Gashinsky},
    title = {{Understanding Switch Buffer Utilization in CLOS Data Center Fabric}},
    booktitle = {Workshop on Buffer Sizing},
    year = {2019},
}

@inproceedings{BS19FAB,
    author = {Apostolaki, Maria and Vanbever, Laurent and Ghobadi, Manya},
    title = {{FAB: Toward Flow-Aware Buffer Sharing on Programmable Switches}},
    booktitle = {Workshop on Buffer Sizing},
    year = {2019},
}

@inproceedings{INFOCOM84BM,
    author={A. K. Thareja and S. K. Tripathi},
    booktitle={IEEE INFOCOM},
    title={{Buffer Sharing in Dynamic Load Environment}},
    year={1984},
}

@inproceedings{INFOCOM88BM,
    author={Tipper, D. and Sundareshan, M.K.},
    booktitle={IEEE INFOCOM},
    title={{Adaptive Policies for Optimal Buffer Management in Dynamic Load Environments}},
    year={1988},
}

@inproceedings{INFOCOM96DT,
    author={Abhijit K. Choudhury and Ellen L. Hahne},
    booktitle={IEEE INFOCOM},
    title={Dynamic Queue Length Thresholds in a Shared Memory ATM Switch},
    year={1996},
}

@inproceedings{INFOCOM98PingPong,
    author={Youngmi Joo and Nick McKeown},
    booktitle={IEEE INFOCOM},
    title={{Doubling Memory Bandwidth for Network Buffers}},
    year={1998},
}

@inproceedings{INFOCOM02HarmonicBM,
    author={Alexander Kesselman and Yishay Mansour},
    booktitle={IEEE INFOCOM},
    title={{Harmonic Buffer Management Policy for Shared Memory Switches}},
    year={2002},
}

@inproceedings{INFOCOM02GA-FT,
  author={Giuseppe Ascia and Vincenzo Catania and Daniela Panno},
    booktitle={IEEE INFOCOM},
    title={{An Efficient Buffer Management Policy based on an Integrated Fuzzy-GA Approach}},
    year={2002},
    pages={1042-1048},
}

@inproceedings{INFOCOM20BCC,
    author={Bai, Wei and Hu, Shuihai and Chen, Kai and Tan, Kun and Xiong, Yongqiang},
    booktitle={IEEE INFOCOM},
    title={{One More Config is Enough: Saving (DC)TCP for High-speed Extremely Shallow-buffered Datacenters}},
    year={2020},
}

@inproceedings{INFOCOM21TDT,
    author={Huang, Sijiang and Wang, Mowei and Cui, Yong},
    booktitle={IEEE INFOCOM},
    title={{Traffic-aware Buffer Management in Shared Memory Switches}},
    year={2021},
}

@inproceedings{INFOCOM22NDT,
    author={Wang, Mowei and Huang, Sijiang and Cui, Yong and Wang, Wendong and Liu, Zhenhua},
    booktitle={IEEE INFOCOM},
    title={{Learning Buffer Management Policies for Shared Memory Switches}},
    year={2022},
}

@inproceedings{INFOCOM23Protean,
    author={Almasi, Hamidreza and Vardekar, Rohan and Vamanan, Balajee},
    booktitle={IEEE INFOCOM},
    title={{Protean: Adaptive Management of Shared-Memory in Datacenter Switches}},
    year={2023},
}

@inproceedings{ICNP21FlashPass,
    author={Zeng, Gaoxiong and Qiu, Jianxin and Yuan, Yifei and Liu, Hongqiang and Chen, Kai},
    booktitle={{IEEE ICNP}},
    title={{FlashPass: Proactive Congestion Control for Shallow-buffered WAN}},
    year={2021},
}

@inproceedings{ASPLOS23Optimus-CC,
    author = {Song, Jaeyong and Yim, Jinkyu and Jung, Jaewon and Jang, Hongsun and Kim, Hyung-Jin and Kim, Youngsok and Lee, Jinho},
    title = {{Optimus-CC: Efficient Large NLP Model Training with 3D Parallelism Aware Communication Compression}},
    year = {2023},
    booktitle = {ACM ASPLOS},
}

@inproceedings{IWQoS21Smartbuf,
  author={Rezaei, Hamed and Almasi, Hamidreza and Vamanan, Balajee},
  booktitle={IEEE/ACM IWQoS},
  title={{Smartbuf: An Agile Memory Management for Shared-Memory Switches in Datacenters}},
  year={2021},
}

@inproceedings{GLOBECOM91Pushout,
    author={Wei, Sherry X. and Coyle, Edward J. Coyle and Hsiao, Man-Tung T.},
    booktitle={IEEE GLOBECOM},
    title={{An Optimal Buffer Management Policy for High-Performance Packet Switching}},
    year={1991},
}

@inproceedings{GLOBECOM93Pushout,
    author={Abhijit K. Choudhury and Ellen L. Hahne},
    booktitle={IEEE GLOBECOM},
    title={{Space Priority Management in a Shared Memory ATM Switch}},
    year={1993},
}

@inproceedings{GLOBECOM99DT,
    author={Ruixue Fan and Ishii, Alexander and Mark, Brian and Ramamurthy, G. and Qiang Ren},
    booktitle={IEEE GLOBECOM},
    title={{An Optimal Buffer Management Scheme with Dynamic Thresholds}},
    year={1999},
}

@inproceedings{HSNMC02PSPP,
  author={Ruey-Bin Yang and Yuan-Sun Chu and Ming-Cheng Liang and Cheng-Shong Wu},
  booktitle={IEEE International Conference on High Speed Networks and Multimedia Communication},
  title={{Dynamic Thresholds Buffer Management in a Shared Buffer Packet Switch}},
  year={2002},
}

@inproceedings{HotChips20Tofino2,
    author={Agrawal, Anurag and Kim, Changhoon},
    booktitle={Hot Chips},
    title={{Intel Tofino2 – A 12.9Tbps P4-Programmable Ethernet Switch}},
    year={2020},
}

@inproceedings{ISSS02Arbiter,
    author = {Eung S. Shin and Vincent J. J. Mooney III and George F. Riley},
    booktitle={ISSS},
    title={{Round-robin Arbiter Design and Generation}},
    year={2002},
}

@article{TON94Pushout,
    author={Leandros Tassiulas and Yao Chung Hung and Shivendra S. Panwar},
    journal={IEEE/ACM Transactions on Networking},
    title={{Optimal Buffer Control During Congestion in an ATM Network Node}},
    year={1994},
    month={8},
    volume={2},
    number={4},
    pages={374-386},
}

@article{TON98DT,
    author={Abhijit K. Choudhury and Ellen L. Hahne},
    journal={IEEE/ACM Transactions on Networking},
    title={{Dynamic Queue Length Thresholds for Shared-memory Packet Switches}},
    year={1998},
    volume={6},
    number={2},
    pages={130-140},
}

@article{TON02DT,
    author={Abhijit K. Choudhury and Ellen L. Hahne},
    journal={IEEE/ACM Transactions on Networking},
    title={{Dynamic Queue Length Thresholds for Multiple Loss Priorities}},
    year={2002},
    month={6},
    volume={10},
    number={3},
    pages={368-380},
}

@article{TON05GA-FT,
  author={Giuseppe Ascia and Vincenzo Catania and Daniela Panno},
  journal={IEEE/ACM Transactions on Networking},
  title={{An Evolutionary Management Scheme in High-Performance Packet Switches}},
  year={2005},
  volume={13},
  number={2},
  pages={262-275},
}

@article{TON21BCC,
    author={Bai, Wei and Hu, Shuihai and Chen, Kai and Tan, Kun and Xiong, Yongqiang},
    journal={IEEE/ACM Transactions on Networking},
    title={{One More Config is Enough: Saving (DC)TCP for High-Speed Extremely Shallow-Buffered Datacenters}},
    year={2021},
    volume={29},
    number={2},
    pages={489-502},
}

@article{JSAC91Pushout,
  author={Kroner, Hans and Hebuterne, Gerard and Boyer, Pierre and Gravey, Annie},
  journal={IEEE Journal on Selected Areas in Communications},
  title={{Priority Management in ATM Switching Nodes}},
  year={1991},
  volume={9},
  number={3},
  pages={418-427},
}

@article{JSAC95Pushout,
    author={Israel Cidon and Leonidas Georgiadis and Roch Guerin and Asad Khamisy},
    journal={IEEE Journal on Selected Areas in Communications},
    title={{Optimal Buffer Sharing}},
    year={1995},
    volume={13},
    number={7},
    pages={1229-1240},
}

@ARTICLE{TCC22LossPass,
  author={Kim, Gyuyeong and Lee, Wonjun},
  journal={IEEE Transactions on Cloud Computing},
  title={{LossPass: Absorbing Microbursts by Packet Eviction for Data Center Networks}},
  year={2022},
  volume={10},
  number={4},
  pages={2717-2728},
}

@article{CACM17Latency,
    author = {Barroso, Luiz and Marty, Mike and Patterson, David and Ranganathan, Parthasarathy},
    title = {{Attack of the Killer Microseconds}},
    journal = {Communications of the ACM},
    year = {2017},
    month = {3},
    pages = {48–54},
    numpages = {7},
}

@article{JHSN94Pushout,
    title={{A Simulation Study of Space Priorities in a Shared Memory ATM Switch}},
    author={Abhijit K. Choudhury and Ellen L. Hahne},
    journal={Journal of High Speed Networks},
    volume={9},
    number={2},
    pages={67--87},
    year={2000},
    publisher={IOS Press}
}

@article{CC96Pushout,
    author = {Abhijit K. Choudhury and Ellen L. Hahne},
    journal = {{Computer Communications}},
    number = {3},
    pages = {245-256},
    title = {{New Implementation of Multi-priority Pushout for Shared Memory ATM Switches}},
    volume = {19},
    year = {1996},
}

@article{TCOM77ST,
    author={Marc A. Rich and Mischa Schwartz},
    journal={IEEE Transactions on Communications},
    title={{Buffer Sharing in Computer-Communication Network Nodes}},
    year={1977},
    month={9},
    volume={25},
    number={9},
    pages={958-970},
}

@article{TCOM78ST,
    author={Marek I. Irland},
    journal={IEEE Transactions on Communications},
    title={{Buffer Management in a Packet Switch}},
    year={1978},
    month={3},
    volume={26},
    number={3},
    pages={328-337},
}

@article{TCOM80ST,
    author={Farouk Kamoun and Leonard Kleinrock},
    journal={IEEE Transactions on Communications},
    title={{Analysis of Shared Finite Storage in a Computer Network Node Environment Under General Traffic Conditions}},
    year={1980},
    month={7},
    volume={28},
    number={7},
    pages={992-1003},
}

@article{TCOM84Pushout,
    author={Ashok K. Thareja and Ashok K. Agrawala},
    journal={IEEE Transactions on Communications},
    title={{On the Design of Optimal Policy for Sharing Finite Buffers}},
    year={1984},
    volume={32},
    number={6},
    pages={737-740},
}

@ARTICLE{TC14MaximumFinder,
    author={Yuce, Bilgiday and Ugurdag, H. Fatih and Gören, Sezer and Dündar, Günhan},
    journal={IEEE Transactions on Computers},
    title={{Fast and Efficient Circuit Topologies for Finding the Maximum of n k-Bit Numbers}},
    year={2014},
    volume={63},
    number={8},
    pages={1868-1881},
}

@article{IJCS07DFT,
    author = {Gazi, B. and Ghassemlooy, Z.},
    title = {{Dynamic Buffer Management using Per-queue Thresholds}},
    journal = {International Journal of Communication Systems},
    volume = {20},
    number = {5},
    pages = {571-587},
    year = {2007}
}

@article{CL97Pushout,
    author={Yu-Sheng Lin and Shung, C.B.},
    journal={IEEE Communications Letters},
    title={{Quasi-Pushout Cell Discarding}},
    year={1997},
    month={9},
    volume={1},
    number={5},
    pages={146-148},
}

@article{COMST00BM,
    author={Mutlu Arpaci and John A. Copeland},
    journal={IEEE Communications Surveys \& Tutorials},
    title={{Buffer Management for Shared-memory ATM Switches}},
    year={2000},
    volume={3},
    number={1},
    pages={2-10},
}

@article{Micro99PPE,
    author={Gupta, Pankaj and McKeown, Nick},
    journal={IEEE Micro},
    title={{Designing and Implementing a Fast Crossbar Scheduler}},
    year={1999},
    volume={19},
    number={1},
    pages={20-28},
}

@article{PC09DoubleBinaryTree,
    author = {Peter Sanders and Jochen Speck and Jesper Larsson Träff},
    title = {{Two-tree Algorithms for Full Bandwidth Broadcast, Reduction and Scan}},
    journal = {Parallel Computing},
    volume = {35},
    number = {12},
    pages = {581-594},
    year = {2009},
}

@inproceedings{INFOCOM15Burst,
    author={Danfeng Shan and Wanchun Jiang and Fengyuan Ren},
    booktitle={IEEE INFOCOM},
    title={{Absorbing Micro-burst Traffic by Enhancing Dynamic Threshold Policy of Data Center Switches}},
    year={2015},
}

@inproceedings{ICDCS23PFC,
    author={Danfeng Shan and Yuqi Liu and Tong Zhang and Yifan Liu and Yazhe Tang and Hao Li and Peng Zhang},
    booktitle={IEEE ICDCS},
    title={{Less is More: Dynamic and Shared Headroom Allocation in PFC-enabled Datacenter Networks}},
    year={2023},
}

@inproceedings {NSDI16MQECN,
    author = {Wei Bai and Li Chen and Kai Chen and Haitao Wu},
    title = {{Enabling ECN in Multi-Service Multi-Queue Data Centers}},
    booktitle = {{USENIX} NSDI},
    year = {2016},
}

@inproceedings{NSDI22BFC,
    author = {Prateesh Goyal and Preey Shah and Kevin Zhao and Georgios Nikolaidis and Mohammad Alizadeh and Thomas E. Anderson},
    title = {{Backpressure Flow Control}},
    booktitle = {{USENIX NSDI}},
    year = {2022},
}

@inproceedings{NSDI23RDMA,
    author = {Wei Bai and Shanim Sainul Abdeen and Ankit Agrawal and Krishan Kumar Attre and Paramvir Bahl and Ameya Bhagat and Gowri Bhaskara and Tanya Brokhman and Lei Cao and Ahmad Cheema and Rebecca Chow and Jeff Cohen and Mahmoud Elhaddad and Vivek Ette and Igal Figlin and Daniel Firestone and Mathew George and Ilya German and Lakhmeet Ghai and Eric Green and Albert Greenberg and Manish Gupta and Randy Haagens and Matthew Hendel and Ridwan Howlader and Neetha John and Julia Johnstone and Tom Jolly and Greg Kramer and David Kruse and Ankit Kumar and Erica Lan and Ivan Lee and Avi Levy and Marina Lipshteyn and Xin Liu and Chen Liu and Guohan Lu and Yuemin Lu and Xiakun Lu and Vadim Makhervaks and Ulad Malashanka and David A. Maltz and Ilias Marinos and Rohan Mehta and Sharda Murthi and Anup Namdhari and Aaron Ogus and Jitendra Padhye and Madhav Pandya and Douglas Phillips and Adrian Power and Suraj Puri and Shachar Raindel and Jordan Rhee and Anthony Russo and Maneesh Sah and Ali Sheriff and Chris Sparacino and Ashutosh Srivastava and Weixiang Sun and Nick Swanson and Fuhou Tian and Lukasz Tomczyk and Vamsi Vadlamuri and Alec Wolman and Ying Xie and Joyce Yom and Lihua Yuan and Yanzhao Zhang and Brian Zill},
    title = {{Empowering Azure Storage with RDMA}},
    booktitle = {{USENIX NSDI}},
    year = {2023},
}

@inproceedings{NSDI23Burst,
    author = {Erfan Sharafzadeh and Sepehr Abdous and Soudeh Ghorbani},
    title = {Understanding the Impact of Host Networking Elements on Traffic Bursts},
    booktitle = {{USENIX NSDI}},
    year = {2023},
}

@inproceedings{NSDI24Credence,
    author = {Vamsi Addanki and Maciej Pacut and Stefan Schmid},
    title = {{Credence: Augmenting Datacenter Switch Buffer Sharing with ML Predictions}},
    booktitle = {USENIX NSDI},
    year = {2024},
}

@inproceedings{NSDI24Reverie,
    author = {Vamsi Addanki and Wei Bai and Stefan Schmid and Maria Apostolaki},
    title = {{Reverie: Low Pass Filter-Based Switch Buffer Sharing for Datacenters with RDMA and TCP Traffic}},
    booktitle = {USENIX NSDI},
    year = {2024},
}

@article{FB,
    author = {Maria Apostolaki and Vamsi Addanki and Manya Ghobadi and Laurent Vanbever},
    title = {{FB: A Flexible Buffer Management Scheme for Data Center Switches}},
    journal   = {CoRR},
    volume    = {abs/2105.10553},
    year      = {2021},
    url       = {https://arxiv.org/abs/2105.10553},
}

@techreport{1981Pushout,
    author={Ashok K. Thareja and Ashok K. Agrawala},
    title={{On the Design of Optimal Policy for Sharing Finite Buffers}},
    institution = {Department of Computer Science, University of Maryland},
    year={1981},
    month={7},
    number = {TR-1081},
}

@report{BroadcomSmartBuffer,
    author = {Das, Sujal and Sankar, Rochan},
    title = {{Broadcom Smart-Buffer Technology in Data Center Switches for Cost-Effective Performance Scaling of Cloud Applications}},
    type = {White Paper},
    institution = {Broadcom},
    year = {2012},
    month = {4},
}

@online{BroadcomTomahawkBuffer,
    title = {{Tomahawk}},
    url = {https://people.ucsc.edu/~warner/Bufs/tomahawk},
}

@report{BroadcomTomahawk4,
    author = {Wheeler, Bob},
    title = {{Tomahawk 4 Switch First to 25.6Tbps}},
    subtitle = {{Broadcom Doubles 400Gbps Ports With Unprecedented 512 Serdes}},
    type = {Microprocessor Report},
    institution = {The Linley Group},
    year = {2019},
    month = {12},
    url = {https://docs.broadcom.com/doc/12398014},
}

@online{BroadcomTomahawk5,
    title = {{Tomahawk 5 / BCM78900 Series}},
    subtitle = {{51.2 Tb/s StrataXGS® Tomahawk® 5 Ethernet Switch Series}},
    institution = {Broadcom},
    url = {https://www.broadcom.com/products/ethernet-connectivity/switching/strataxgs/bcm78900-series},
}

@report{BroadcomTrident3,
    author = {Arcilla, Alex and Palmer, Tony},
    title = {{Broadcom Trident 3 Platform Performance Analysis}},
    type = {White Paper},
    institution = {Broadcom},
    year = {2019},
    month = {5},
    url = {https://docs.broadcom.com/doc/12395356},
}

@manual{BCM88800TM,
    title = {{BCM88800 Traffic Management Architecture}},
    type = {Design Guide},
    organization = {Broadcom},
    year = {2021},
    month = {2},
    url = {https://docs.broadcom.com/doc/88800-DG1-PUB},
}

@report{CiscoNexus3100,
    title = {{Cisco Nexus 3100 Platform Switch Architecture}},
    type = {White Paper},
    institution = {Cisco},
    year = {2013},
    month = {10},
    url = {https://www.cisco.com/c/dam/assets/events/i/interop-ny-Cisco-Nexus3100-Switch-Architecture-Whitepaper.pdf},
}

@manual{CiscoNexus9000ConfigGuide,
    title = {{Cisco Nexus 9000 Series NX-OS Quality of Service Configuration Guide, Release 6.x}},
    organization = {Cisco},
    date = {2020-04-22},
    url = {https://www.cisco.com/c/en/us/td/docs/switches/datacenter/nexus9000/sw/6-x/qos/configuration/guide/b_Cisco_Nexus_9000_Series_NX-OS_Quality_of_Service_Configuration_Guide.pdf},
}

@online{CiscoSiliconOneG202,
    title = {{Cisco Silicon One G202 Data Sheet}},
    organization = {Cisco},
    url = {https://www.cisco.com/c/en/us/solutions/collateral/silicon-one/silicon-one-g202-ds.html},
}

@report{CiscoSiliconOneHBM,
    title = {{Converged Web Scale Switching and Routing Becomes a Reality}},
    subtitle = {{Cisco Silicon One and HBM Memory Change the Paradigm}},
    type = {White Paper},
    institution = {Cisco},
    year = {2020},
    url = {https://www.cisco.com/c/dam/en/us/solutions/collateral/silicon-one/white-paper-sp-hybrid-buffer-architecture.pdf},
}

@online{MellanoxDT,
    title = {{Understanding the Alpha Parameter in the Buffer Configuration of Mellanox Spectrum Switches}},
    organization = {Mellanox},
    year = {2018},
    month = {12},
    url = {https://support.mellanox.com/s/article/howto-configure-mellanox-spectrum-switch-for-lossless-roce},
}

@online{NVIDIASpectrum4,
    title = {{NVIDIA Spectrum-4}},
    subtitle = {{51.2 Tb/s Ethernet Switch ASIC}},
    organization = {NVIDIA},
    year = {2022},
    month = {4},
    url = {https://nvdam.widen.net/s/v6skdhzdrh/ethernet-switches-product-brief-gtc22-spring-spectrum-4-2169045-r8},
}

@manual{JuniperQFXTM,
    title = {{Traffic Management User Guide (QFX Series Switches and EX4600 Switches)}},
    type = {User Guide},
    organization = {Juniper},
    year = {2024},
    month = {6},
    url = {https://www.juniper.net/documentation/us/en/software/junos/traffic-mgmt-qfx/traffic-mgmt-qfx.pdf},
}

@report{ExtremeBuffer,
    title = {{Congestion Management and Buffering in Data Center Networks}},
    institution = {Extreme Networks},
    type = {White Paper},
    year = {2014},
    url = {http://learn.extremenetworks.com/rs/extreme/images/Congestion-Management-and-Buffering-wp.pdf},
    % urldate = {2022-06-05},
}

@report{Arista7050X3,
    title = {{Arista 7050X3 Series Switch Architecture}},
    type = {White Paper},
    institution = {Arista},
    year = {2013},
    month = {10},
    url = {https://www.arista.com/assets/data/pdf/Whitepapers/7050X3_Architecture_WP.pdf},
}

@report{Stanford01SharedMemory,
    title={{Techniques for Fast Shared Memory Switches}},
    author={Iyer, Sundar and McKeown, Nick},
    type={{Stanford HPNG Technical Report}},
    number={TR01-HPNG-081501},
    year={2001},
}

@online{AlveoU50,
    title={{AMD Alveo U50 Data Center Accelerator Card}},
    url = {https://www.amd.com/en/products/accelerators/alveo/u50/a-u50-p00g-pq-g.html},
}

@online{Vivado,
    title={{AMD Vivado Design Suite}},
    url = {https://www.amd.com/en/products/software/adaptive-socs-and-fpgas/vivado.html},
}

@online{DesignCompiler,
    title={{Synopsys Design Compiler}},
    url = {https://www.synopsys.com/implementation-and-signoff/rtl-synthesis-test/dc-ultra.html},
}

@online{45nmASICLib,
    title={{FreePDK45}},
    url = {https://eda.ncsu.edu/freepdk/freepdk45/},
}

@online{pktgen-dpdk,
    title = {{Pktgen-DPDK}},
    url = {https://github.com/pktgen/Pktgen-DPDK},
}

@online{BroadcomTomahawk5Ship22,
    title = {{Broadcom Ships Tomahawk 5, Industry’s Highest Bandwidth Switch Chip to Accelerate AI/ML Workloads}},
    url = {https://investors.broadcom.com/news-releases/news-release-details/broadcom-ships-tomahawk-5-industrys-highest-bandwidth-switch},
    date = {2022-08-16},
    organization = {Broadcom},
}

@online{BroadcomTomahawk5Ship23,
    title = {{Broadcom Now Shipping World’s First 51.2 Tbps Switch in Production Volume}},
    url = {https://investors.broadcom.com/news-releases/news-release-details/broadcom-now-shipping-worlds-first-512-tbps-switch-production},
    date = {2023-03-15},
    organization = {Broadcom},
}

@online{BroadcomDCSwitchChip21,
    title = {{Driving the Data Center into the Future}},
    url = {https://www.broadcom.com/blog/driving-the-data-center-into-the-future},
    date = {2021-06-30},
    organization = {Broadcom},
}

@online{CiscoSiliconOneG202Announce,
    title = {{Cisco Silicon One Breaks the 51.2 Tbps Barrier}},
    author = {Rakesh Chopra},
    url = {https://blogs.cisco.com/sp/cisco-silicon-one-breaks-the-51-2-tbps-barrier},
    date = {2023-06-20},
    organization = {Cisco},
}

@online{NVIDIASpectrum4Announce,
    title = {{NVIDIA Announces Spectrum High-Performance Data Center Networking Infrastructure Platform}},
    url = {https://nvidianews.nvidia.com/news/nvidia-announces-spectrum-high-performance-data-center-networking-infrastructure-platform},
    date = {2022-03-22},
    organization = {NVIDIA},
}

@online{NVIDIASpectrum4Datasheet,
    title = {{NVIDIA Spectrum-4 Datasheet}},
    url = {https://resources.nvidia.com/en-us-accelerated-networking-resource-library/ethernet-switches-pr?xs=425229#page=1},
    date = {2022-04-22},
    organization = {NVIDIA},
}

@online{MarvellTeralynx10Announce,
    title = {{Marvell Announces Cloud-Optimized 51.2 Tbps Networking Platform for AI/ML and Data Center Networks}},
    url = {https://www.marvell.com/company/newsroom/marvell-next-gen-data-center-512t-networking-solution.html},
    date = {2023-03-02},
    organization = {Marvell},
}

@inproceedings{SIGCOMM10DCTCP,
    author = {Alizadeh, Mohammad and Greenberg, Albert and Maltz, David A. and Padhye, Jitendra and Patel, Parveen and Prabhakar, Balaji and Sengupta, Sudipta and Sridharan, Murari},
    title = {{Data Center TCP (DCTCP)}},
    booktitle = {ACM SIGCOMM},
    year = {2010},
}

@inproceedings{SIGCOMM15Jupiter,
    author = {Singh, Arjun and Ong, Joon and Agarwal, Amit and Anderson, Glen and Armistead, Ashby and Bannon, Roy and Boving, Seb and Desai, Gaurav and Felderman, Bob and Germano, Paulie and Kanagala, Anand and Provost, Jeff and Simmons, Jason and Tanda, Eiichi and Wanderer, Jim and H\"{o}lzle, Urs and Stuart, Stephen and Vahdat, Amin},
    title = {{Jupiter Rising: A Decade of Clos Topologies and Centralized Control in Google's Datacenter Network}},
    booktitle = {ACM SIGCOMM},
    year = {2015},
}

@inproceedings{SIGCOMM16RDMA,
    author = {Guo, Chuanxiong and Wu, Haitao and Deng, Zhong and Soni, Gaurav and Ye, Jianxi and Padhye, Jitu and Lipshteyn, Marina},
    title = {{RDMA over Commodity Ethernet at Scale}},
    booktitle = {ACM SIGCOMM},
    year = {2016},
}

@inproceedings{SIGCOMM19HPCC,
    author = {Li, Yuliang and Miao, Rui and Liu, Hongqiang Harry and Zhuang, Yan and Feng, Fei and Tang, Lingbo and Cao, Zheng and Zhang, Ming and Kelly, Frank and Alizadeh, Mohammad and Yu, Minlan},
    title = {{HPCC: High Precision Congestion Control}},
    booktitle = {ACM SIGCOMM},
    year = {2019},
}

@inproceedings{SIGCOMM20Swift,
    author = {Kumar, Gautam and Dukkipati, Nandita and Jang, Keon and Wassel, Hassan M. G. and Wu, Xian and Montazeri, Behnam and Wang, Yaogong and Springborn, Kevin and Alfeld, Christopher and Ryan, Michael and Wetherall, David and Vahdat, Amin},
    title = {{Swift: Delay is Simple and Effective for Congestion Control in the Datacenter}},
    booktitle = {ACM SIGCOMM},
    year = {2020},
}

@inproceedings{SIGCOMM21AIFO,
    author = {Yu, Zhuolong and Hu, Chuheng and Wu, Jingfeng and Sun, Xiao and Braverman, Vladimir and Chowdhury, Mosharaf and Liu, Zhenhua and Jin, Xin},
    title = {{Programmable Packet Scheduling with a Single Queue}},
    booktitle = {ACM SIGCOMM},
    year = {2021},
}

@inproceedings{SIGCOMM22ABM,
    author = {Addanki, Vamsi and Apostolaki, Maria and Ghobadi, Manya and Schmid, Stefan and Vanbever, Laurent},
    title = {{ABM: Active Buffer Management in Datacenters}},
    booktitle = {ACM SIGCOMM},
    year = {2022}
}

% \input{data/appendix}

%%%%%%%%%%%%%%%%%%%%%%%%%%%%%%%%%%%%%%%%%%%%%%%%%%%%%%%%%%%%%%%%%%%%%%%%%%%%%%%%
\end{document}